\documentclass[fleqn,usenatbib]{mnras}

\usepackage{newtxtext,newtxmath}

\usepackage[T1]{fontenc}
\usepackage{ae,aecompl}

\usepackage{graphicx}	
\usepackage{amsmath}	
\usepackage{braket}
\usepackage{mathtools}
\usepackage{amsmath}

\usepackage{multirow}
\usepackage{rotating}
\usepackage{tabularx}
\usepackage{subcaption}
\usepackage{outlines}
\usepackage{booktabs}

\title[Redshifted XLF method for AGN]{A New Method to Determine X-ray Luminosity Functions of AGN and their Evolution with Redshift}

\author[A. Alqasim et al.]{
Ahlam Alqasim,$^{1}$\thanks{E-mail: ahlam.alqasim.17@ucl.ac.uk}
Mat J. Page,$^{1}$
\\
$^{1}$UCL Mullard Space Science Laboratory, Holmbury Hill Road, Dorking, Surrey, RH5 6NT, UK
}

\date{Accepted 2022 December 22. Received 2022 December 15; in original form 2022 February 8}

\pubyear{2022}

\begin{document}
\label{firstpage}
\pagerange{\pageref{firstpage}--\pageref{lastpage}}
\maketitle

\begin{abstract}
Almost all massive galaxies today are understood to contain supermassive black holes (SMBH) at their centers. 
SMBHs grew by accreting material from their surroundings, emitting X-rays as they did so. 
X-ray Luminosity Functions (XLFs) of Active Galactic Nuclei (AGN) have been extensively studied in order to understand the AGN population's cosmological properties and evolution. 
We present a new fixed rest-frame method to achieve a more accurate study of the AGN XLF evolution over cosmic time. 
Normally, XLFs are constructed in a fixed observer-frame energy band, which can be problematic because it probes different rest-frame energies at different redshifts. 
In the new method, we construct XLFs in the fixed rest-frame band instead, by varying the observed energy band with redshift. 
We target a rest-frame 2$-$8 keV band using \textit{XMM-Newton} and \textit{HEAO 1} X-ray data, with 7 observer-frame energy bands that vary with redshift for $0 < z < 3$.
We produce the XLFs using two techniques; one to construct a binned XLF, and one using a Maximum Likelihood (ML) fit, which makes use of the full unbinned source sample. 
We find that our ML best-fit pure luminosity evolution (PLE) results for both methods are consistent with each other, suggesting that performing XLF evolution studies with the high-redshift data limited to high-luminosity AGN is not very sensitive to the choice of fixed observer-frame or rest-frame energy band, which is consistent with our expectation that high-luminosity AGN typically show little absorption. 
We have demonstrated the viability of the new method in measuring the XLF evolution.
\end{abstract}

\begin{keywords}
galaxies: active --
quasars: supermassive black holes -- 
X-rays: galaxies -- 
galaxies: nuclei -- 
galaxies: luminosity function -- 
methods: data analysis
\end{keywords}



\section{Introduction}\label{sec:intro}

It is now understood that almost all massive galaxies today contain Supermassive Black Holes (SMBH) at their centers \citep{kormendy2013smbh}, with masses ranging between $\sim 10^6 - 10^9 M_{\odot}$. 
Their massive growth is mainly due to accretion of matter from their surroundings, shining as Quasi-Stellar Objects (QSOs) and emitting X-rays as they did so. 
QSOs belong to a larger population of Active Galactic Nuclei (AGN). 
SMBHs can also grow due to other processes such as black hole mergers and the tidal capture of stars. 
Most of them have been observed to be dormant in present-day galaxies, exhibiting luminosities largely below their Eddington limits, leaving only $\sim 10\%$ of galaxies hosting AGN \citep{ho2008nuclear}. 
Hence, AGN studies are very important in understanding the evolution of galaxies over cosmic time, mainly because they are strongly linked to the growth of SMBHs and can thus tell us useful information about the accretion history of the universe \citep{brandt2015cosmic}. 
It is now also well-known that the evolution of SMBHs and the evolution of their host galaxies are strongly connected, or that they co-evolve \citep{symeonidis2013agn}. 
AGN feedback plays a significant role in quenching star formation and stopping the growth of their host galaxy after a certain point \citep{bongiorno2016agnfeedback}, so understanding how they evolve with time can provide very useful insights into the evolution of galaxies.

Most of the observed luminosity in AGN is radiated in the optical-UV, but the X-ray flux of AGN shows the fastest variability on short timescales in all of the wavelength ranges \citep{netzer2013physics}. 
This suggests that the X-rays originate from a small region close to the central object, now known to be a SMBH \citep{mushotzky1993x}. 

Studies of Seyfert galaxies showed that one of the most plausible physical processes driving the X-ray emission was inverse Compton radiation \citep{haardt1991two}. 
In this scenario, the intrinsic high energy X-rays seen in AGN originate from a hot corona plasma (consisting of relativistic electrons) close to the accretion disk \citep{beckmann2013active}. 
The hot corona scatters the optical-UV photons coming from the inner regions of the accretion disk to X-ray energies. 
This mechanism drives the shape observed in the X-ray spectra of AGN \citep{netzer2013physics}. 
Further out from the central engine of the AGN is a thick torus surrounding the central SMBH and accretion disk, responsible for obscuring the X-ray emission due to photoelectric absorption \citep{morrison1983photoelectric, antonucci1993unified}. 
Seyfert I galaxies give a clear view of the active nucleus because the line of sight is unobstructed, while Seyfert II galaxies have a line of sight that is obscured by the torus, causing them to appear to have less evidence of activity. 
\textit{XMM-Newton} surveys particularly helped constrain the evolution of X-ray absorption, as well as other physical properties of AGN, by studying faint X-ray sources \citep{hasinger2001xmm}. 

The luminosity function (LF) of galaxies is generally defined as the number of objects per unit volume (i.e. Mpc$^3$) per unit logarithmic luminosity interval. 
For AGN, the best model shape that describes their X-ray LF (XLF) is a double power-law modified by a factor for evolution \citep[e.g.][]{boyle1988qsoevolution,miyaji2000rosatxlf}.
XLFs of AGN can be constructed using extensive X-ray surveys in order to understand their cosmological properties and study their evolution over cosmic time. 
Many techniques have been published to quantify the cosmological evolution of AGN, including using the $\braket{V/V_{max}}$ method \citep{schmidt1968space},
the $V_{e}/V_{a}$ method \citep{avni1979simultaneous} for combined samples,
Monte Carlo simulations \citep{cristiani1990composite}, 
and the $1/V_{a}$ method \citep{maccacaro1991properties}.
The $\braket{V/V_{a}}$ (or $\braket{V/V_{max}}$) approach is mainly used for determining whether there is evolution with redshift, while the $1/V_{a}$ (or $1/V_{max}$) approach is used for calculating a binned luminosity function.
The $1/V_{a}$ is more commonly used because it is simpler, incorporating a binned differential luminosity function within a redshift interval. 
Improved methods to $1/V_{a}$ for constructing binned XLFs have also been demonstrated \citep{page2000improved, cara2008method}.

Some works have shown that the cosmic evolution of AGN is consistent with models in which the luminosity varies with redshift, e.g. Pure Luminosity Evolution (PLE) model \citep{barger2005cosmic}.
In such models, the shape of the XLF evolution remains the same and the luminosity evolution causes the model curve to simply shift right or left on the luminosity plane as the XLF evolves.
Other models try to explain the XLF evolution by varying the density with redshift, e.g. Pure Density Evolution (PDE) model \citep{fotopoulou2016xlf}.
In such models, the XLF shape also remains the same and the density evolution causes the model curve to simply shift up or down on the density plane as the XLF evolves.
Other works have proposed models that combine both luminosity and density evolution, e.g. Independent Luminosity Density Evolution (ILDE) model \citep{yencho2009hardxlfagn} and Luminosity and Density Evolution (LADE) model \citep{aird2010hardxlf}.
In such models, the XLF shape is kept the same, and the combined luminosity and density evolution causes the model curve to be shifted in any direction across the luminosity-density plane as the XLF evolves.
Some studies favor a more complicated model in which the shape of the XLF evolution changes while also varying the density and luminosity, e.g. Luminosity-Dependent Density Evolution (LDDE) model \citep{hasinger2005luminosity, ueda2014evo}.
In such models, the curves are not only shifted in any direction, but their shapes can also change when moving around the luminosity-density plane as the XLF evolves.
Most of these models include a critical redshift $z_{c}$ value (also referred to as cutoff redshift), after which the XLF evolution either changes or stops \citep{fotopoulou2016xlf}. 
Some papers also use a Bayesian approach to explore the AGN evolution in a model-independent way \citep{georgakakis2015xlf, fotopoulou2016xlf}.

Previous XLF studies of AGN show that the number of AGN per unit volume per unit luminosity has been observed to change strongly with redshift \citep{ebrero2009xmm}.
One of the main issues to be addressed in XLF studies of AGN (especially with \textit{Chandra} and \textit{XMM-Newton} surveys) is the impact of absorption, which can suppress the observed X-ray flux (especially for redshifts $z \leq 1$), and is even more problematic for Compton-thick sources \citep{aird2015nustar}. 
Most studies try to correct for this absorption and model its evolution with redshift. 
Normally, luminosity functions for AGN are constructed in a fixed observed energy band, and there is still no consensus on what the best approach is to model how they evolve with redshift when looking at the rest-frame energy bands. 
How an observed energy band, $E_{obs}(z)$, relates to the rest-frame energy band, $E_{rf}$, for a given redshift $z$ is described as
	
	\begin{equation}\label{eq:observedenergy}
	E_{obs}(z) = \frac{E_{rf}}{1+z}.
	\end{equation}

The most common X-ray bands normally used to study XLFs of AGN are 0.5$-$2 keV \citep{miyaji2000softxlf, hasinger2005luminosity, ebrero2009xmm} and 2$-$10 keV \citep{aird2010hardxlf, miyaji2015xlfevolution, georgakakis2015xlf}.
The 2$-$8 keV band has also been studied in some cases instead of 2$-$10 keV \citep{barger2005cosmic, silverman2008luminosity}. 
Some other studies have focused on the 5$-$10 keV band \citep{fotopoulou2016xlf} to avoid correcting for the absorbed part of the AGN spectrum. 
Since hard X-rays ($\sim$ E $>$ 2 keV) are significantly less affected by absorption, the soft X-ray band (0.5$-$2 keV) should mainly sample unabsorbed AGN (and the derived XLF should only include the unabsorbed population). 
However, since the rest-frame energy band changes with redshift, the population will include absorbed AGN at higher redshifts, which affects the binned luminosity function because the K-correction doesn't take that effect into account. 
As a result, the sample will start to gain more absorbed sources as you move up in the rest-frame energy band with redshift, even if the study is conducted in a harder observed X-ray band \citep{aird2015nustar}. 
This problem was mitigated to some degree by \cite{cowie2003xlfzevol} and \cite{barger2005cosmic}, who used the 2$-$8 keV observer-frame to study lower redshift sources ($z<\sim1.5$), and used the flux from the 0.5$-$2 keV observer-frame to calculate the 2$-$8 keV luminosity in the rest-frame for the high redshift sources ($z>\sim1.5$).

In this paper, we present a new method that aims to tackle this problem by varying the observed energy band with redshift, allowing us to fix the X-ray energy band in the rest-frame. 
This eliminates the need to model the redshift-dependence of X-ray absorption from material surrounding the SMBHs. 
We make use of X-ray data from \textit{XMM-Newton} and \textit{HEAO 1} in this work to produce X-ray luminosity functions in the fixed observed band (the standard method) and the fixed rest-frame band (the new method) for the 2$-$8 keV band. 
The fixed rest-frame band requires the analysis of several observer-frame bands that correspond to each redshift bin, which will be described in more detail in Section~\ref{sec:redshiftmethod}.

This paper is structured as follows: 
in Section~\ref{sec:redshiftmethod}, we introduce the new method used to produce the fixed rest-frame XLF along with the corresponding redshifted observed energy bands.
We then present the X-ray data used in this sample and its selection criteria in Section~\ref{sec:data} for both the fixed observed band and the fixed rest-frame band. 
In Section~\ref{sec:opticalid}, we describe the optical identifications used to study the completeness of the X-ray sample and obtain redshifts for them. 
In Section~\ref{sec:xlfmethod}, we describe the two techniques used to compute our XLFs; one using the method of \cite{page2000improved} to construct a binned XLF, and one using a Maximum Likelihood (ML) fit on the full unbinned source sample. 
Both techniques are computed for the standard method (fixed observed band) and the new method (fixed rest-frame band) introduced in this paper. 
We present the results of the XLFs in Section~\ref{sec:results}.
In Section~\ref{sec:discussion}, we discuss the performance of the new method compared with the standard method, as well as with previous XLF studies. 
The conclusions of this work are then summarized in Section~\ref{sec:conclusions}.

The cosmological parameters that were assumed in this paper were $H_{0} = 70$ km s$^{-1}$ Mpc$^{-1}$, $\Omega_{M} = 0.3$ and $\Omega_{\Lambda} = 0.7$.

\section{The Fixed Rest-frame XLF Method}\label{sec:redshiftmethod}
 	
\begin{figure}
		\centering
  		\includegraphics[width=\linewidth]{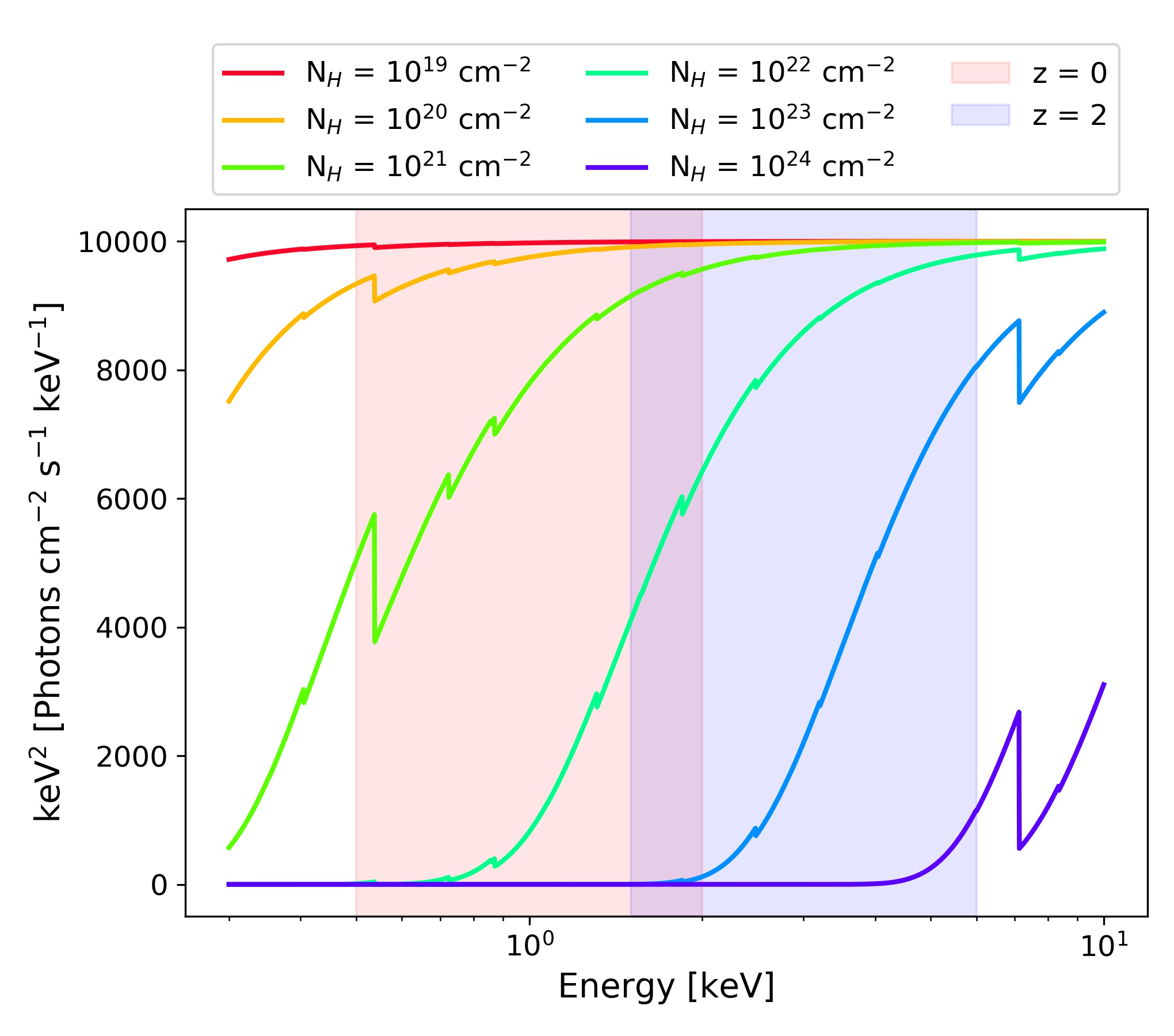}
  		\caption{AGN model spectrum in the rest-frame, constructed for a wide range of column densities $N_{H}$. The spikes correspond to photoelectric absorption edges, which occur when the X-ray emission passes through a highly absorbing material, usually the torus in the case of AGN.}
  		\label{fig:agnmodel_allnh}
	\end{figure}

In this section, we describe our new method that allows us to vary the observed energy band with redshift, and to fix the X-ray energy band in the rest-frame.
Fig.~\ref{fig:agnmodel_allnh} shows AGN model spectra in the rest-frame, constructed using PyXspec\footnote{\protect\url{https://heasarc.gsfc.nasa.gov/xanadu/xspec/python/html/index.html}} (the Python interface for Xspec).
The model curves are produced assuming a model with a cold photoelectric absorber and a powerlaw component with $\Gamma = 2.0$. 
We explored a wide range of column densities $N_{H}$ to account for AGN that display different levels of intrinsic absorption.
The spikes correspond to photoelectric absorption edges, which occur when the X-ray emission passes through a highly absorbing material, usually the torus in the case of AGN.
Looking at Fig.~\ref{fig:agnmodel_allnh}, if we study an AGN at 0.5$-$2 keV in a fixed observed band, a source at redshift $z=0$ (red shaded region) would be in the correct rest-frame energy band. 
However, if the source is redshifted at $z=2$ (blue shaded region), the part of the spectrum that is observed corresponds to 1.5$-$6 keV in the rest-frame. 
This means that we are inherently looking at very different parts of the AGN rest-frame spectrum at different redshifts. 
As a result, studying AGN in a fixed observed energy band can be problematic since it probes very different rest-frame energies at different redshifts.

To construct a fixed rest-frame XLF for a given survey, a fixed rest-frame energy band $E_{rf}$ must be chosen to cover a redshift range $z_{i} < z < z_{f}$. 
The redshift range determines what redshifted, observed energy bands $E_{obs}(z)$ will be used to generate images and sourcelists for the X-ray data.
In this new method, one can choose to do this in as many $E_{obs}(z)$ bands as desired, depending on how many redshift intervals are chosen. 
The more $E_{obs}(z)$ bands there are, the more precise the fixed rest-frame XLF will be, but a good middle ground will need to be established to minimize computational time spent depending on how large the data sample is.
Once the $E_{obs}(z)$ bands are determined, a sourcelist is produced for each given $E_{obs}(z)$ band at the end of the data reduction process. 
The X-ray fluxes obtained from the sourcelists are used to convert to X-ray luminosity.
In this work, the X-ray luminosity is for the 2$-$8 keV rest-frame band, and is corrected for Galactic absorption, but not for absorption occurring within the AGN and its host galaxy. 
As can be seen in Fig.~\ref{fig:agnmodel_allnh}, this energy band will only be sensitive to sources with column densities of up to $10^{23}$~cm$^{-2}$.
Since the degree of attenuation due to photoelectric absorption is strongly energy dependent (as shown in Fig.~\ref{fig:agnmodel_allnh}), it is important to use a fixed rest-frame energy band when producing our X-ray sourcelists.
The sourcelists are filtered to only contain sources with measured redshifts corresponding to the $E_{obs}(z)$ band of the data sample.
A separate flux limit $F_{lim}$ is then derived for each $E_{obs}(z)$ band. 
So in essence, the fixed rest-frame XLF uses observed bands that vary with redshift, and uses a flux limit that is not constant but is also a function of redshift $F_{lim}(z)$.

	\subsection{Binned XLF}\label{subsec:zmethodbinned}
	
\begin{figure}
    	\centering
      	\includegraphics[width=\linewidth, trim={0.5cm 0.5cm 3cm 2cm}, clip]{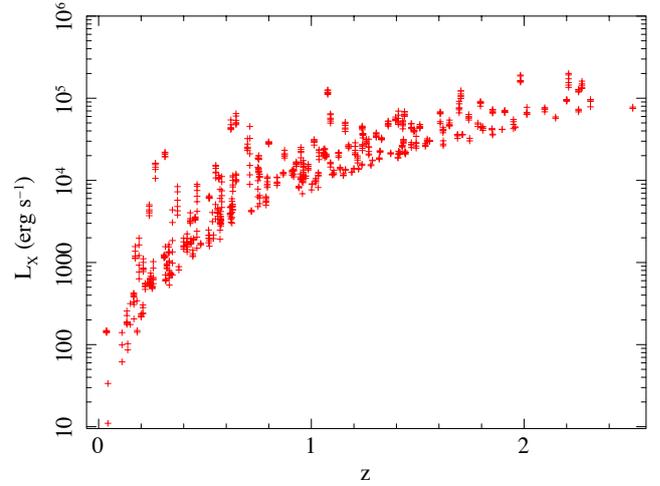}
      	\caption{Plot of Redshift vs. Luminosity ($L_{X} - z$ plane) of the XMS survey in the fixed rest-frame 2$-$8 keV band.}
     	\label{fig:redlum}
    	\end{figure}

    When applying the new method to a binned XLF (see Section~\ref{subsec:binnedxlf} for more details), redshift bins $z_{bin}$ first need to be determined. 
    A plot of the $L_{X} - z$ plane is made for the targeted $E_{rf}$, where $L_{X}$ is the X-ray luminosity of the sources, shown in Fig.~\ref{fig:redlum}.
    This figure shows the redshift distribution of the data used in this paper and at which redshifts most AGN with luminosities lie in at 2$-$8 keV.
    Looking at the diagram, we have enough sources from our sample spread over the luminosity-redshift plane to use bins of $\sim$ 0.5 in redshift.

    Once the redshift bins are determined, the midpoint of each redshift bin $z_{mid}$ is then used in equation (\ref{eq:observedenergy}) with $z = z_{mid}$ to determine what the corresponding $E_{obs}(z)$ band will be. 
    The minimum number of $E_{obs}(z)$ bands needed will be set by the number of $z_{bin}$ chosen for the XLF. 
    In principle, the number of $E_{obs}(z)$ bands need not be restricted by the number of redshift bins used. 
    Each $z_{bin}$ can include multiple $E_{obs}(z)$ bands that correspond to smaller redshift intervals within the same bin.
    For each $z_{bin}$, an XLF is then produced using the data from its corresponding $E_{obs}(z)$ band (or set of bands). 
    For the binned XLF, $F_{lim}(z)$ is a function that maps the discrete redshift intervals to discrete flux limit values, and each $E_{obs}(z)$ will have its own flux limit.

	\subsection{Model-fitted XLF}\label{subsec:zmethodfitted}  
	
    When applying the new method to produce a model fitted XLF using the ML technique (see Section~\ref{subsec:mlfit} for more details), the number of $E_{obs}(z)$ bands used is not determined by any redshift binning.
    The more $E_{obs}(z)$ bands used, the more precisely the energy range is fixed in the rest-frame for each source for the XLF.
    As was the case in the binned XLF, $F_{lim}(z)$ is used to derive a flux limit for each $E_{obs}(z)$ band used for the model-fitted XLF.

	\subsection{Redshifted Energy Bands}\label{subsec:energybands}  

	For this work, a fixed rest-frame energy band of 2$-$8 keV was chosen to cover a redshift range of $0 < z < 3$.
    We use redshift intervals of 0.5 for the binned XLF, with good coverage over the $L_{X} - z$ plane (see Fig.~\ref{fig:redlum}).
    To obtain the redshifted, observed energy bands in which sourcelists are produced, the midpoint of each redshift interval was used, giving a total of 6 redshifted energy bands (see Table \ref{table:observedbands}).
        
        \setlength{\extrarowheight}{3pt}
        \begin{table}
        \caption{The $E_{obs}(z)$ bands for which images and sourcelists are produced to construct the fixed rest-frame XLF. $z_{bin}$ gives the redshift interval, for which the midpoint ($z_{mid}$) is used to determine the observer-frame energy range ($E_obs(z)$) that corresponds to $2-8$ keV in the rest-frame band.}
        \centering
        \begin{tabular}{ccc}
            \hline
            $z_{bin}$ & $z_{mid}$ & $\mathrm{E_{obs}}(z)$ \\
             & & [eV] \\
            \addlinespace[4pt]
            \hline
            0.0 $-$ 0.5 & 0.25 & 1600 $-$ 6400 \\
            0.5 $-$ 1.0 & 0.75 & 1142 $-$ 4571 \\
            1.0 $-$ 1.5 & 1.25 & 888  $-$ 3555 \\
            1.5 $-$ 2.0 & 1.75 & 727  $-$ 2909 \\
            2.0 $-$ 2.5 & 2.25 & 615  $-$ 2461 \\
            2.5 $-$ 3.0 & 2.75 & 533  $-$ 2133 \\
            \addlinespace[4pt]
            \hline
            \end{tabular}
        \label{table:observedbands}
        \end{table}

\section{X-ray Data}\label{sec:data}

In the following sections, we describe the targets and observations used for the \textit{XMM-Newton} data in this paper. 
We then describe the data processing methods used to reduce the \textit{XMM-Newton} data, and the source selection criteria used when producing the final X-ray sourcelists. 
Finally, we describe the flux-limited X-ray catalogue from \cite{piccinotti1982heao1xlf}, from which the \textit{HEAO 1} X-ray data was taken. 

For the \textit{XMM-Newton} data reduction and processing, c-shell scripts were used to run tasks using \texttt{sas-18.0.0}, \texttt{heasoft-6.27}, \texttt{wcstools-3.9.5}, and \texttt{SAOImageDS9-8.1}. 
Python scripts were run using \texttt{python-3.7.6}.

	\subsection{\textit{XMM-Newton} Observations}\label{subsec:xmmobs}  
	
	The data used in this work were taken from the \textit{XMM-Newton} Satellite, which carries multiple telescopes to study X-ray sources.
    \textit{XMM-Newton} is sensitive up to 10 keV with a spatial resolution of $\sim$5 arcsec \citep{jansen2001xmm}. 
    The \textit{XMM-Newton} data used were extracted from the three EPIC (European Photon Imaging Camera) cameras on \textit{XMM-Newton}. 
    The EPIC cameras are placed at the focus points of the X-ray mirror assemblies.
    Two of the cameras use EPIC-MOS CCDs \citep{turner2001european}, while the third camera uses EPIC-PN CCDs \citep{struder2001european}. 
    Each EPIC instrument is fitted with a filter wheel carrying X-ray transparent light blocking filters to block out background light outside the desired X-ray band.
	
	In this work, we use 25 \textit{XMM-Newton} target fields adopted from the the \textit{XMM-Newton} Medium Sensitivity Survey (XMS).
	The XMS is a survey built using a sample of the AXIS survey \citep{carrera2007xmm} covering a geometric sky area of 3.33 deg$^2$ \citep{ebrero2009xmm}. 
	The luminosity distribution of the entire sample shows that the survey contains Seyfert-like AGN as well as QSOs.
	The XMS is sensitive to AGN with intrinsic column densities up to $10^{23}$~cm$^{-2}$ \citep{mateos2005x}.
	
	We have used the redshifted energy bands $E_{obs}(z)$ listed in Table \ref{table:observedbands} for the methods and analysis performed in this work.
    For each $E_{obs}(z)$, the data reduction method from Section~\ref{subsec:datareduc} was performed for a total of 29 \textit{XMM-Newton} observations.
	A list of the \textit{XMM-Newton} observations used and their general properties can be seen in Table \ref{table:xmmfields}).
    Taking into account the excluded areas from our masks (see Section~\ref{subsec:datareduc}), the total geometric sky area covered amounts to 2.61 deg$^2$ for the \textit{XMM-Newton} X-ray data.
	
    \begin{table*}
    \caption{List of the \textit{XMM-Newton} fields used in this work, with their general properties. This includes the \textit{XMM-Newton} observation number, the RA/Dec coordinates at the center of the field (RA$_{Field}$, Dec$_{Field}$), the Galactic column density in the direction of the field ($N_{H,Gal}$), the filter used for the EPIC cameras, the "clean" exposure time for the M1, M2 and PN cameras (T$_{exp, M1}$, T$_{exp, M2}$, T$_{exp, PN}$, respectively), and the M1, M2 and PN rate thresholds (M1$_{th}$, M2$_{th}$, PN$_{th}$, respectively) used to filter out intervals of flaring particle background rate lightcurves (produced at $E$ > 5 keV using a time bin size of 20 s).}
    \centering
    \setlength{\extrarowheight}{5pt}
    \begin{tabular}{lllllllllll}
        \hline
                       Observation &           RA &          Dec &               $N_{H}$\protect\footnotemark &    Filter & T$_{exp}$ & T$_{exp}$ &  T$_{exp}$ &          M1$_{th}$ &        M2$_{th}$ &        PN$_{th}$ \\
                                   &   $_{Field}$ &   $_{Field}$ &         $_{Galactic}$ &           &   $_{M1}$ &    $_{M2}$ &     $_{PN}$ &                  &                  &                  \\
                                   &        [deg] &        [deg] & [10$^{20}$ cm$^{-2}$] &           &      [ks] &       [ks] &        [ks] & [count s$^{-1}$] & [count s$^{-1}$] & [count s$^{-1}$] \\
        \addlinespace[4pt]
        \hline               
                        0012440301 &  331.2908325 &   -1.9216667 &                  5.94 &      Thin &      29.1 &       29.3 &        24.3 &              2.0 &              2.0 &              8.0 \\
                        0081340901 &   342.955833 &  -17.8731111 &                  2.27 &    Medium &      22.3 &       22.3 &        17.9 &              1.2 &              1.2 &              4.6 \\
                    0092850201$^a$ &   324.438501 &  -14.5487222 &                  4.15 &    Medium &      41.3 &       41.2 &        36.6 &             14.0 &             14.0 &             20.0 \\
                        0100240801 &   233.095833 &   -8.5347222 &                  8.39 &    Medium &      26.5 &       26.6 &        19.3 &              2.0 &              2.0 &              7.0 \\
                        0100440101 &  337.1266665 &   -5.3152778 &                  4.98 &     Thick &      45.3 &       45.5 &        36.7 &              1.5 &              1.5 &              5.0 \\
                        0102040201 &   172.789167 &   31.2352777 &                  1.88 &  Thin$^d$ &      17.6 &       23.2 &        11.3 &              2.0 &              1.2 &              7.0 \\
                        0102040301 &  157.7462505 &   31.0488889 &                  1.67 &  Thin$^d$ &      25.4 &       26.0 &        20.6 &              1.2 &              1.2 &              5.0 \\
                        0103060101 &   322.300833 &  -15.6447222 &                  4.45 &    Medium &      20.4 &       20.5 &        13.7 &              1.5 &              1.5 &              8.0 \\
                        0106460101 &  145.7500005 &   46.9916666 &                   1.1 &      Thin &      47.4 &       47.6 &        36.7 &              2.0 &              2.0 &              6.5 \\
                        0109910101 &     210.3945 &      -11.127 &                  4.26 &      Thin &      48.6 &       48.7 &        39.2 &              1.2 &              1.2 &              4.0 \\
                        0111000101 &    4.6375005 &   16.4383333 &                  3.77 &    Medium &      31.3 &       31.1 &        23.6 &              1.2 &              1.2 &              4.0 \\
                        0111220201 &    93.894375 &   71.0353333 &                  9.29 &    Medium &      48.5 &       49.4 &        40.4 &              2.0 &              2.0 &             12.5 \\
                        0112260201 &   44.6041665 &         13.3 &                   9.9 &      Thin &      18.1 &       18.3 &        12.4 &              1.2 &              1.2 &              4.0 \\
                        0112260201 &   44.6041665 &         13.3 &                   9.9 &      Thin &      18.1 &       18.3 &        12.4 &              1.2 &              1.2 &              4.0 \\
                        0112370301 &   34.9000005 &         -5.0 &                   2.0 &      Thin &      44.5 &       44.5 &        34.3 &              2.2 &              2.2 &              8.0 \\
                        0112371001 &         34.5 &         -5.0 &                  2.06 &      Thin &      43.6 &       43.8 &        35.8 &              1.2 &              1.2 &              6.0 \\
                        0112620101 &       130.35 &   70.8947222 &                  2.81 &    Medium &      27.6 &       27.9 &        23.9 &              2.5 &              2.5 &             14.0 \\
                        0112650401 &    16.100001 &         -6.4 &                  6.19 &  Thin$^d$ &      23.6 &       23.6 &        15.2 &              1.1 &              1.1 &              4.0 \\
                        0112650501 &   16.0000005 &         -6.7 &                  6.26 &  Thin$^d$ &      20.2 &       22.3 &        14.8 &              1.3 &              1.3 &              7.0 \\
                        0112880301 &   352.958334 &   19.9380555 &                  3.96 &     Thick &      14.4 &       14.5 &        10.8 &              1.7 &              1.7 &              8.5 \\
                    0124110101$^b$ &   185.433333 &   75.3102778 &                   2.9 &    Medium &      33.8 &       33.9 &        29.8 &               -- &               -- &               -- \\
                        0124900101 &   187.883334 &   64.2391666 &                  2.52 &      Thin &      29.7 &       30.1 &        24.8 &              2.0 &              2.0 &              8.0 \\
         \parbox{2cm}{\vspace{1.5mm} 0112370401 + $^c$ \\ 0112371501 \vspace{1.5mm}} &   34.6999995 &   -4.6536111 &                  2.03 &      Thin &      23.1 &       23.1 &        15.2 &              2.0 &              2.0 &              8.0 \\
         \parbox{2cm}{\vspace{1.5mm} 0123100101 + $^c$ \\ 0123100201 \vspace{1.5mm}} &  116.0187495 &     74.56375 &                  3.68 &      Thin &      58.8 &       57.2 &        32.2 &              2.5 &              2.5 &              7.0 \\
         \parbox{2cm}{\vspace{1.5mm} 0100240101 + $^c$ \\ 0100240201 \vspace{1.5mm}} &   202.695834 &   24.2330556 &                 0.999 &    Medium &      65.2 &       65.2 &        44.7 &              2.5 &              2.5 &              6.0 \\
         \parbox{2cm}{\vspace{1.5mm} 0106660101 + $^c$ \\ 0106660601 \vspace{1.5mm}} &   333.881958 &  -17.7349166 &                  1.85 &      Thin &     151.1 &      151.4 &       126.2 &              1.5 &              1.5 &              6.0 \\
        \hline
        \multicolumn{9}{l}{$^a$ Different exposures were merged within the same \textit{XMM-Newton} observation (see Table \ref{table:mergedevtlist}).}\\
        \multicolumn{9}{l}{\multirow{2}{*}{\parbox{14cm}{$^b$ Different exposures (with different frame modes) within the same \textit{XMM-Newton} observation were reduced separately and the final images were summed (see Table \ref{table:summedobs}).}}}\\
        \\
        \multicolumn{9}{l}{$^c$ Different \textit{XMM-Newton} observations were merged for the same target (see Table \ref{table:mergedobs}).}\\
        \multicolumn{9}{l}{\multirow{2}{*}{\parbox{14cm}{$^d$ Listed filter corresponds to EMOS1 and EPN. The EMOS2 filter was different (Thick for the first two and Medium for the second two).}}}\\
        \end{tabular}
    \label{table:xmmfields}
    \end{table*}
    
    \footnotetext{Obtained using the HEASOFT tool \texttt{nh}.}

    \subsection{\textit{XMM-Newton} Data Reduction}\label{subsec:datareduc}

	The \textit{XMM-Newton} mission provides the Science Analysis System (SAS) pipeline software \citep{gabriel2004xmm} specifically designed to reduce and analyze data collected by the \textit{XMM-Newton} observatory. 
	The SAS tasks \texttt{epproc} and \texttt{emproc} are used to produce a calibrated event list for each instrument.
	For the EPIC-PN event lists, we excluded the PN readout outer-edge regions at the top and bottom of the detector chips \citep[as done in][]{carrera2007xmm}.
	
	The data was first reduced in in the $0.5-2$ keV energy band, as described in Section~\ref{subsubsec:reductionprocess}, for the purpose of correcting the offset positions of sources within the attitude file of the observation and the event lists obtained from \texttt{epproc} and \texttt{emproc}.
	This only needs to be done once for each \textit{XMM-Newton} observation.
	This makes it more efficient when reducing the data in the multiple energy bands required for this work without needing to correct the source position offsets for each energy band at a later stage in the processing.
	Once the position offsets were corrected, the new attitude and event list files were used to generate images and sourcelists in the desired $E_{obs}(z)$ bands, as described in Section~\ref{subsubsec:sourcelistprocess}.

        \subsubsection{Reduction Process to Correct Source Position Offsets}\label{subsubsec:reductionprocess}
    
    	To filter out intervals of particle background flares from EPIC event lists, a high energy light curve ($E$ > 5 keV) was extracted from the event file. 
    	A background rate threshold was then determined by where the light curve is steady with low background intervals (see Table \ref{table:xmmfields} for the rate thresholds used for each \textit{XMM-Newton} observation).
    	This threshold varies with each observation, depending on the background.
    	All the work in this section after this point was done using a $0.5-2$ keV energy band.
    	
    	We first produce X-ray images separately for each EPIC instrument using \texttt{evselect} in the targeted energy range.
    	For EPIC-PN, ``Out-Of-Time" (OOT) images were also produced to take into account OOT events that end up being mixed within the read-out direction in the CCD frame, which then gets added to the PN background map once it's scaled out, and accounts for the bright streaks that tend to be seen in PN images.    	
    	To combine information from all three instruments, the X-ray images were then summed up together into a final X-ray image using the \texttt{ftools}\footnote{\url{https://heasarc.gsfc.nasa.gov/ftools/}} task \texttt{farith}.
    	
    	Along with X-ray images, we produce exposure maps for each of the 3 EPIC instruments in the relevant energy band.
    	MOS exposure maps were multiplied by the ratio of MOS/PN countrates assuming a power-law spectrum with a photoelectric absorption component.
    	This was done using PyXspec with the Galactic hydrogen column density $N_{H,Gal}$ (see Table \ref{table:xmmfields}) and a photon index $\Gamma = 1.9$ \citep{mateos2005x}.
    	Energy channels outside the targeted energy range were excluded.
    	The MOS exposure maps were then added to the PN exposure map to make a summed exposure map.

    	An Energy Conversion Factor (ECF), defined as the ratio of the count rate to flux, was then calculated using PyXspec to determine how to convert EPIC band count rates to fluxes in a given energy band, which is then used for further SAS tasks.
    	Since the MOS1 and MOS2 images are added on top of the PN image, the final combined EPIC image is in the format of an EPIC-PN image. 
    	Hence, the ECF is calculated assuming a PN image.
        The summed exposure map was then used to make a mask (using the SAS task \texttt{emask}) that filters out sources in areas of the image where there were CCD gaps or bad pixels. 
        
        Finally, we produce background maps with our own background script for each EPIC instrument.
    	This background script uses 2 models, a vignetted background model and a flat background model, as well as an OOT component for EPIC-PN. 
    	The flat background model accounts for the particle background.
    	The script used here then performs an ML fit to the background in a similar manner to the method described in \cite{loaring2005xmm13h}.
        For each EPIC instrument, we ran our background script using the list of sources from the \texttt{eboxdetect} SAS task, along with the EPIC exposure map and X-ray image, producing 3 separate background maps.
        The background maps were then summed up into a final background map using \texttt{farith}.
        
        The output from \texttt{eboxdetect} was also used in the SAS task \texttt{emldetect} with an ML threshold of 8 to make an X-ray sourcelist.
        We corrected the astrometry of our event lists by correlating the positions of the X-ray sources with optical sources from the SDSS Photometric DR12 Catalogue, as well as the Pan-STARRS Survey, according to the method described in \cite{traulsen2020xmmstacked}.
        This correction only needed to be done once for each \textit{XMM-Newton} observation, after which the event lists and attitude files were used to generate images and sourcelists for our set of $E_{obs}(z)$ energy bands (see Section~\ref{subsubsec:sourcelistprocess}).
        New (position-corrected) images, background maps and exposure maps in the $0.5-2$ keV energy band were finally reproduced and subsequently summed up in the same manner described in this section.
        
        Using the position-corrected $0.5-2$ keV exposure maps, we produce 2 masks using \texttt{emask}, which are used when producing X-ray sourceslists in the $E_{obs}(z)$ bands (see Table \ref{table:parvals} for parameter values used in all the masks produced).
        The first initial mask $M_{det}$ covers the full field of view (FOV) detector image, constructed using the summed exposure map.
        We then run the mask through the \texttt{ftools} program \texttt{fgauss}, which convolves the image with a circular Gaussian function to produce a smoothed image.
        The smoothed image was then used to produce a modified version of the initial mask, which was then used to make the \textit{XMM-Newton} X-ray sourcelist in Section~\ref{subsubsec:sourcelistprocess}.
        
        The second and final mask $M_{final}$ has excluded regions from the image rather than retaining the full FOV.
        This mask is used in Section~\ref{subsubsec:sourcelistprocess} to filter out sources from the X-ray sourcelist for each of the $E_{obs}(z)$ bands.
        $M_{final}$ is the mask constructed by multiplying the $M_{filter}$ and the $M_{axis}$ masks together, using the task \texttt{farith}.
        We describe what these are as follows.
        We applied two constraints when producing the mask $M_{axis}$.
        We used the PN exposure map rather than the summed exposure map when making the mask. 
        We also adjusted the \texttt{fgauss} and \texttt{emask} parameters to add an extra blur to the detector chip edges and exclude pixels falling too close to them (see Table \ref{table:parvals}). 
        AXIS only used PN data when processing their \textit{XMM-Newton} observations, and removed an extra 5-7 pixels from the edges of their detector chips \citep{carrera2007xmm}.
        Without limiting the mask to PN and incorporating the extra blurring to account for the 5-7 pixels that were excluded by AXIS, we ended up with a lot of unidentified sources. 
	    This brought our completeness statistics down, not necessarily because the sources were spurious, but simply because they were not included in the X-ray source list in AXIS, and as a result they weren't part of their optical identification campaign.
	    Thus, we chose to include these two constraints in $M_{axis}$ to be consistent with the methods adopted in the AXIS survey, given that we are using their optical identification campaign.
        When producing the mask $M_{filter}$, we set a minimum exposure threshold of 10 ks from the summed exposure map to filter out regions with low-exposure pixels.  
        This was done to avoid the occurrence of spurious sources (see Section~\ref{subsec:srcselect} for more details).
        
        After multiplying the $M_{filter}$ and the $M_{axis}$ masks together, additional regions were excluded from the resulting $M_{final}$ mask, which we describe as follows.
        The XMS survey did a serendipitous search rather than a blind search, and previously verified sources were taken as a target around which sources were searched for, and the target source itself was excluded (or masked out) during this search. 
        There were a total of 25 target sources that were excluded in the XMS, which we also remove from the final mask.
        We adopt the RA, Dec and target exclusion radius used by AXIS from Table 1 in \cite{carrera2007xmm}. 
        One \textit{XMM-Newton} observation (0112260201) is pointed between two cluster targets (A 399 and A 401). 
        AXIS provided an exclusion region for one of these cluster targets (A 399). 
        For that, we adopted a more conservative approach and extended the AXIS exclusion radius to 307$^{\prime\prime}$ to remove extra cluster sources.
        We also added an additional exclusion region to remove the second cluster target (A 401) as well. 
        We additionally excluded rectangular OOT regions using the widths provided from Table 1 in \cite{carrera2007xmm}, and we provide the RA, Dec, angle and height of the OOT rectangular region.
        All details regarding the exclusion areas applied to the final mask can be found in Table \ref{table:exclusionareas}.

        \begin{table}
        \caption{Parameter values for the Gaussian sigma and mask thresholds used in the XMM-SAS tasks \texttt{fgauss} and \texttt{emask}, respectively. $M_{det}$ is the full FOV detector mask constructed using the summed exposure map; $M_{filter}$ is the mask constructed using the summed exposure map with low exposure pixels filtered out (set to a minimum exposure threshold of 10 ks); $M_{axis}$ is the mask constructed using the PN exposure map, where \texttt{fgauss} and \texttt{emask} parameters were adjusted to add an extra blur to the detector chip edges and exclude pixels falling too close to them.}
        \centering
        \begin{tabularx}{\linewidth}{XXXXX}
        	\hline
        	Task            &     Task Parameter &                 \multicolumn{3}{c}{Task Parameter Values}       \\
                            &                    &       $M_{det}$ &                  $M_{filter}$ &    $M_{axis}$ \\
            \hline
            \texttt{fgauss} &              sigma &             4.0 &                           4.0 &           6.0 \\
            \texttt{emask}  &    \parbox{2cm}{\vspace{1.5mm} threshold1 \\ threshold2 \vspace{1.5mm}} &    \parbox{2cm}{\vspace{1.5mm} 0.93 \\ 0.5 \vspace{1.5mm}} &    \parbox{2cm}{\vspace{1.5mm} 0.93 \\ 0.5 \vspace{1.5mm}} &    \parbox{2cm}{\vspace{1.5mm} 0.98 \\ 0.5 \vspace{1.5mm}} \\
            \hline  
            \end{tabularx}
        \label{table:parvals}
        \end{table}

        \subsubsection{Reduction Process for $E_{obs}(z)$ Bands}\label{subsubsec:sourcelistprocess}

        The reduction process in the fixed rest-frame method follows a similar procedure as described in Section~\ref{subsubsec:reductionprocess}.
        The position-corrected attitude files and event lists were used to generate images and sourcelists for the \textit{XMM-Newton} observations in 6 redshifted energy band $E_{obs}(z)$, as listed in Table \ref{table:observedbands}. 
        Some changes and additions were incorporated, which will be described in this section. 
        
        The X-ray images and exposure maps were produced in the given $E_{obs}(z)$ energy band and summed up together using the same process described in Section~\ref{subsubsec:reductionprocess}. 
    	To check the dependence of the ECF and the MOS/PN ratios on intrinsic AGN absorption, we recalculated them for each XMM observation assuming $N_{H} = 0$ and $N_{H} = 10^{22.5}$ cm$^{-2}$, and derived the fractional change for each parameter.
            We find that the ECF decreases by 11\% between $N_{H} = 0$ and $N_{H}=10^{22.5}$~cm$^{-2}$, so luminosities of absorbed sources will be slightly underestimated. 
            The MOS/PN ratio is much less affected by absorption, decreasing by just 1.6\% between $N_{H} = 0$ and $N_{H}=10^{22.5}$~cm$^{-2}$.
    	Since the MOS/PN ratio and ECF are both a function of energy, they were derived separately for each $E_{obs}(z)$ using the same methods described previously. 
    	When making the summed exposure map, the MOS/PN ratios derived for the $E_{obs}(z)$ energy band were used. 
    	To make the summed background map in the targeted $E_{obs}(z)$ (as described in the previous section), we run the background script with 4 iterations to get the best background map.
    	The multiple iterations allow us to get rid of most of the bright sources, and help us produce maps that are very close to the background for the final summed background map.

        The final X-ray sourcelist is then produced using the final summed images, background maps and exposure maps, with an ML threshold of 4 in \texttt{eboxdetect}, an ML threshold of 9 for \texttt{emldetect}, and the ECF derived for the given $E_{obs}(z)$ energy band. 
        The mask used when making the sourcelist (in both \texttt{eboxdetect} and \texttt{emldetect}) was the initial first mask described in Section~\ref{subsubsec:reductionprocess}.
        This means that the sourcelist includes all sources detected from the X-ray image, since the initial mask covers the full FOV of the detector. 
        The second and final mask (described in Section~\ref{subsubsec:reductionprocess}) was then used to remove sources that lie outside of the mask from the X-ray sourcelist.
        This gives us our final X-ray sourcelist for a given $E_{obs}(z)$ band.

    \subsection{Combining \textit{XMM-Newton} Event Lists and Observations}\label{subsec:merging}
	Given that we're not going down to fluxes below 10$^{-14}$ ergs s$^{-1}$ cm$^{-2}$, a 50 ks exposure depth is more than sufficient to have good X-ray measurements well below the limit of the optical identifications.
	If a given \textit{XMM-Newton} observation included more than one science exposure for any of the EPIC instruments (i.e. extra S00 or U00 exposures), we sought to merge the top two event lists with the longest exposure times using the SAS task \texttt{merge} to achieve a decent exposure time (see Table \ref{table:mergedevtlist}). 
	We placed the criteria that the event lists being merged had to have the same filter and the same frame mode.
	This resulted in 3 \textit{XMM-Newton} observations (0092850201, 0112370401 and 0123100101) that contained merged event lists from multiple exposures within the same observation.
	
	One \textit{XMM-Newton} observation (0124110101) contained multiple science exposures that were taken in different frame modes, and thus could not be merged.
	Instead, we reduced the science exposures separately for each EPIC instrument, and summed up their X-ray images, exposure maps and background maps at the end of the data reduction process.
	We then used these summed images, exposure maps and background maps when running the source detection chain in Section~\ref{subsubsec:sourcelistprocess} and making the final X-ray sourcelist (see Table \ref{table:summedobs}).
	
	If there was more than one \textit{XMM-Newton} observation targeted towards the same field, AXIS had a preference for those that were public at an earlier period in the programme or those that were part of the SSC Guaranteed Time Program \citep{carrera2007xmm}.
	Instead, to make use of all the data, we sought to merge together the top two observations with the longest exposure times using the SAS task \texttt{merge} to achieve a better exposure time (see Table \ref{table:mergedobs}).
	The same criteria was placed on merging different \textit{XMM-Newton} observations, requiring the event lists being merged to have the same filter and the same frame mode.
	If any of these observations had more than one science exposure, they were merged internally first before being merged with another \textit{XMM-Newton} observation.
	For these merged observations, we used the exclusion area properties from the AXIS-chosen observation for the target field when modifying the final mask described in Section~\ref{subsubsec:reductionprocess}.
    These properties are listed in Table \ref{table:exclusionareas}.

    \subsection{\textit{XMM-Newton} Source Selection}\label{subsec:srcselect}
    
\begin{figure}
        \centering
          \includegraphics[width=\linewidth]{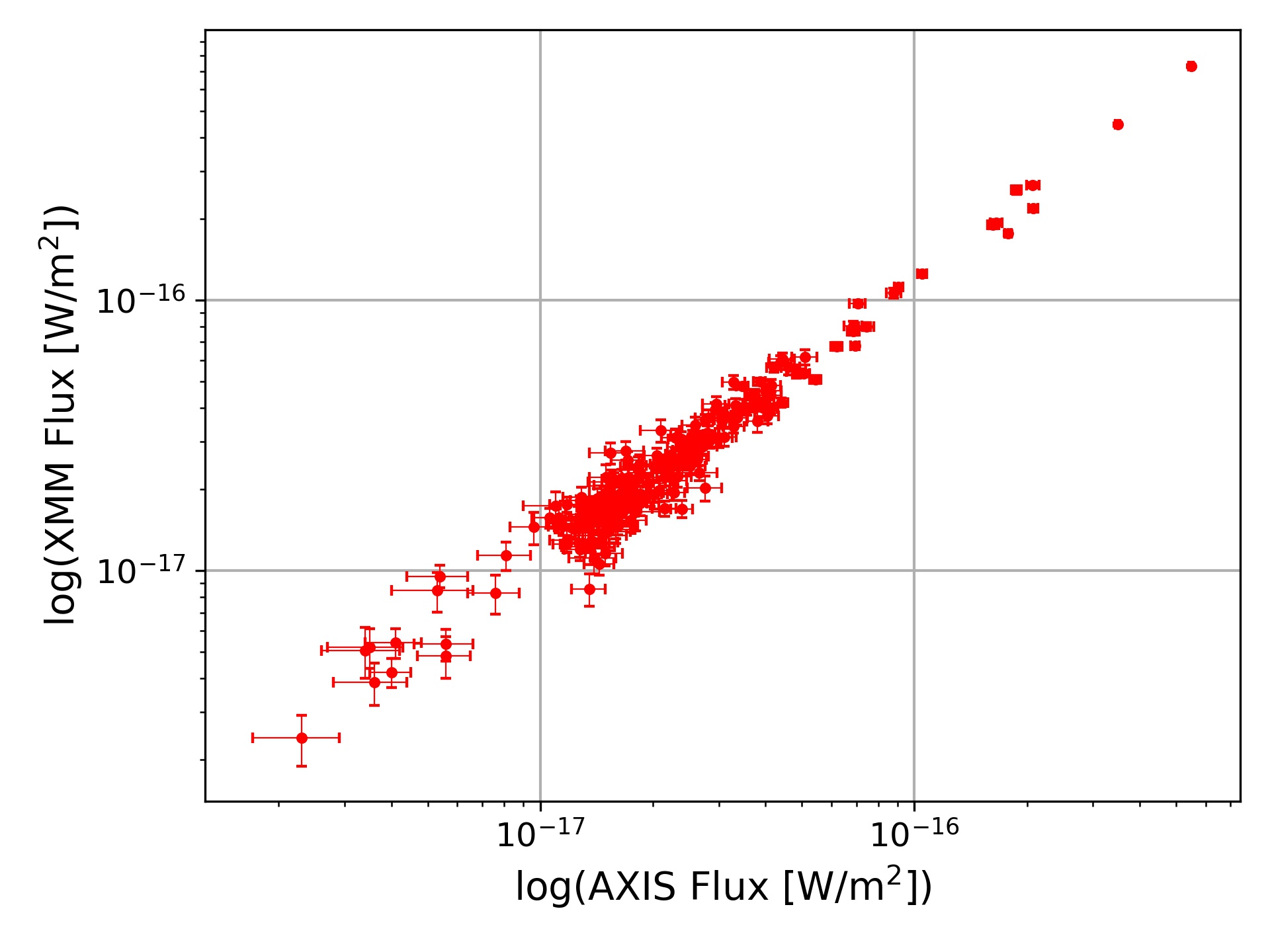}
          \caption{Comparison of the measured \textit{XMM-Newton} fluxes to the published AXIS fluxes in the $0.5-2$ keV band.}
         \label{fig:axisvsxmm}
        \end{figure}

    
    In this work, a separate sourcelist was produced for each $E_{obs}(z)$ band, as well as the $E_{rf}$ band ($2-8$ keV), per \textit{XMM-Newton} observation.
    This gives us a total of 7 X-ray sourcelists for each observation.
    We then combined the sourcelists from all the \textit{XMM-Newton} observations, resulting in a comprehensive X-ray soureclist spanning the entire dataset for each of the 7 energy bands.
    Extended sources were removed from these sourcelists.
    To make sure our X-ray fluxes produced through the data reduction process were reasonable, we compared our $0.5-2$ keV X-ray fluxes to the published $0.5-2$ keV AXIS fluxes.
    They were found to be in good agreement (see Fig.~\ref{fig:axisvsxmm}).
    
\begin{figure*}
    \centering
    	\begin{subfigure}{0.5\textwidth}
    		\centering
      		\includegraphics[width=\linewidth]{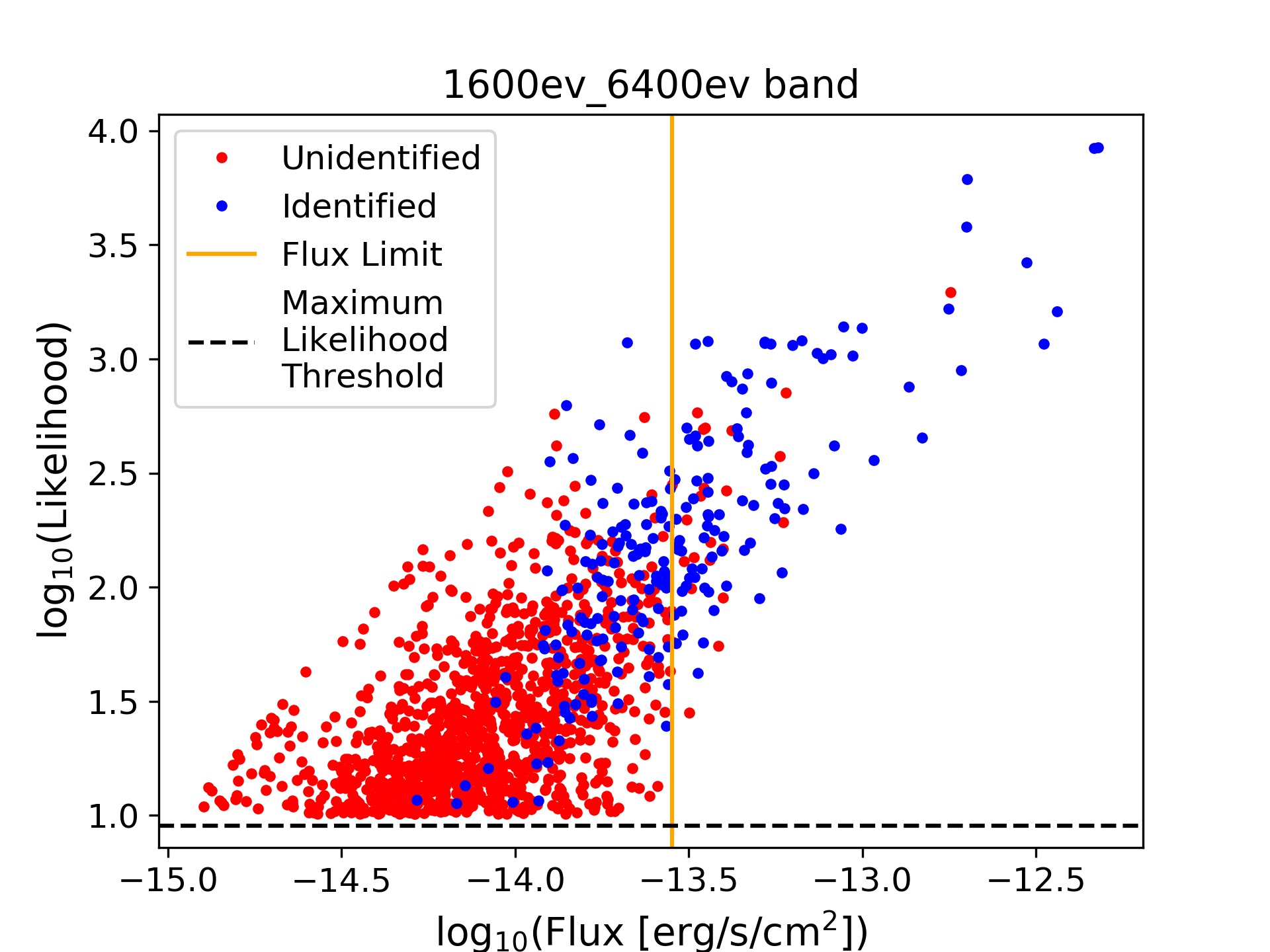}
    	\end{subfigure}%
    	\begin{subfigure}{0.5\textwidth}
    		\centering
      		\includegraphics[width=\linewidth]{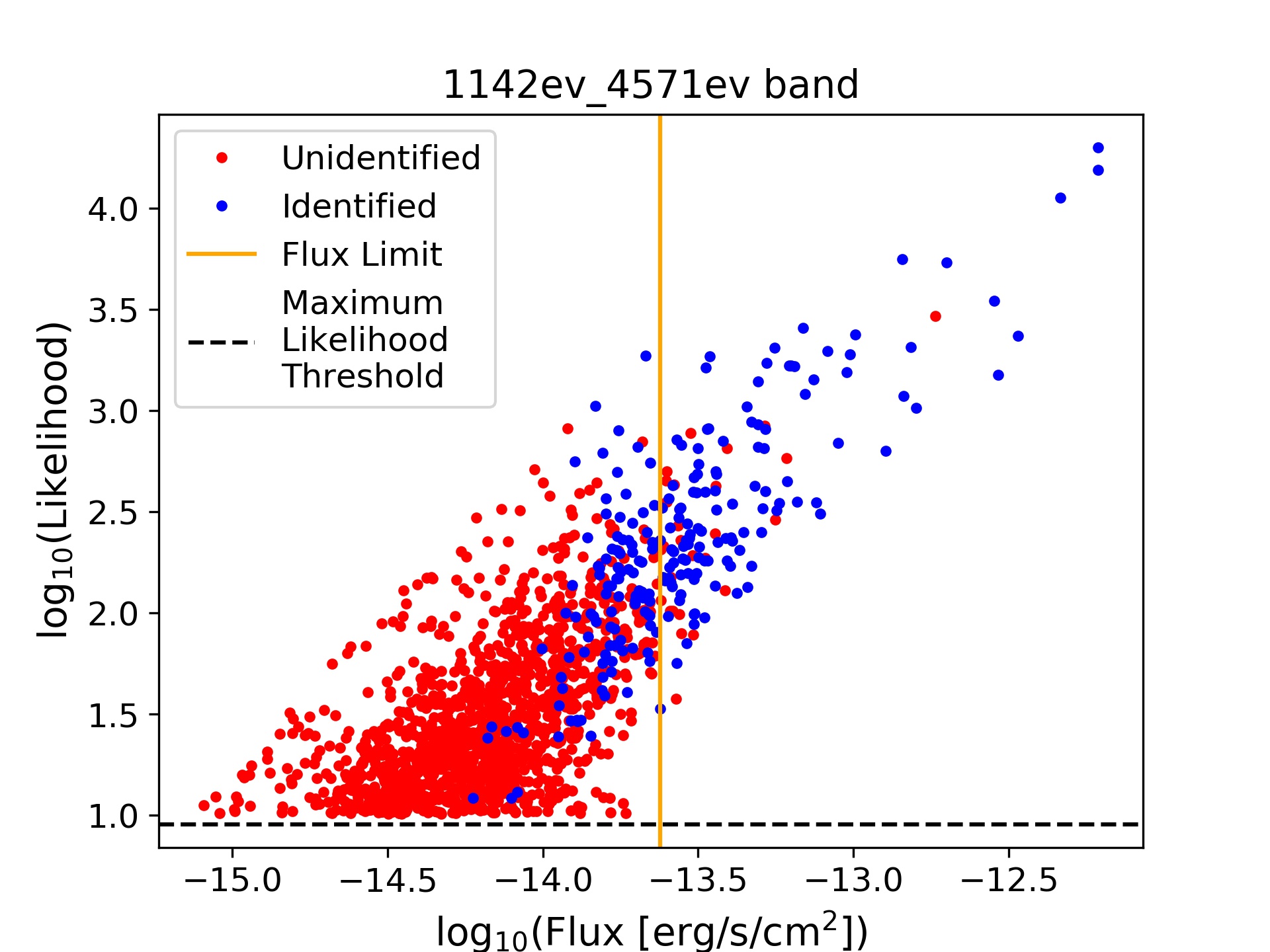}
    	\end{subfigure}%
    
    	\begin{subfigure}{0.5\textwidth}
     		\centering
      		\includegraphics[width=\linewidth]{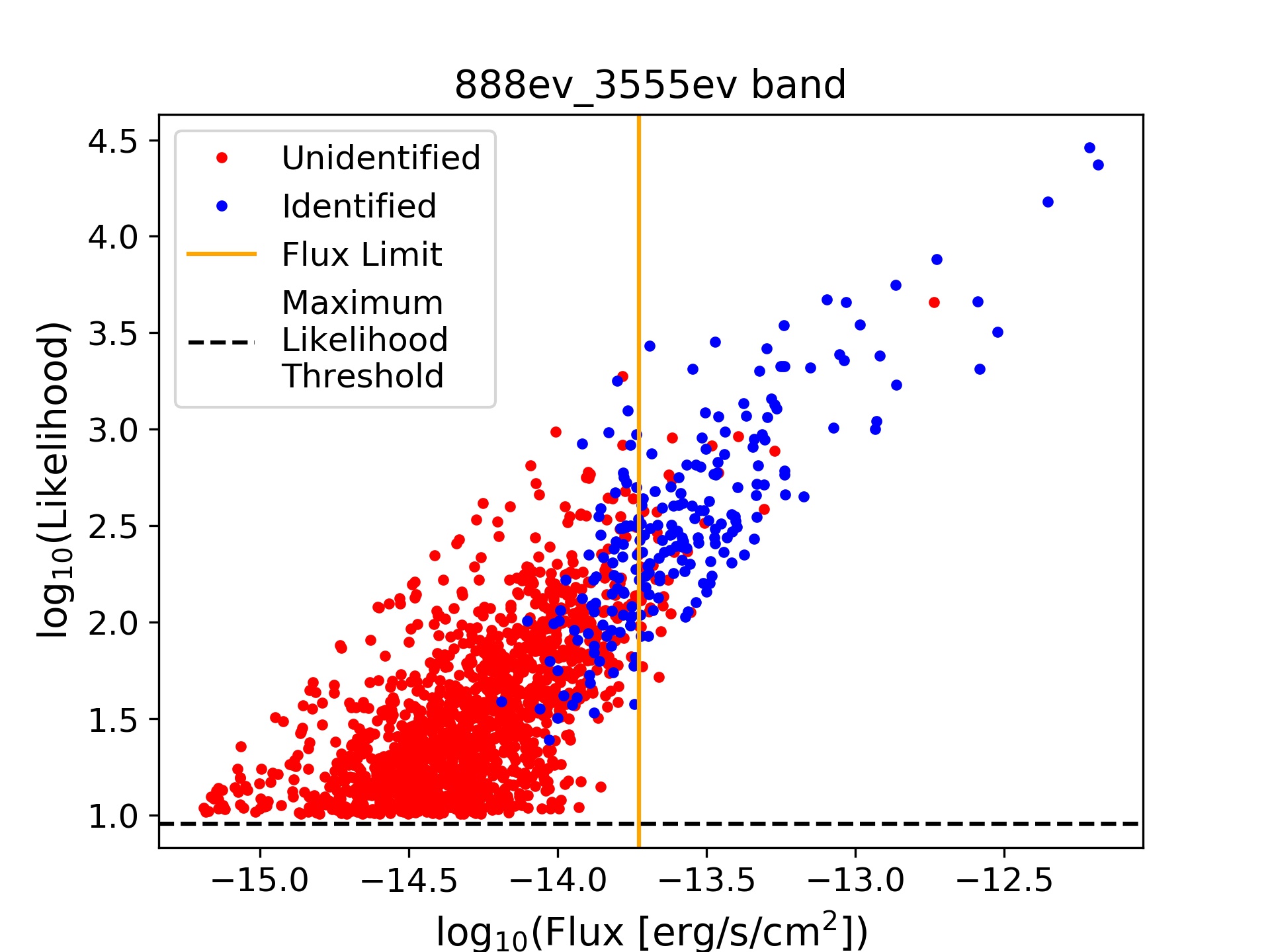}
    	\end{subfigure}%
    	\begin{subfigure}{0.5\textwidth}
     		\centering
      		\includegraphics[width=\linewidth]{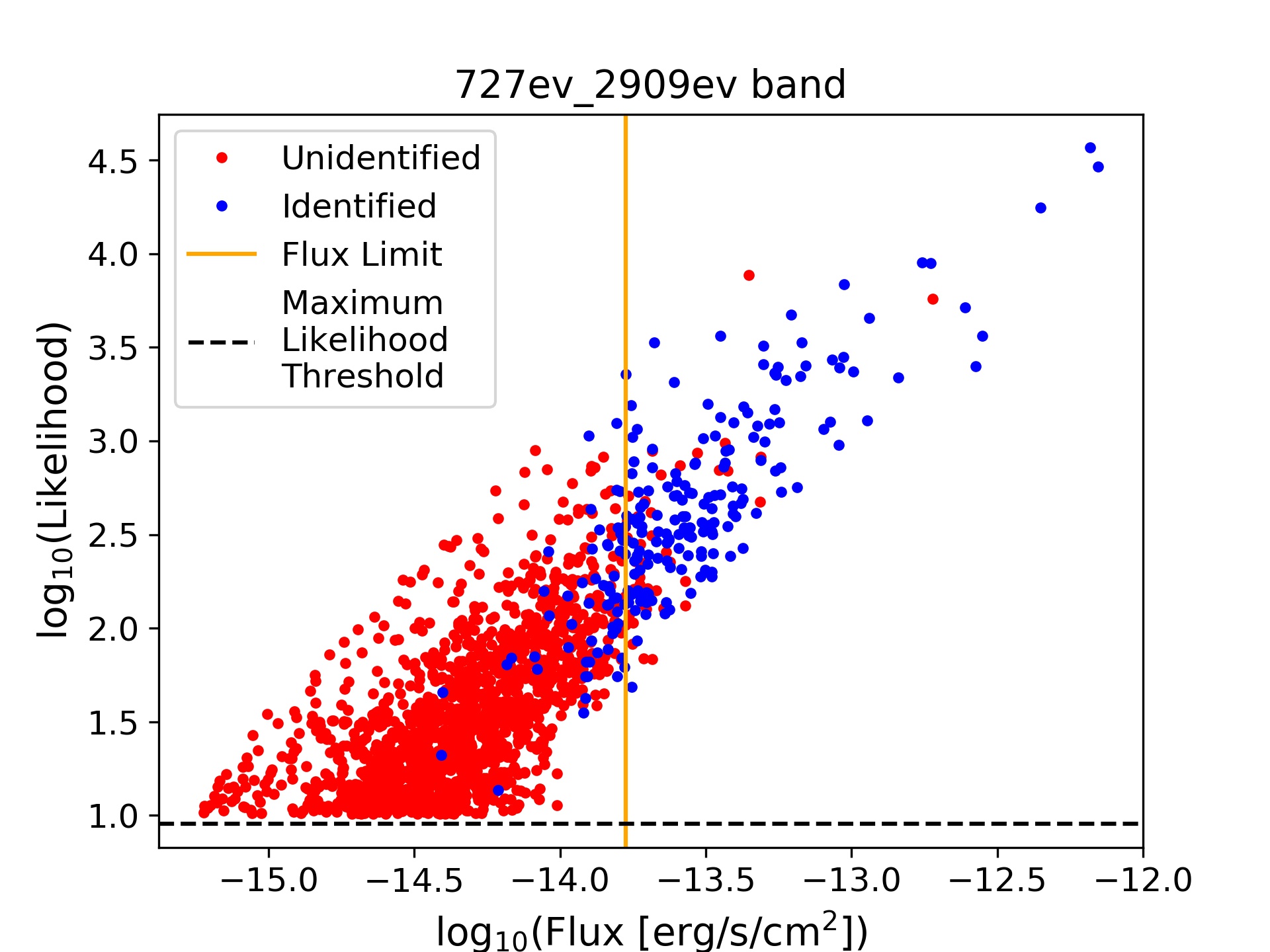}
    	\end{subfigure}%
    
    	\begin{subfigure}{0.5\textwidth}
     		\centering
      		\includegraphics[width=\linewidth]{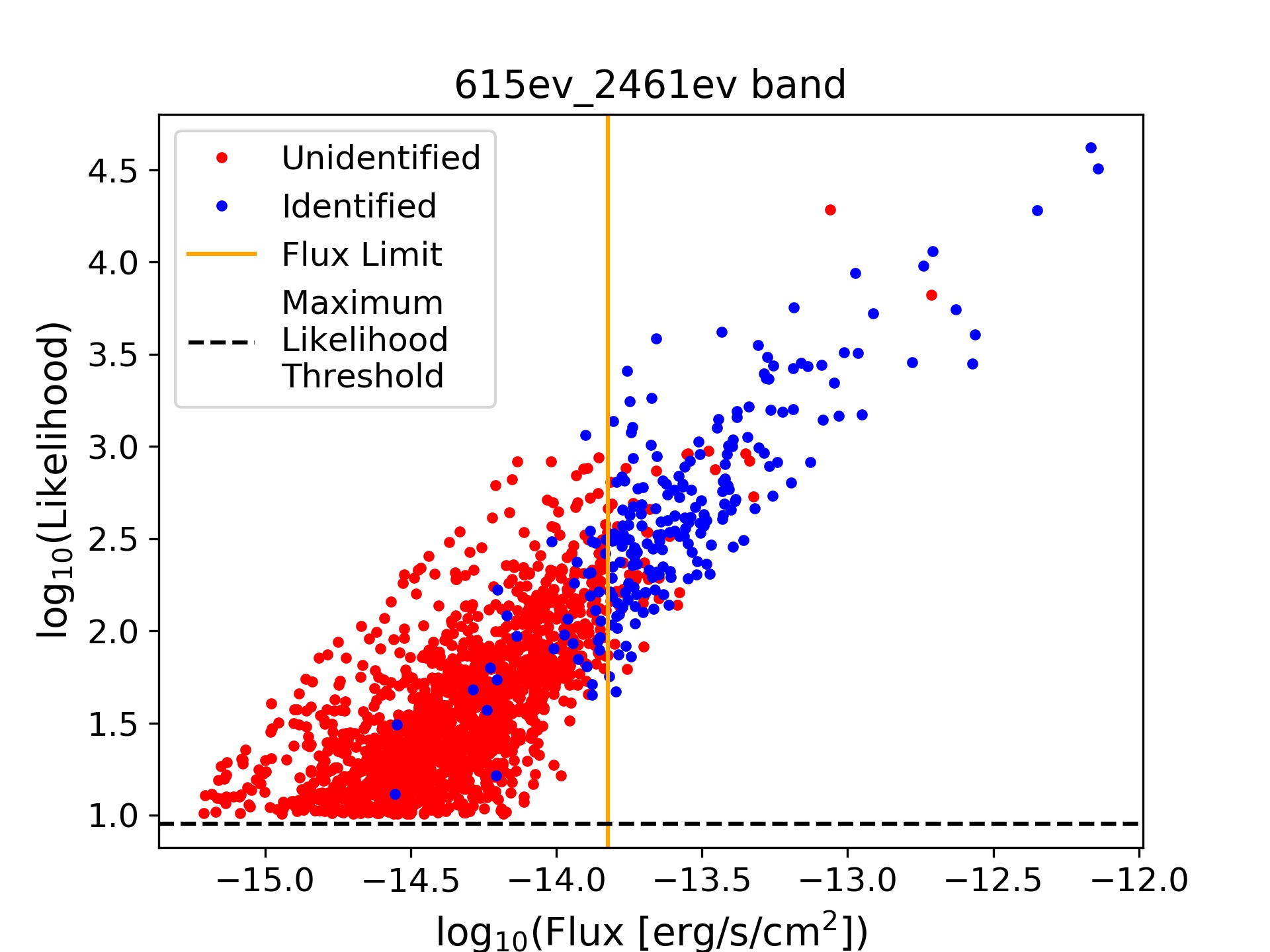}
    	\end{subfigure}%
    	\begin{subfigure}{0.5\textwidth}
     		\centering
      		\includegraphics[width=\linewidth]{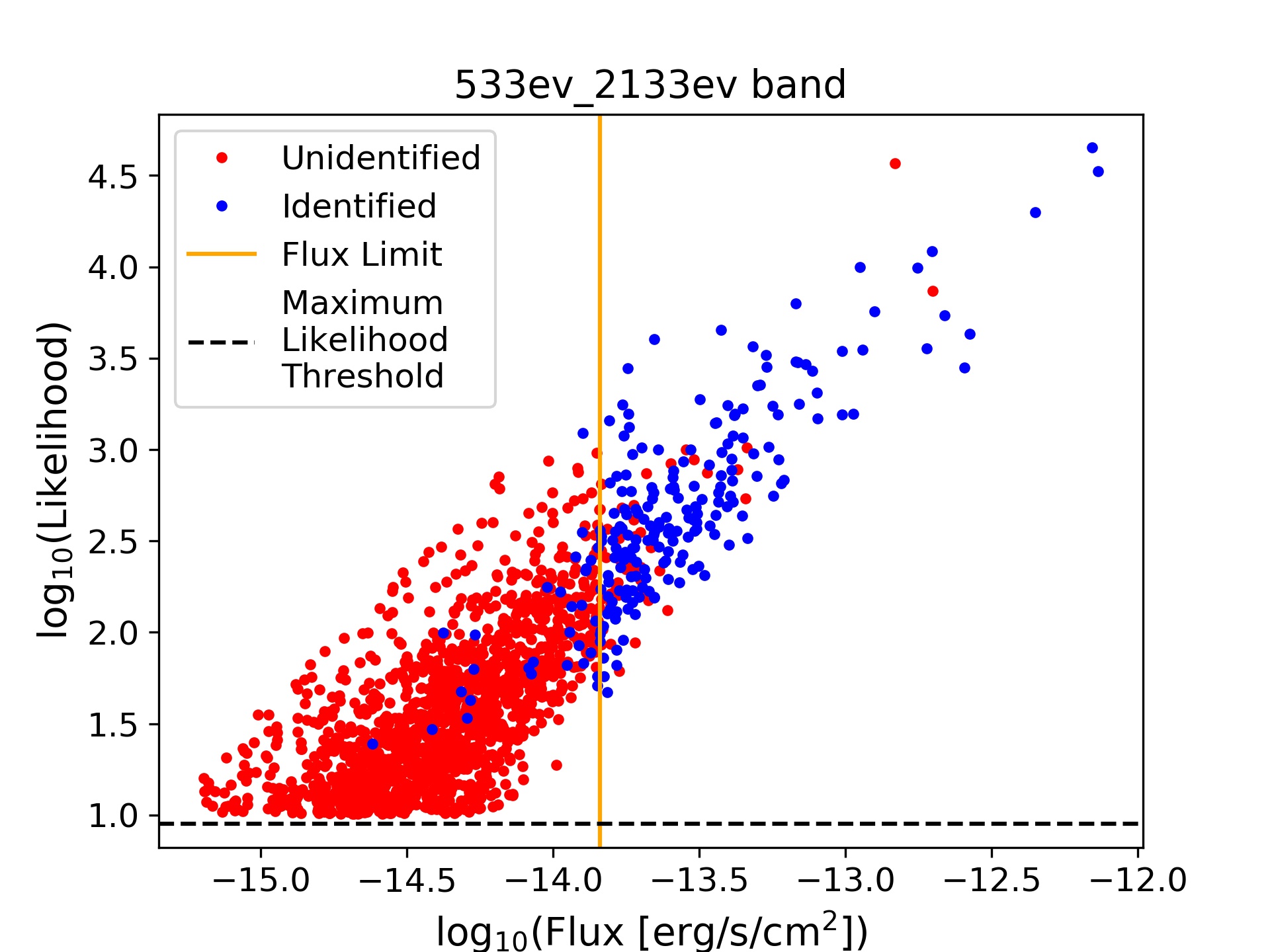}
    	\end{subfigure}%
    \caption{X-ray source flux vs. Detection Likelihood for each of the $E_{obs}(z)$ bands. Red data points are unidentified X-ray sources, and blue data points are X-ray sources identified via optical spectroscopy from the AXIS-XMS survey. The black dashed line indicates the maximum likelihood threshold of 9 used in \texttt{emldetect}. The vertical solid line is the flux limit of the X-ray sample, below which sources are not included when making the XLF.}
    \label{fig:flux_vs_ml}
    \end{figure*}

    Spurious sources resulting from data artifacts and systematic effects were being detected through the source detection chain, which posed some issues. 
    Many of these sources had low detection likelihoods but very high fluxes (above the flux limit of the $E_{obs}(z)$ band, determined in Section~\ref{subsec:completeness}). 
    To reduce the occurrence of these types of sources, we filtered out low exposure pixels (using a 10 ks exposure threshold) from the $0.5-2$ keV summed exposure map in Section~\ref{subsubsec:reductionprocess}, excluding them from the final mask used to filter out the X-ray sourcelists.
    The ML threshold for \texttt{emldetect} was also set to 9 in Section~\ref{subsubsec:sourcelistprocess} when making the final sourcelist to avoid sources with low detection likelihoods.
    This reduced the number of spurious sources significantly.
    Fig.~\ref{fig:flux_vs_ml} displays the X-ray source fluxes vs. the detection likelihood of the sources from the final filtered sourcelists for each $E_{obs}(z)$ band. 
    The black dashed line indicates the maximum likelihood threshold of 9 used in \texttt{emldetect}.
    The vertical solid line is the flux limit of the X-ray sample, below which sources are not included when making the XLF. 
    
    Despite the measures taken, some spurious sources were still found above the flux limit of the $E_{obs}(z)$ sourcelists.
    These sources were individually followed-up and investigated.
    Upon visual inspection of the X-ray images, some of them were not real sources and had a very low signal-to-noise ratio (SNR, defined by the X-ray flux of the source divided by the flux error of the source).
    These were likely due to systematic errors resulting from the data reduction pipeline process.
    To avoid sources like these, we set a condition to filter out X-ray sources that had a SNR < 2.5.
    
    In the XMS survey, there were two sets of \textit{XMM-Newton} observations that partially overlapped over the same region of the sky, as listed below:
    \begin{outline}
        \1 G133-69 Pos$\_$1 and G133-69 Pos$\_$2
        \1 SDS-1, SDS-2, and SDS-3
    \end{outline}
    
    \noindent AXIS dealt with this by masking out portions of the overlapping regions in the second (and third) fields.
    Instead of doing this, we use equation (\ref{eq:wavg}) to take the weighted average $\bar{x}$ of the RA and Dec positions for any detected sources that were overlapped for a given energy band.
    We consider sources to be overlapped if they have an angular separation distance $\theta_{D}$ $\leq$ 10$^{\prime\prime}$.
    We search for these overlapped sources within the combined sourcelist across all \textit{XMM-Newton} observations for each energy band and filter them out, replacing them with the weighted average,
    
		\begin{equation}\label{eq:wavg}
		\bar{x} = \left( \frac{x_{i}}{\epsilon_{i}^{2}} + \frac{x_{j}}{\epsilon_{j}^{2}} \right) \left( \frac{1}{\epsilon_{i}^{2}} + \frac{1}{\epsilon_{j}^{2}} \right)^{-1}
		\end{equation}
	
	\noindent where $\bar{x}$ is the weighted average, $x$ is the RA/Dec/flux of the $i^{th}$ and $j^{th}$ overlapping sources, and $\epsilon$ is the RA/Dec/flux error of the $i^{th}$ and $j^{th}$ overlapping sources.

	\subsection{HEAO 1 A-2 X-Ray Sample}\label{subsec:heao1survey}  
    
    To fill in the gaps in our X-ray data for high luminosity sources at low redshifts ($z < 0.2$), we also include the flux-limited X-ray sample by \cite{piccinotti1982heao1xlf}.
    This is a complete catalogue of X-ray sources at Galactic latitudes produced in the 2$-$10 keV band using data from the \textit{HEAO 1} experiment A-2 X-ray survey \citep{rothschild1979heao1survey}.
    The survey goes down to a limiting sensitivity of $3.1 \times 10^{-11}$ ergs cm$^{-2}$ s$^{-1}$ and covers a sky area of $2.7 \times 10^{4}$ deg$^2$.
    The survey reports their measurements in two separate scans (1st scan and 2nd scan).
    Since the 1st scan is deeper, \cite{piccinotti1982heao1xlf} treat it as the primary measurement for deriving their best-fit parameters, and the 2nd scan fluxes were only used for independent confirmation. 
    Hence, we adopt the 1st scan flux values as the X-ray flux measurement when adding the data to our XLF.
    
    \cite{piccinotti1982heao1xlf} report their X-ray fluxes in units of R15, which is a counting rate derived using the 1.5$^\circ$ $\times$ 3$^\circ$ FWHM fields of view of the layers of the X-ray counters in the \textit{HEAO 1} A-2 experiment.
    They also list conversion factors for each R15 flux measurement corresponding to each X-ray source.
    We multiply each conversion factor by the first-scan R15 flux measurement to convert the fluxes to units of 10$^{-11}$ ergs cm$^{-2}$ s$^{-1}$.
    
    Given the slight difference in the $2-8$ keV and $2-10$ keV energy bands, the $2-10$ keV source fluxes from the \textit{HEAO 1} sample had to be corrected as follows. 
	The flux between two energies (for a given energy range $E_{i}$ < $E$ < $E_{f}$) is given by equation (\ref{eq:flux_1}), which is the X-ray spectral form of the majority of Seyfert galaxies.

		\begin{equation}\label{eq:flux_1}
		F_{E_{i} - E_f} = \int_{E_{i}}^{E_{f}} kE^{1-\Gamma} dE
		\end{equation}
		
	\noindent where $\Gamma$ is the photon index, $k$ is a normalization constant, and $E_{i}$ and $E_{f}$ define the energy range. 
	
	When making the model spectrum in PyXspec in Section~\ref{subsubsec:reductionprocess}, a photon index of $\Gamma = 1.9$ was used. 
	To remain consistent, the same photon index was assumed when converting the X-ray fluxes. 
	To account for the differences between the $2-8$ keV and $2-10$ keV fluxes, equation (\ref{eq:flux_1}) was evaluated for the two energy ranges, and the ratio between them $R_{F}$ ($R_{F} = F_{2-8} / F_{2-10}$) was taken.
	This gives us $R_{F} = 0.852$, which was then multiplied by the \textit{HEAO 1} $2-10$ keV fluxes to give us the $2-8$ keV fluxes for our work.

\section{Optical Identifications}\label{sec:opticalid}

To construct an XLF, redshifts are required along with source X-ray fluxes. 
In this section, we describe the optical identifications used to obtain redshifts for the X-ray sources in our work.
We take the optical identifications from the XMS survey for the \textit{XMM-Newton} X-ray sources, and the optical identifications from the \cite{piccinotti1982heao1xlf} catalogue for the \textit{HEAO 1} X-ray sources.
We restrict our redshifts to be from spectroscopically identified optical counterparts.

	\subsection{\textit{XMM-Newton} Medium Sensitivity Survey}\label{subsec:xmssurvey}  

	The XMS survey is comprised of four overlapping samples in the $0.5-2$ keV, $0.5-4.5$ keV, $2.0-10$ keV and $4.5-7.5$ keV energy bands with flux limits well above the sensitivity of the data. 
	XMS covers a total of 318 distinct X-ray sources, and counterparts for each source were searched for in optical catalogues within 5 arcsec from the position of the X-ray source.
	Redshifts were measured by matching emission and absorption features to the sliding wavelengths of these features \citep{barcons2007xmm}. 
	The XMS has a high identification completeness, giving us 255 AGN sources with counterparts that are positively identified via optical spectroscopy.

    To identify the X-ray sources in each of the 7 filtered sourcelists, the RA/Dec positions of our sources were matched with the XMS optical counterpart source positions, taken from Table 5 in \cite{barcons2007xmm}. 
	For this work, only sources that were identified via optical spectroscopy within the XMS sample were considered for our identifications.
	This criteria was also applied for $N_{id}$ in equation (\ref{eq:completenessfraction}) when calculating the completeness fraction.
	Since \textit{XMM-Newton} has a $\sim$ 5$^{\prime\prime}$ spatial resolution \citep{jansen2001xmm}, we matched sources within an angular distance of 5$^{\prime\prime}$.
	Matched X-ray sources were then assigned corresponding redshifts from their matched optical counterparts.
	These were then used when conducting the completeness studies of our sample, described in more detail in Section~\ref{subsec:completeness}.

	\subsection{HEAO 1 A-2 Survey}\label{subsec:heao1id}
	
	The optical identifications for the X-ray sources from the \textit{HEAO 1} Survey were taken from the \cite{piccinotti1982heao1xlf} catalogue.
	We included X-ray sources that were classified as either of the following: Seyfert-1; Seyfert-2, NELG, N or other active galaxy; BL Lacerate object; and QSO. 
    Of these, we only use sources with an ID quality of ``certain" or ``possible", with a redshift measurement $z < 0.2$.
    This gives us a total of 29 spectroscopically identified AGN adopted from the \cite{piccinotti1982heao1xlf} catalogue.

	\subsection{Flux Limits}\label{subsec:fluxlimits}  
	
	To understand what the appropriate flux limits are to use for our \textit{XMM-Newton} X-ray sources, the completeness of the data pool was studied for each redshifted energy band (see Section~\ref{subsec:completeness} for more details).
	We list the derived flux limits in Table \ref{table:fluxlimits} for each $E_{obs}(z)$ band.
	
    The \textit{HEAO 1} sample is defined by sources brighter than a countrate of 1.25 R15 in the 1st scan, so we use this as the flux limit for these X-ray sources.
    We convert the flux limit to units of 10$^{-11}$ ergs cm$^{-2}$ s$^{-1}$ using a conversion factor of 2.175, the value most used by \cite{piccinotti1982heao1xlf} for converting their X-ray fluxes.
    Correcting this to the $2-8$ keV band using $R_{F}$ gives us a flux limit of $2.315 \times 10^{-11}$ ergs cm$^{-2}$ s$^{-1}$ for the X-ray sources used from the \textit{HEAO 1} survey.

	\subsection{Completeness Studies of the \textit{XMM-Newton} Data}\label{subsec:completeness}  

\begin{figure*}
		\centering
			\begin{subfigure}{0.5\textwidth}
			\includegraphics[width=\linewidth]{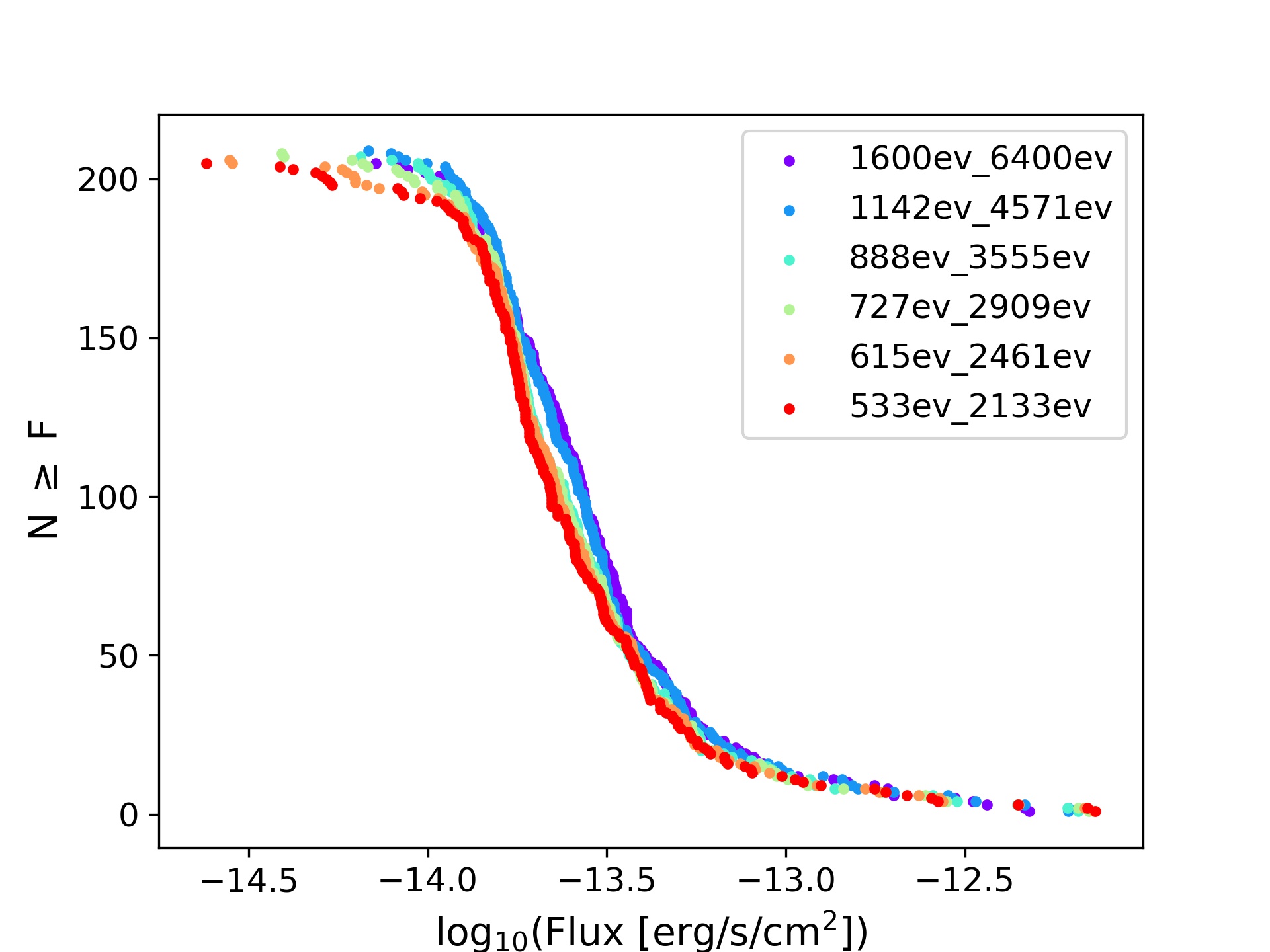}
			\caption{N$_{id} \in z_{all}$}
 			\label{fig:flux_vs_Nztot_energy}
			\end{subfigure}%
			\begin{subfigure}{0.5\textwidth}
  			\includegraphics[width=\linewidth]{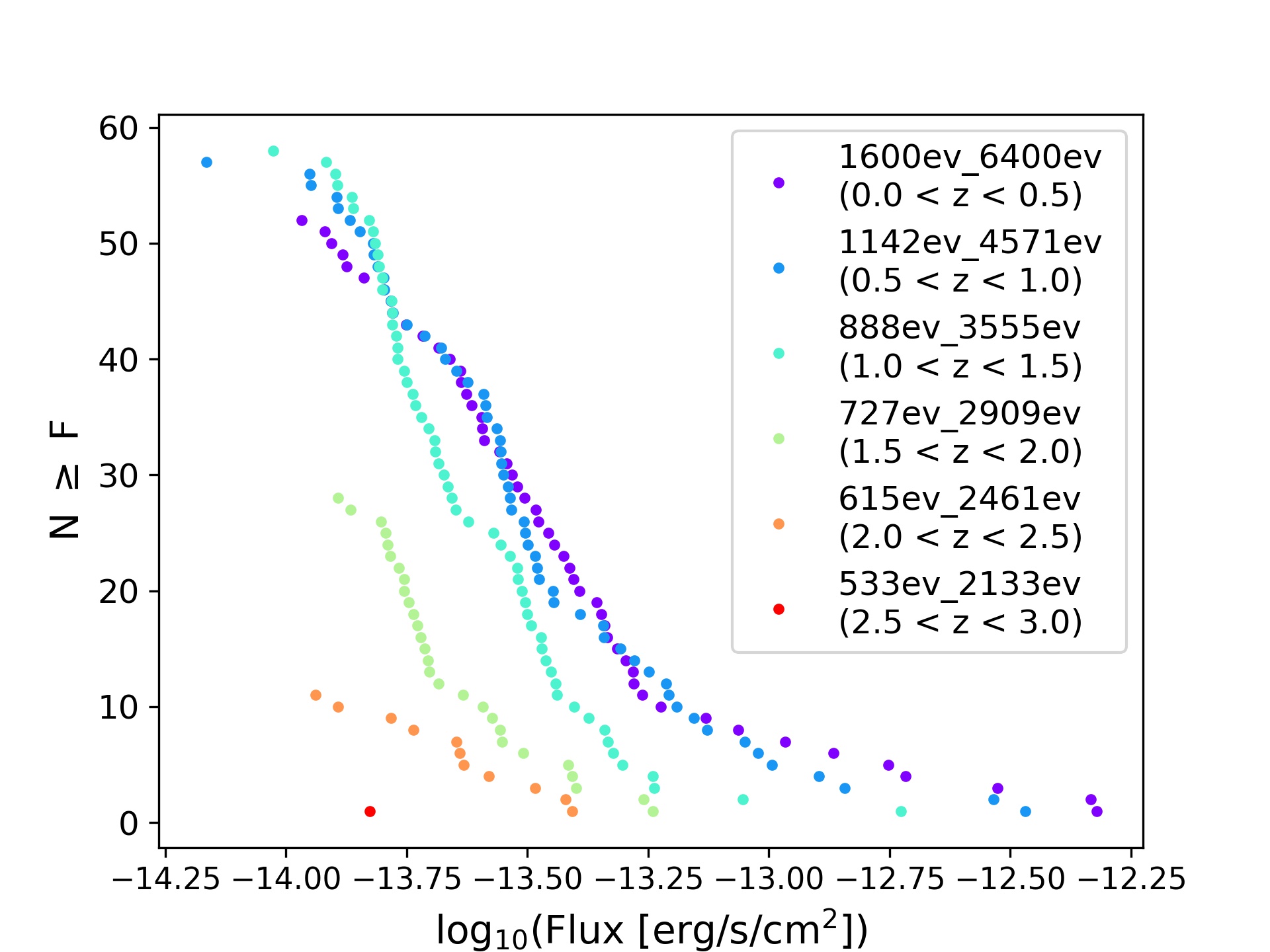}
  			\caption{N$_{id} \in z_{bin}$}
 			\label{fig:flux_vs_Nz_energy}
		    \end{subfigure}%
		\caption{Plot of flux $F$ vs. the number of identified sources $N_{id}$ with a flux greater than or equal to a given X-ray source flux $F_{x}$ ($N_{id}(F \geq F_{x})$) for the 7 energy bands. \textit{Left:} $N_{id}$ includes sources within the entire redshift range $0 < z < 3$. \textit{Right:} $N_{id}$ only includes sources that have redshifts within their corresponding $E_{obs}(z)$.}
		\label{fig:flux_vs_Nz}
		\end{figure*}

\begin{figure}
        \centering
          \includegraphics[width=\linewidth]{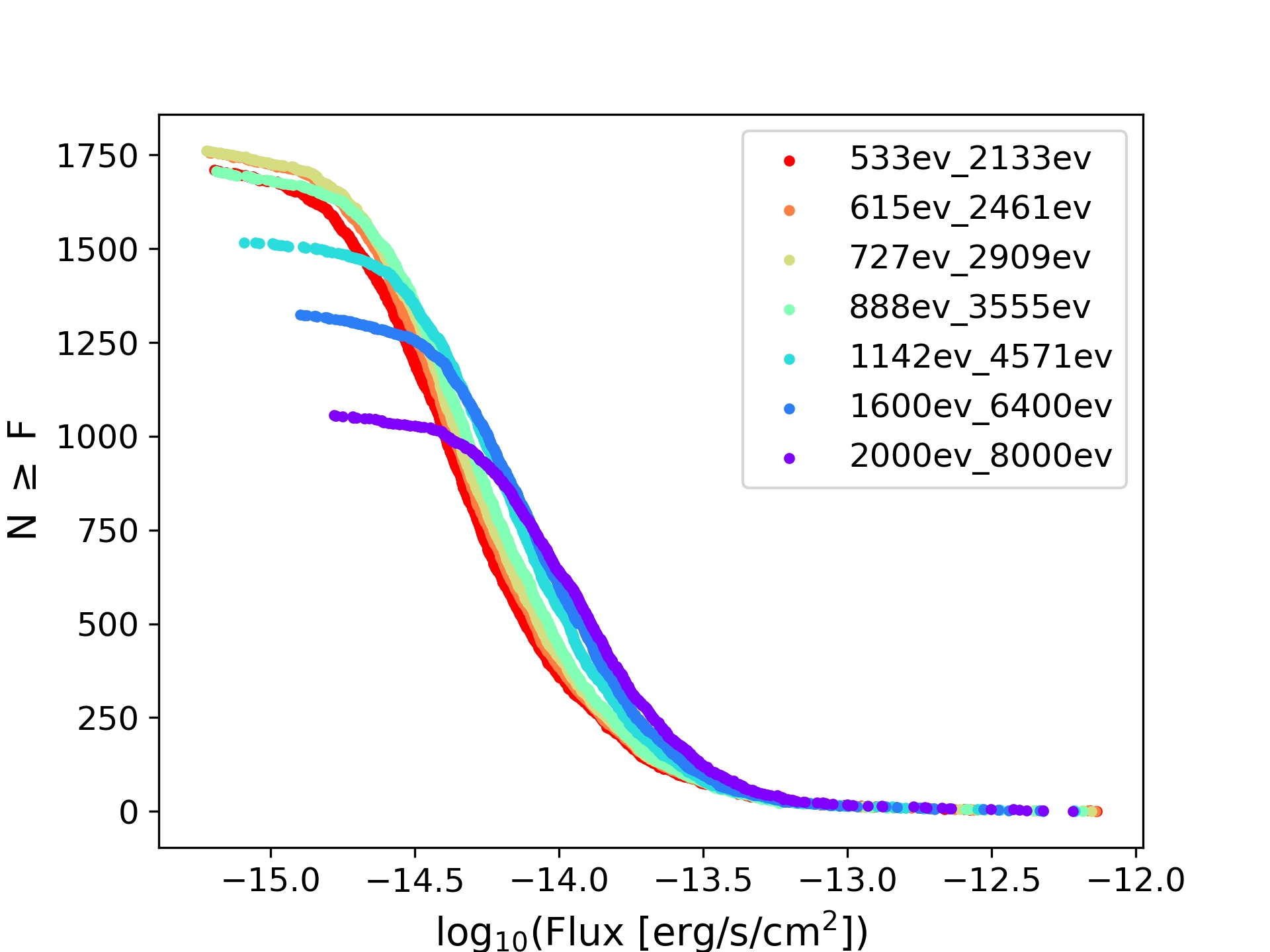}
          \caption{Plot of flux $F$ vs. the number of total sources $N_{total}$ with a flux greater than or equal to a given X-ray source flux ($N_{total}(F) \geq F_{x}$) for the 7 energy bands. The purple data points mark the $E_{rf}$ band, and the rest of the colors mark the $E_{obs}(z)$ bands.}
        \label{fig:flux_vs_N_energy}
        \end{figure}

    The completeness of the \textit{XMM-Newton} X-ray data was studied for each of the 7 energy bands (6 $E_{obs}(z)$ bands and 1 $E_{rf}$ band). 
    This allows us to assign the appropriate flux limits when making the XLF. 
    The completeness fraction $f_{c}(F_{x})$ as a function of X-ray source flux $F_{x}$, from brightest to faintest, is given by 
	
    	\begin{equation}\label{eq:completenessfraction}
    	f_{c}(F_{x}) = \frac{N_{id}(F \geq F_{x})}{N_{total}(F \geq F_{x})},
    	\end{equation}
    
    \noindent where the numerator term $N_{id}(F \geq F_{x})$ represents the number of optically-identified X-ray sources with a flux, $F$, greater than or equal to $F_{x}$, and the denominator term $N_{total}(F \geq F_{x})$ represents the total number of X-ray sources with a flux greater than or equal to $F_{x}$. 
    The $N_{id}$ term in equation (\ref{eq:completenessfraction}) thus represents the number of X-ray sources identified in the XMS survey.
    We map out this numerator term, $N_{id}(F \geq F_{x})$, as a function of flux $F_{x}$ in Fig.~\ref{fig:flux_vs_Nz}, as well as the denominator term, $N_{total}(F \geq F_{x})$, as a function of flux $F_{x}$ in Fig.~\ref{fig:flux_vs_N_energy}.

    The identification criteria for $N_{id}$ required X-ray sources to be matched with XMS sources that were positively identified via optical spectroscopy, including AGN, clusters, and stars. 
    Additional sources were found to have published spectroscopic redshifts that were not included in the optical identifications of the XMS survey.
    These were included in $N_{id}$ for the X-ray source identifications in this work, listed in Table \ref{table:extraidsources}.

	\begin{table*}
	\caption{Spectroscopically identified sources from the literature that were not included in the optical identifications of the XMS survey. These were used in $N_{id}$ for the X-ray source identifications in this work. The table lists the catalogue from which the source was taken, the name of the source, the RA/Dec position of the source, and the spectroscopic redshift of the source.}
	\centering
    	\begin{tabular}{ccccc} 
    	\hline
    	Source Name                  & RA         & Dec         & Redshift   & Redshift Origin                      \\
    	                             & [deg]      & [deg]       &            &                                      \\
    	\addlinespace[4pt]
    	\hline
     	MS 0737.0+7436               & 115.802083 & 74.493333   & 0.312      & EMSS \citep{stocke1991zEMSS}         \\
     	GALEXASC J074202.51+742625.5 & 115.512068 & 74.440213   & 0.599      & XBS \citep{caccianiga2008zXBS}       \\
     	2MASS J21300228-1534131      & 322.509363 & -15.570248  & 0.562      & 2MASS \citep{caccianiga2004zHBS28}   \\
     	J133120.3+242304             & 202.835000 & 24.384472   & 0.753      & BUXS \citep{mateos2015revisiting}    \\
    	\addlinespace[4pt]
    	\hline
    	\label{table:extraidsources}
    	\end{tabular}
	\end{table*}

    Fig.~\ref{fig:flux_vs_Nz} displays the source flux $F_{x}$ vs. $N_{id}(F \geq F_{x})$ for each energy band.  
    The figure includes two plots: Fig. \ref{fig:flux_vs_Nztot_energy} displays $N_{id}$ sources over the entire completeness sample, and Fig. \ref{fig:flux_vs_Nz_energy} only displays $N_{id}$ sources that have redshifts within their corresponding $E_{obs}(z)$ redshift bin $z_{bin}$.

\begin{figure*}
		\centering
			\begin{subfigure}{0.5\textwidth}
			\includegraphics[width=\linewidth]{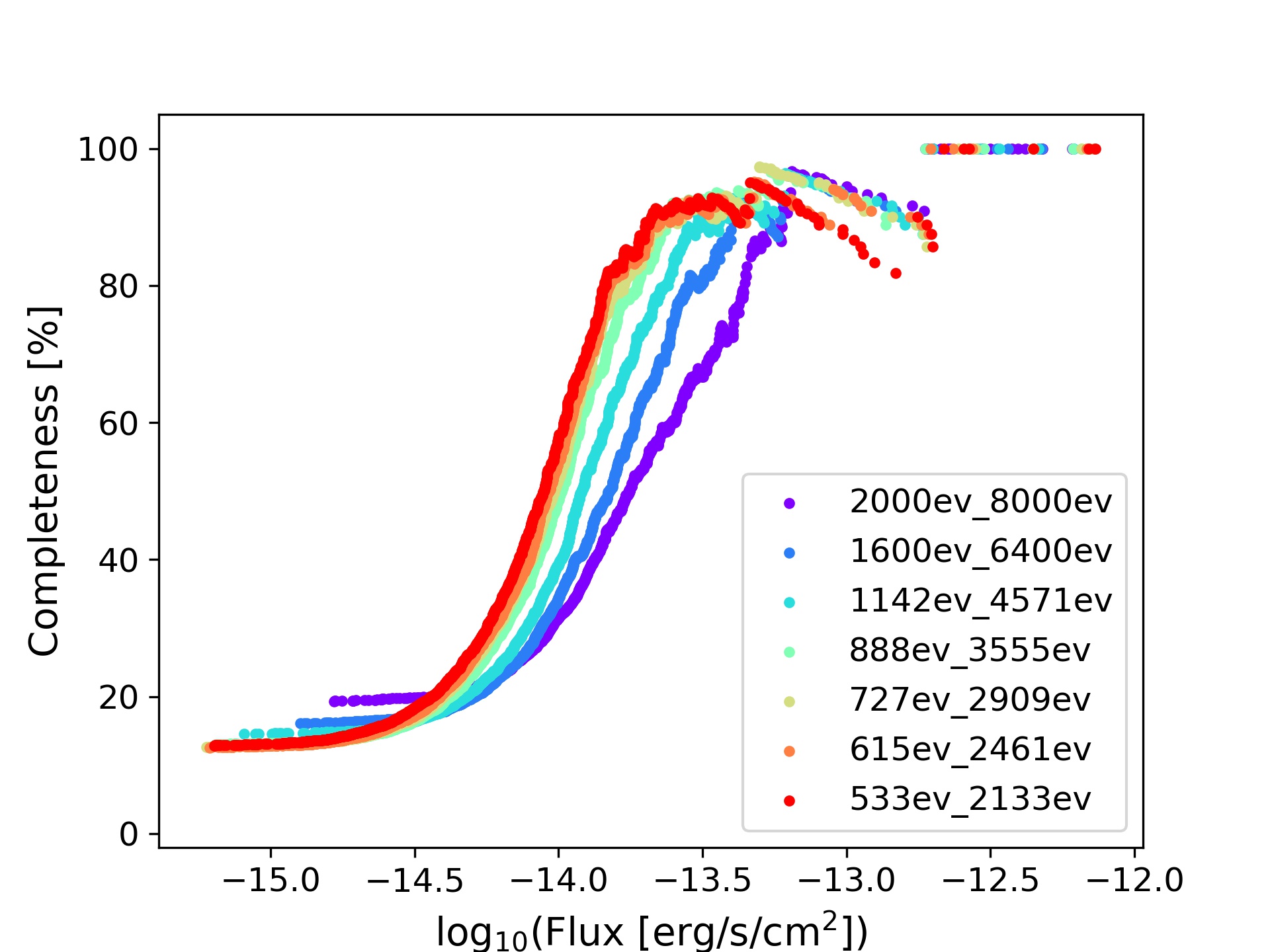}
 			\label{fig:flux_vs_cf}
			\end{subfigure}%
			\begin{subfigure}{0.5\textwidth}
  			\includegraphics[width=\linewidth]{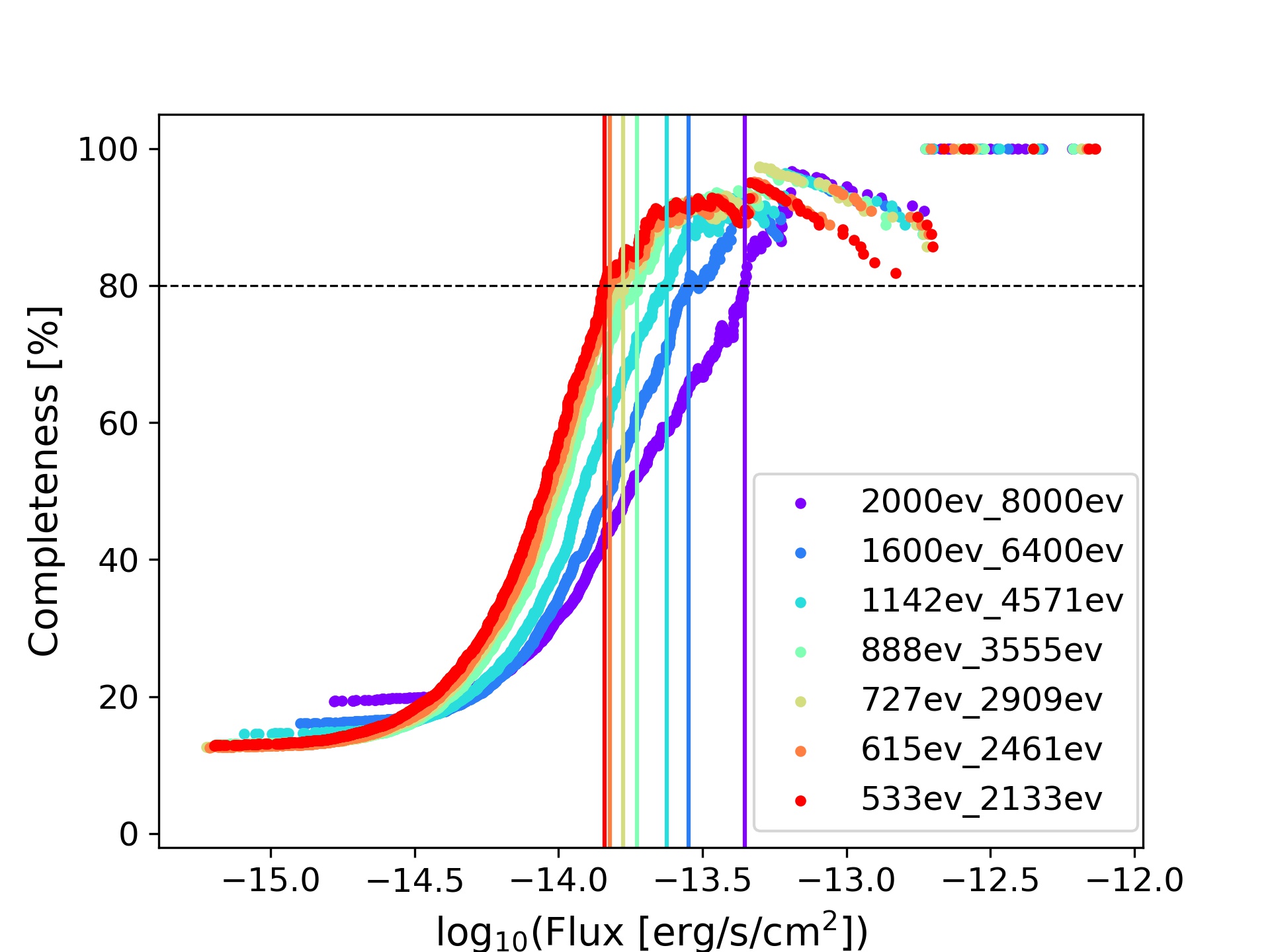}
 			\label{fig:flux_vs_cf_cutoff}
		    \end{subfigure}%
		\caption{Plot of flux $F$ vs. the completeness fraction $f_{c}$ for each energy band. The purple data points mark the $E_{rf}$ band, and the rest of the colors mark the $E_{obs}(z)$ bands. \textit{Right:} Completeness plot displaying the completeness flux limit threshold at $f_{c} = 80\%$, marked by the horizontal black dashed line. The flux limits for each $E(z)$ band are also plotted as vertical solid lines, with the same color corresponding to their data points. \textit{Left:} For clarity, we show the same plot without the threshold and flux limit lines.}
		\label{fig:flux_vs_completeness}
		\end{figure*}

\begin{figure}
		    \centering
  			\includegraphics[width=\linewidth]{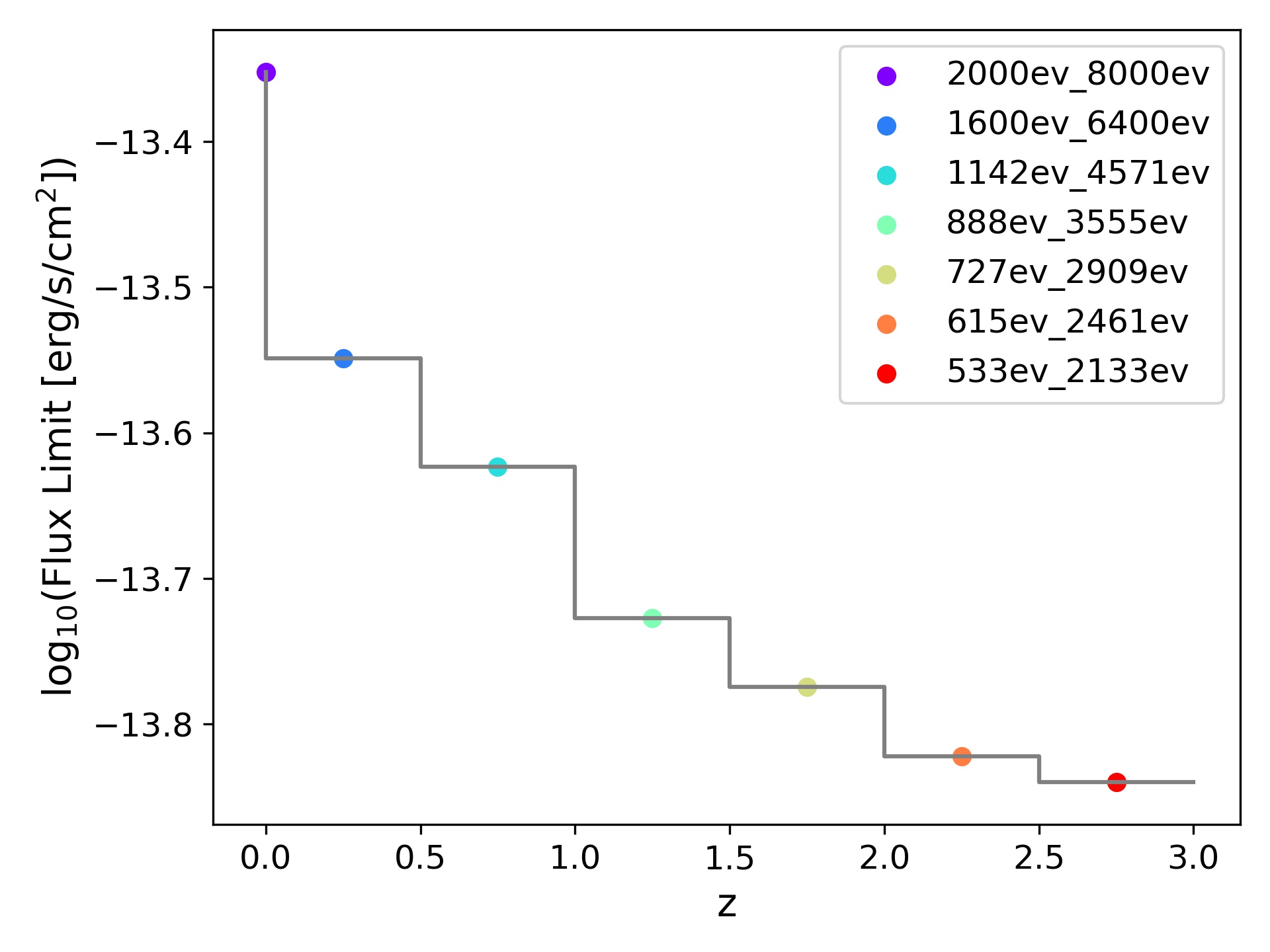}
		    \caption{Plot of redshift $z$ vs. the flux limit $F_{lim}(z)$ for each energy band, based on the values reported in Table \ref{table:fluxlimits}. The points correspond to the flux limits of each energy band used in this work. The purple data point marks the $E_{rf}$ band, and the rest of the colors mark the $E_{obs}(z)$ bands. The step-function of $F_{lim}(z)$ (marked with a gray solid line) is a function that maps the discrete redshift intervals to discrete flux limit values.}
		\label{fig:flimit_vs_redshift}
		\end{figure}

    To study the completeness of our X-ray sourcelists across all energy bands, a plot of $F_{x}$ vs. $f_{c}$ was produced (see Fig.~\ref{fig:flux_vs_completeness}).
    This also allowed us to derive flux limits needed to make the redshifted XLF for each energy band.
    To derive the required $F_{lim}(z)$ to make the fixed rest-frame XLF, a separate $F_{lim}$ is needed for each $E_{obs}(z)$ band.
    Each $F_{lim}(z)$ was determined based on where the completeness curve reaches $f_{c} = 80\%$ for each $E_{obs}(z)$ band.
    A pragmatic choice was made in choosing a limit of $80\%$, as there is a trade off between achieving high completeness and having good number of sources to produce the XLF. 
    In Fig.~\ref{fig:flux_vs_completeness}, we mark this completeness flux limit threshold with a horizontal dashed black line, and the set of derived $F_{lim}(z)$ for the redshifted energy bands are marked by the vertical solid lines.
    Table \ref{table:fluxlimits} lists the derived flux limit for each energy band.
    A plot of these flux limits as a function of redshift is shown in Fig.~\ref{fig:flimit_vs_redshift}.
    On the same figure, we also plot the discreet function of $F_{lim}(z)$ used when making the fixed rest-frame XLF in this work.

		\begin{table}
		\caption{Derived flux limits for each energy band in the XMS survey. This includes the $E_{rf}$ band ($2-8$ keV) and the $E_{obs}(z)$ bands. The $F_{lim}(z)$ was derived for each energy band based on where the completeness curve in Fig.~\ref{fig:flux_vs_completeness} reaches $f_{c} = 80\%$ (indicated by the black dashed line). The converted $2-8$ keV flux limit from the \textit{HEAO 1} survey is also reported (see Section~\ref{subsec:fluxlimits} for more details).}
		\centering
    		\begin{tabular}{cc} 
        		\hline
        		$E(z)$           & $F_{lim}(z)$ \\
        		$\mathrm{[eV]}$  & $\mathrm{[10^{-14} \ ergs \ s^{-1} \ cm^{-2}]}$ \\
        		\addlinespace[4pt]
        		\hline
        		XMS Survey                  \\
        		\addlinespace[4pt]
        		\hline
        		2000 $-$ 8000    & 4.44186  \\
        		1600 $-$ 6400    & 2.82524  \\
        		1142 $-$ 4571    & 2.38016  \\
        		888 $-$ 3555     & 1.87509  \\
        		727 $-$ 2909     & 1.68052  \\
        		615 $-$ 2461     & 1.50592  \\
        		533 $-$ 2133     & 1.44649  \\
        		\addlinespace[4pt]
        		\hline
        		HEAO 1 Survey               \\
        		\addlinespace[4pt]
        		\hline
        		2000 $-$ 8000    & 2315.18  \\
        		\hline
    		\label{table:fluxlimits}
    		\end{tabular}
		\end{table}

    Once the flux limits were derived as a function of redshift, a fixed rest-frame XLF could be constructed. 
    Only AGN sources with spectroscopic redshifts were used when constructing the XLFs. 
    Clusters, stars and AGN with no redshifts or ones that only had photometric redshifts were excluded.
    See Section~\ref{subsec:binnedxlf} and Section~\ref{subsec:mlfit} for more details.

\section{The X-ray Luminosity Function}\label{sec:xlfmethod}

In this section, we present two techniques for calculating XLFs of AGN in the $E_{rf}$ and $E_{obs}(z)$ energy bands. 
The first technique is used to produce a binned XLF over discreet luminosity and redshift bins (Section~\ref{subsec:binnedxlf}), while the second technique is used to compute the XLF using an ML fit to the full set of sources in the sample (Section~\ref{subsec:mlfit}). 
We also describe the analytical function we use to fit the XLF (Section~\ref{subsec:analyticalmodel}).

	\subsection{Binned XLF}\label{subsec:binnedxlf}  
	
	To construct a binned luminosity function of a sample of objects, we divide the Luminosity$-$Redshift plane into $L-z$ bins. 
	For this paper, the binned XLF was constructed using the method of \cite{page2000improved}.
	Binned luminosity functions are by their nature averaged over a luminosity and redshift bin, and hence where $\phi$ varies significantly with luminosity and/or redshift within the bin (as for example, at high $L$ where $\phi$ changes rapidly with $L$) the value expected from a binned estimator may be somewhat different to the value of the model at the center of the bin. 
	We have examined the magnitude of this difference in our survey by comparing the expectation values for the model bin (as defined in Section 5 of \cite{page2000improved}) to the value of the model evaluated at the midpoint of the bin. 
        For example, we looked at the difference in the $1.0 < z < 1.5$ range at $\log$ $L_X$ = 44.98 ergs/s, and find that the difference between the model and expectation values of $\log \phi$ is about 0.05.
		
	We define $\phi$, the differential luminosity function, in terms of $\log L$ rather than in terms of $L$, because it is easier to use when dealing with a large span of luminosities \citep{cara2008method},
	
		\begin{equation}\label{eq:xlfeq}
		\phi(\log L,z) = \frac{d^2N}{dV d \log L},
		\end{equation}
	
	\noindent where $N$ is the number of objects, $z$ is the redshift, $L$ is the luminosity and $V$ is the comoving volume. Note that $\phi(L,z)$ and $\phi(\log L,z)$ are related to each other by a factor of $L \, \ln(10)$.
		
	The binned estimate of the luminosity function using the \cite{page2000improved} method can be obtained for $N$ objects found over any volume-luminosity region, described by:
	
		\begin{equation}\label{eq:binnedxlf}
		\phi(\log L,z) \approx \frac{N}{\int_{\log L_{min}}^{\log L_{max}} \int_{z_{min}}^{z_{max(\log L)}} \frac{dV}{dz} dz \, d \log L},
		\end{equation}
	
	\noindent where $\braket{N}$ is the expectation value of the number of objects, $L$ is the luminosity, $z_{min}$ is the minimum redshift in $\Delta z$, and $z_{max(L)}$ is the maximum possible redshift for an object of luminosity $L$ to be detected and remain contained within $\Delta z$.
	We have adopted a uniform bin width in $\Delta$~log~$L$ of 0.3 for our binned XLFs.
	
	For each redshifted energy band $E_{obs}(z)$, an XLF was produced within its corresponding redshift bin $z_{bin}$, as listed in Table \ref{table:observedbands}. 
	For example, the XLF for the $2.5 < z < 3.0$ bin was done using the X-ray sources in the $0.5-2.1$ keV band. 
	Since we use discreet redshift intervals to construct the binned XLF, the equations and methods used are the same as \cite{page2000improved}.
	The key difference in the way we construct the binned XLF is that the flux limit $F_{lim}(z)$ is different for each of the redshift shells (see Table \ref{table:fluxlimits} and Fig.~\ref{fig:flimit_vs_redshift}).
	Together, these make up the $2-8$ keV XLF, fixed in the rest-frame.
	
	The data processed in the $2-8$ keV band (the $E_{rf}$ band) was then used to construct the fixed observer-frame XLF for the entire redshift range ($0 < z < 3$) to compare with the fixed rest-frame XLF (using the $E_{obs}(z)$ bands). 
	The data points from the binned XLFs follow the trend of a double power-law, with the break luminosity evolving with redshift. 
	This illustrates the expected AGN evolution with redshift.
	Due to the limited number of sources in our higher redshift bins, this trend becomes harder to see and analytical models are required to understand how AGN evolve.

	\subsection{Analytical Model}\label{subsec:analyticalmodel}  
	
	The X-ray luminosity in a pure luminosity evolution (PLE) model evolves with redshift, and can be expressed by 
	
		\begin{equation}\label{eq:xlf_ple}
		\phi(\log L,z) = \frac{d\phi(L / e(z), 0)}{d \log L}.
		\end{equation}
	
		\noindent The evolution factor $e(z)$ of the PLE is expressed by
	
		\begin{equation}\label{eq:e_z}
    		e(z) =
		\begin{dcases}
   		 	(1+z)^{p_{1}} & \text{if } z < z_{c},\\
    			e(z_{c})\bigg(\frac{1+z}{1+z_{c}}\bigg)^{p_{2}} & \text{if } z \geq z_{c},
		\end{dcases}
		\end{equation}
	
	\noindent where $z_c$ is the cut-off redshift, $p_1$ is is the parameter that accounts for the evolution below $z_c$, and $p_2$ is the parameter that accounts for the evolution above $z_c$ \citep{miyaji2000rosatxlf}. 
	The shape of the present-day XLF for which we adopt a smoothly connected double power law can then be expressed by
		
		\begin{equation}\label{eq:xlf_powerlaw}
		\phi(\log L,0) = A \Bigg[\bigg(\frac{L}{L_{0}} \bigg)^{\gamma_{1}} + \bigg(\frac{L}{L_{0}}\bigg)^{\gamma_{2}}\Bigg]^{-1},
		\end{equation}
	
	\noindent where $\gamma_1$ and $\gamma_2$ are the slopes, $L_0$ is the luminosity value where the change of slope occurs, and $A$ is the normalization constant \citep[e.g.][]{boyle1988qsoevolution,miyaji2000rosatxlf}. 
	
	Equation (\ref{eq:xlf_powerlaw}) was then used to plot the PLE analytical model on the binned XLF data in the $2-8$ keV energy range for $0 < z < 3$. 
	This was done for the fixed rest-frame $2-8$ keV band, as well as the fixed observed $2-8$ keV band for comparison with the new method.
	
	To be able to compare our results with that of the AXIS survey, we first constructed the PLE model curves using parameters taken from the PLE fit in the $2-10$ keV band in \cite{ebrero2009xmm}. 
	Given the slight difference in the $2-8$ keV and $2-10$ keV energy bands, the $\log_{10} L_{0}$ parameter ($43.60\pm0.13$ $h^{-2}_{70}$ erg s$^{-1}$) had to be corrected, as was done for the \textit{HEAO 1} X-ray fluxes in Section~\ref{subsec:heao1survey}. 
	As before, we assume a photon index of $\Gamma = 1.9$.
	We use $R_{F}$ in log space to convert the $\log_{10} L_{0}$ parameter, giving us a final corrected parameter of $\log_{10} L_{0} = 43.53$ to be used in the binned XLF.

	The rest of the parameters adopted from \cite{ebrero2009xmm} were $\gamma_1 = 0.81\pm0.06$, $\gamma_2 = 2.37^{+0.19}_{-0.18}$, $A = 17.96^{+9.97}_{-6.09}$ in units of $10^{-6}$ $h^{3}_{70}$ Mpc$^{-3}$, $z_c =$ 1.9 (fixed), and $p_1 =$ 2.04.
	The parameter $p_{2}$ was fixed to 0, as done in \cite{ebrero2009xmm} for the $2-10$ keV XLF, with the evolution stopping after the cut-off redshift.
	It is important to note that their fitting also took into account the amount of absorption in the modeling, which is not done when using the new method introduced in this paper.
	We thus regenerate the PLE models for the binned XLFs after performing our own model fitting and using our derived best-fit parameters (see Section~ \ref{subsec:mlfit} for more details).

	\subsection{Maximum Likelihood Fit}\label{subsec:mlfit}  
	
	We use the ML method \citep{crawford1970mlmethod} to fit the PLE model directly to the sources and obtain the best-fit evolution parameters.
	We perform this technique for XLFs in the fixed rest-frame $2-8$ keV band using the new method, as well as the fixed observer-frame $2-8$ keV band for comparison. 
	The ML method takes into account properties from each individual source, and no information is lost as a result of binning the XLF.
		
	The likelihood function is defined as the product of the probabilities of the the X-ray sources used in the XLF.
	This gives us the overall probability density for the observed distribution of objects.
	This is normally easier to compute over a logarithmic scale, as it allows us to sum over the logarithms of the probabilities.
	We follow the method from \cite{page2021uvlf} to maximize the likelihood by minimizing the expression $C$, when in logarithmic scale, as described by,

		\begin{align}\label{eq:mlminfunc3}
		C = & 2N \ln \bigg(\int_{\log L_{min}}^{\log L_{max}} \int_{z_{min}}^{z_{max}(\log L)} \phi(\log L,z) \frac{dV}{dz} dz \, d \log L \bigg)  \\
		&- 2 \sum_{i=1}^{N}\ln \phi(\log L_{i},z_{i}). \nonumber
		\end{align}
		
	We solve this using using the \texttt{amoeba} routine described in \cite{press1997numrecipe}. When applying this method to fitting an XLF, we are not just maximizing the values of the luminosity function, but we are also turning it into the probability of having observed each source. 
	This requires taking into account flux limits when generating the probabilities.
	In the new fixed rest-frame method, the flux limit is a function of redshift $F_{lim}(z)$. 
	This is incorporated into the ML-fitting routine through $z_{max}(\log L)$ (see eq. \ref{eq:mlminfunc3}), because the maximum redshift that you can see an object of given luminosity depends on the flux limit.
	$z_{max}(\log L)$ is thus the redshift at which the flux limit is equal to the flux, which is the maximum redshift to which a source could be detected.
	In previous works which used the ML method to model the XLF \citep[e.g.][]{ebrero2009xmm,ueda2014evo}, the flux limit used to determine $z_{max}(\log L)$ was not dependent on redshift, whereas in our method the flux limit does depend on redshift.
    
    The ML method changes the shape of the model distribution function to match as best as possible the distribution of the observed sources in the sample.
    The model parameters we fit for are the luminosity break $\log_{10}(L_{0})$, the slope before the break $\gamma_{1}$, slope after the break $\gamma_{2}$, and the evolution parameter $p_{1}$. 
    To be consistent with \cite{ebrero2009xmm}, we fix $p_{2}$ to 0, and we also fix the cut-off redshift $z_{c}$ to 1.9.
    This means that the evolution law we are fitting stops at $z = z_c$, so that equation (\ref{eq:e_z}) becomes $e(z) = e(z_{c})$ for $z \geq z_{c}$.
    We use these best-fit parameters to generate the PLE curves in our binned XLFs for both the fixed rest-frame and fixed observer-frame bands.

\section{Results}\label{sec:results}

\begin{figure*}
    \centering
    	\begin{subfigure}{0.5\textwidth}
    		\centering
      		\includegraphics[width=\linewidth, trim={0.5cm 0.5cm 3cm 0.3}, clip]{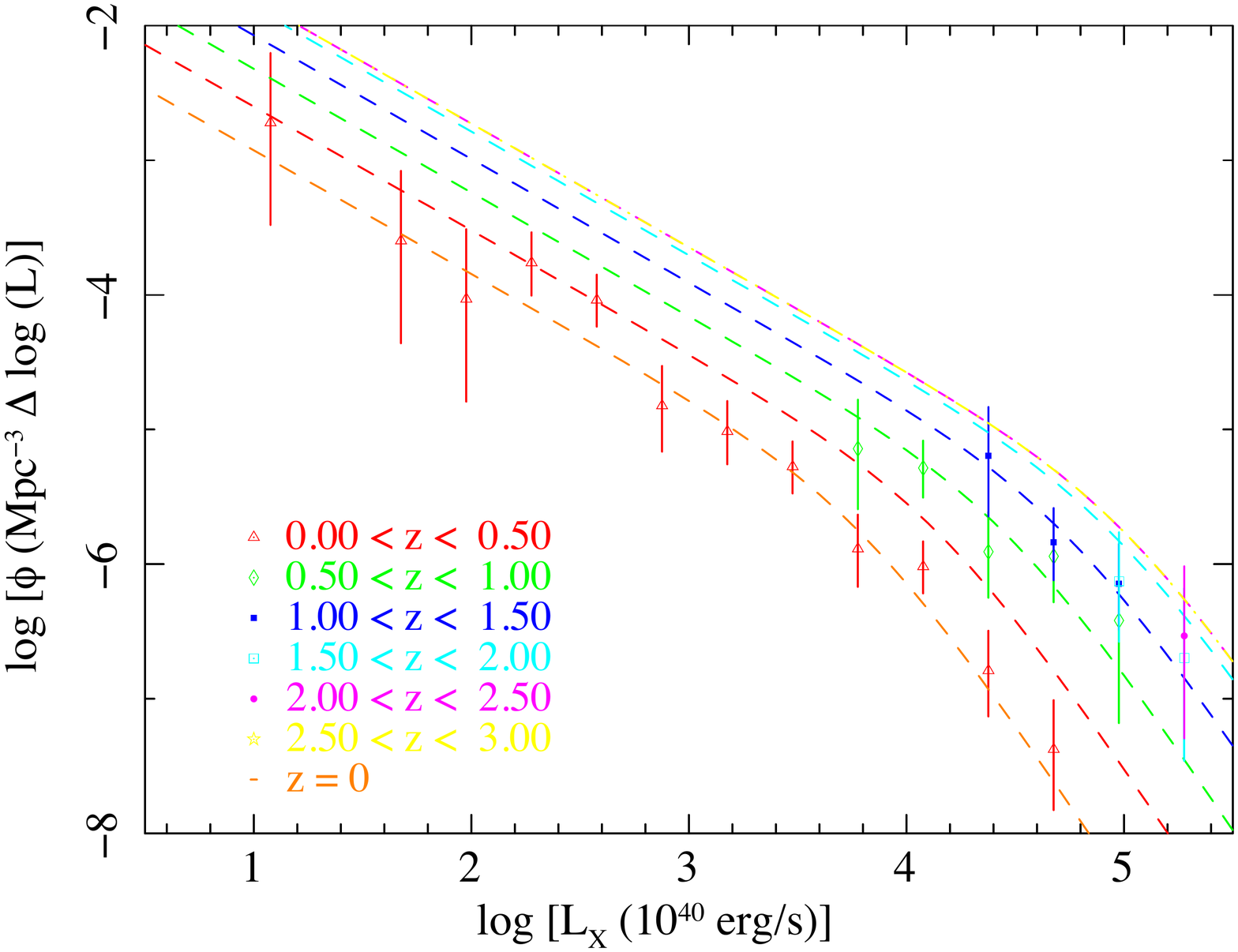}
      		\label{fig:loglumi_2000ev_8000ev}
      		\caption{Fixed Observer-frame XLF}
    	\end{subfigure}%
    	\begin{subfigure}{0.5\textwidth}
    		\centering
      		\includegraphics[width=\linewidth, trim={0.5cm 0.5cm 3cm 0.3cm}, clip]{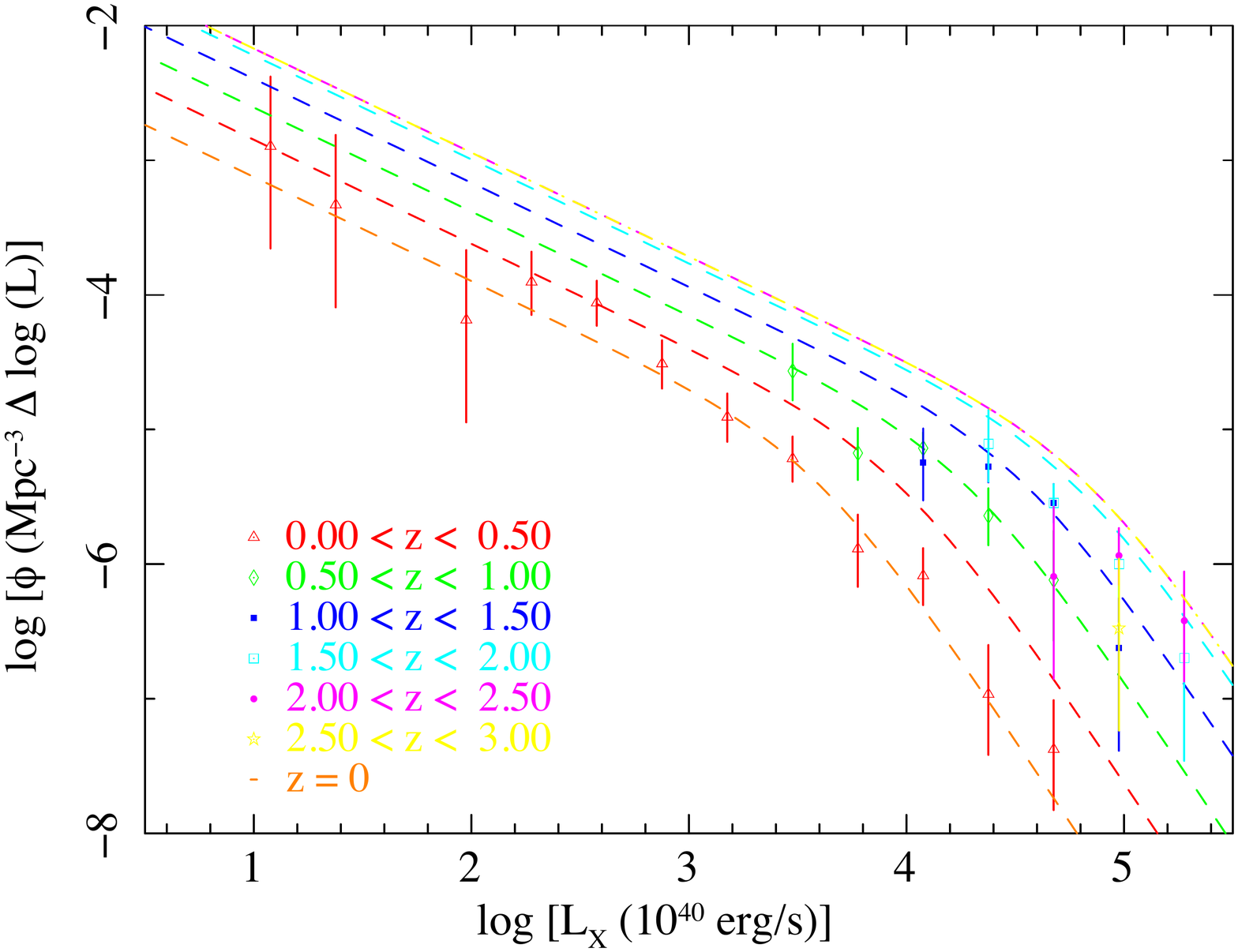}
      		\label{fig:loglumi_all}
      		\caption{Fixed Rest-frame XLF}
    	\end{subfigure}%
    \caption{X-ray Luminosity Functions in the 2$-$8 keV band for $0 < z < 3$. The data points represent the results from the binned XLF and the dashed lines are the curves of the analytical PLE model (eq. \ref{eq:xlf_powerlaw}), produced using our own ML best-fit parameters. The orange dashed lines correspond to the PLE model evaluated at $z=0$. \textit{Left:} Standard XLF in the fixed observed band. \textit{Right:} New XLF in the fixed rest-frame band.}
    \label{fig:xlfplots}
    \end{figure*}

The results of the binned XLFs can be seen in Fig.~\ref{fig:xlfplots}, constructed as described in Section~\ref{subsec:binnedxlf}. 
In the same figure, we display the XLFs in both the fixed rest-frame (produced using the combined $E_{obs}(z)$ band data) and the fixed observer-frame (produced using the $E_{rf}$ band data).
The PLE model curves were also plotted on the binned XLFs using our ML best-fit parameters.

\begin{figure*}
    \centering
    	\begin{subfigure}{0.5\textwidth}
    		\centering
      		\includegraphics[width=\linewidth, trim={0 0 3cm 2cm}, clip]{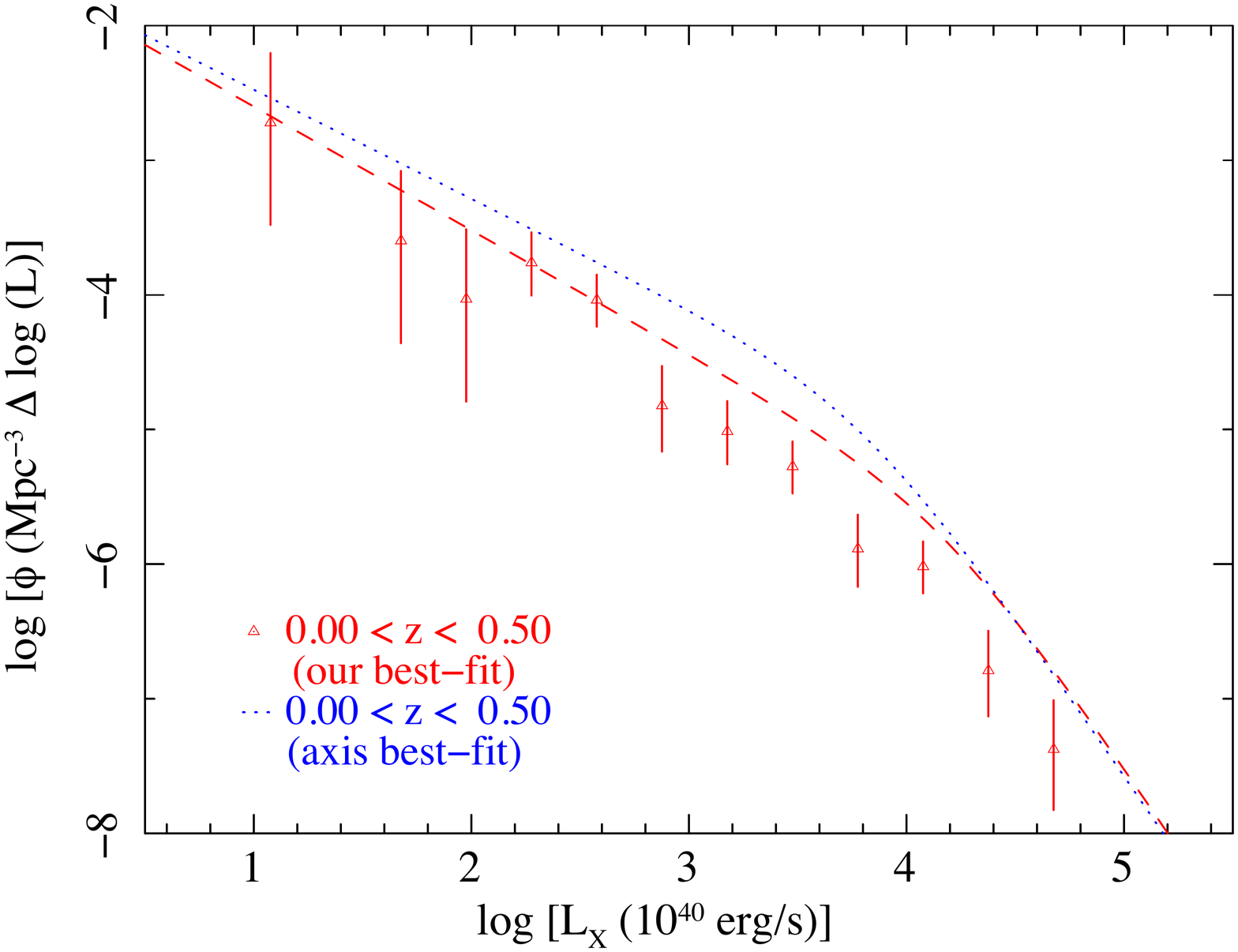}
      		\label{fig:loglumi_2000ev_8000ev_z1}
    	\end{subfigure}%
    	\begin{subfigure}{0.5\textwidth}
    		\centering
      		\includegraphics[width=\linewidth, trim={0 0 3cm 2cm}, clip]{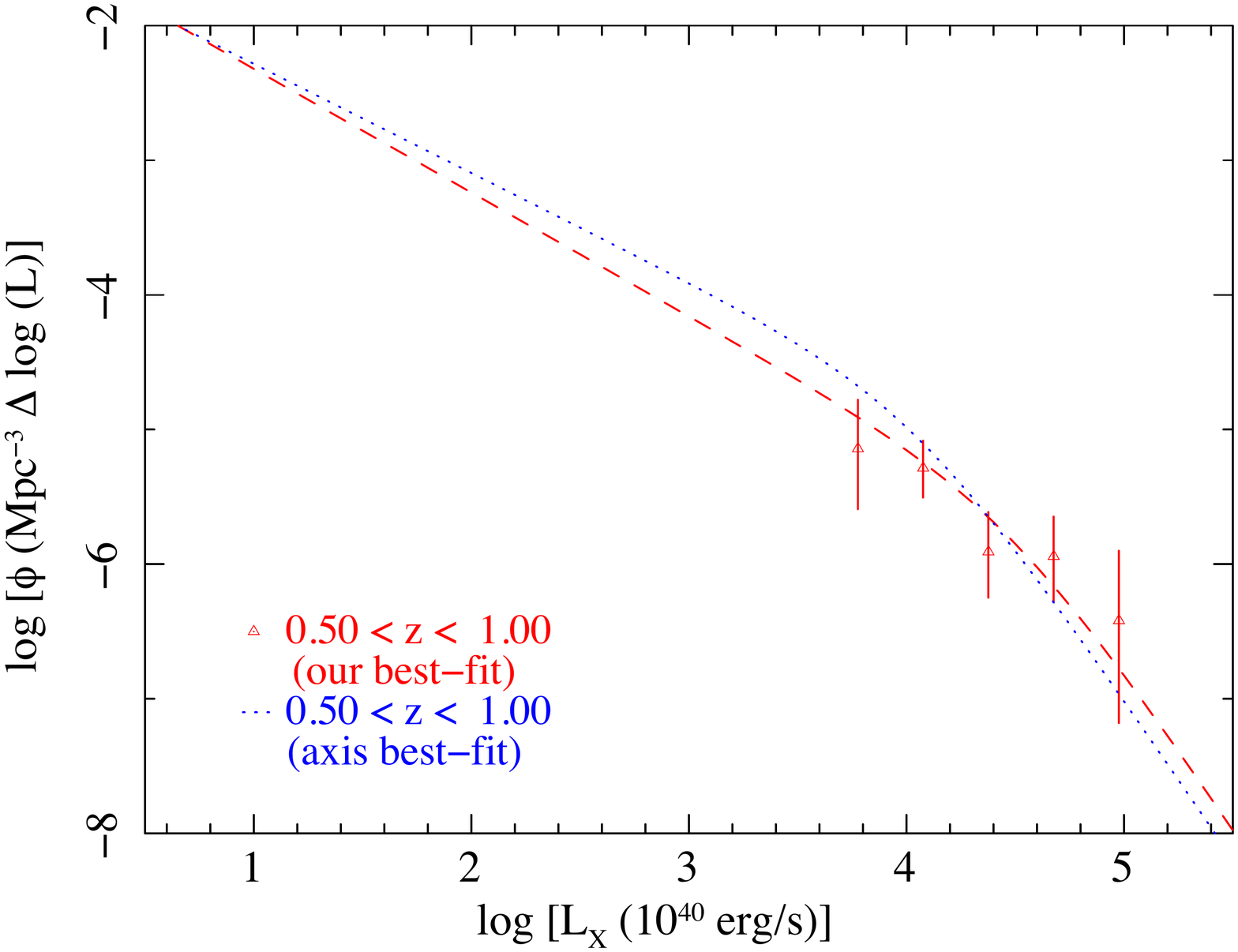}
      		\label{fig:loglumi_2000ev_8000ev_z2}
    	\end{subfigure}%
    
    	\begin{subfigure}{0.5\textwidth}
     		\centering
      		\includegraphics[width=\linewidth, trim={0 0 3cm 2cm}, clip]{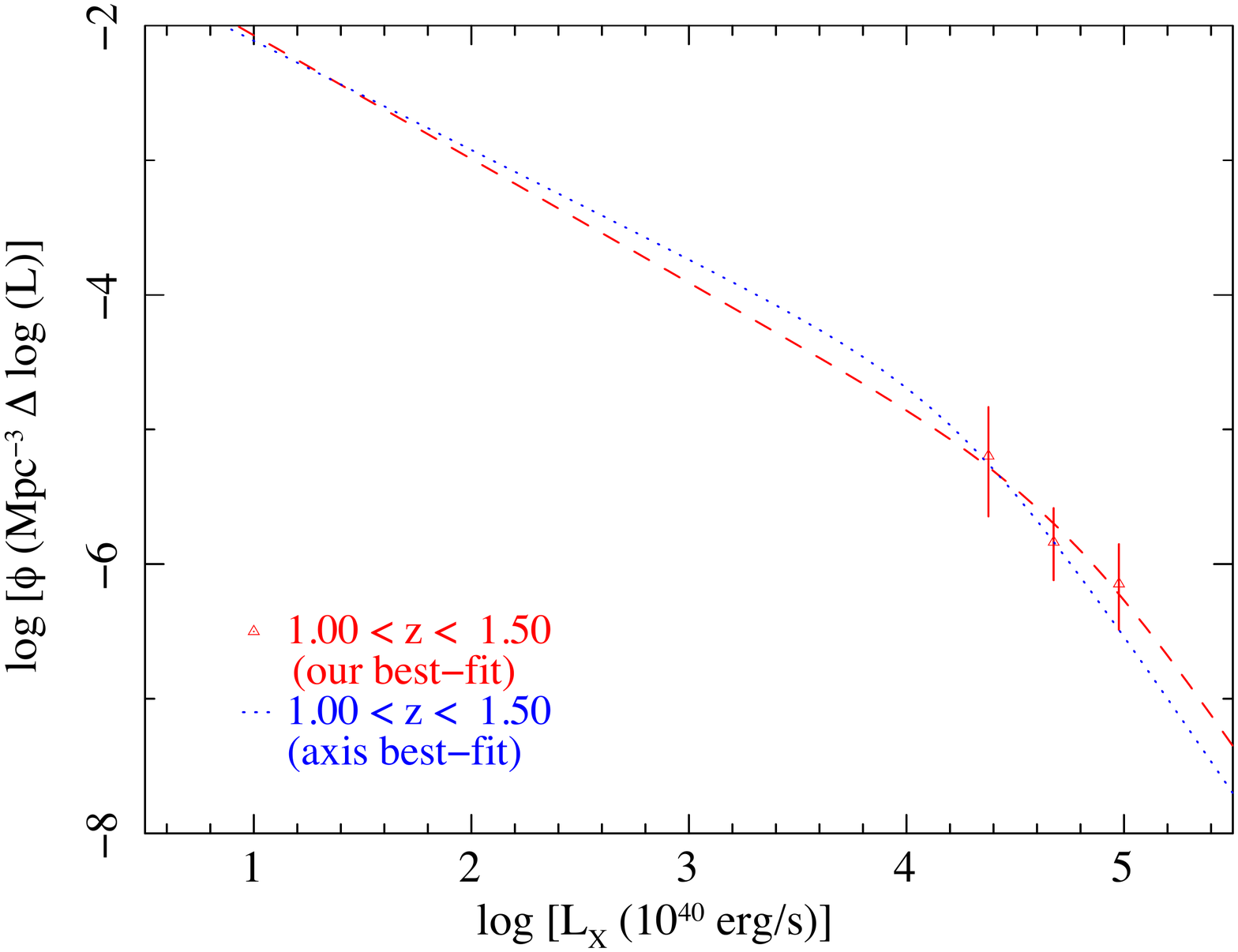}
      		\label{fig:loglumi_2000ev_8000ev_z3}
    	\end{subfigure}%
    	\begin{subfigure}{0.5\textwidth}
     		\centering
      		\includegraphics[width=\linewidth, trim={0 0 3cm 2cm}, clip]{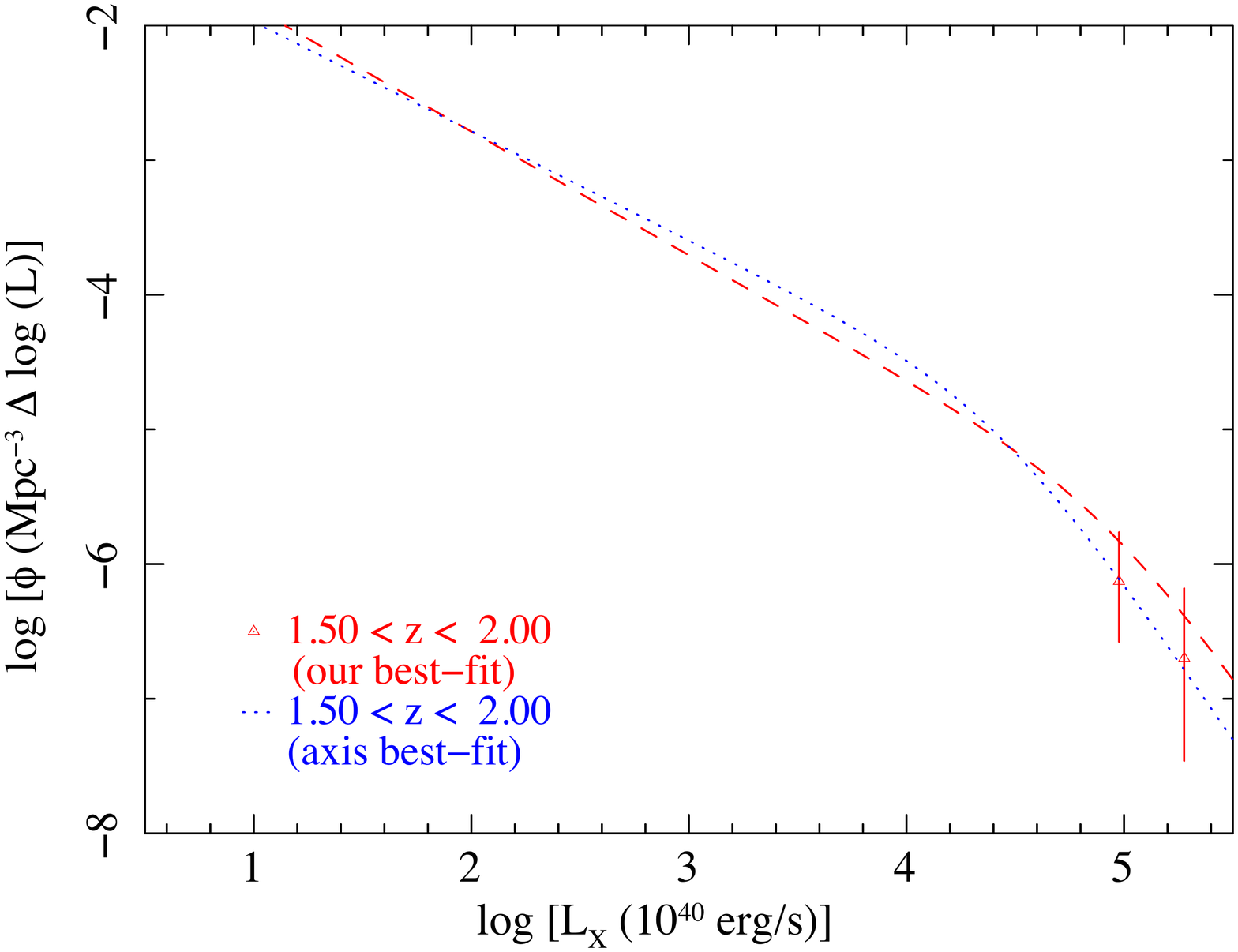}
      		\label{fig:loglumi_2000ev_8000ev_z4}
    	\end{subfigure}%
    
    	\begin{subfigure}{0.5\textwidth}
     		\centering
      		\includegraphics[width=\linewidth, trim={0 0 3cm 2cm}, clip]{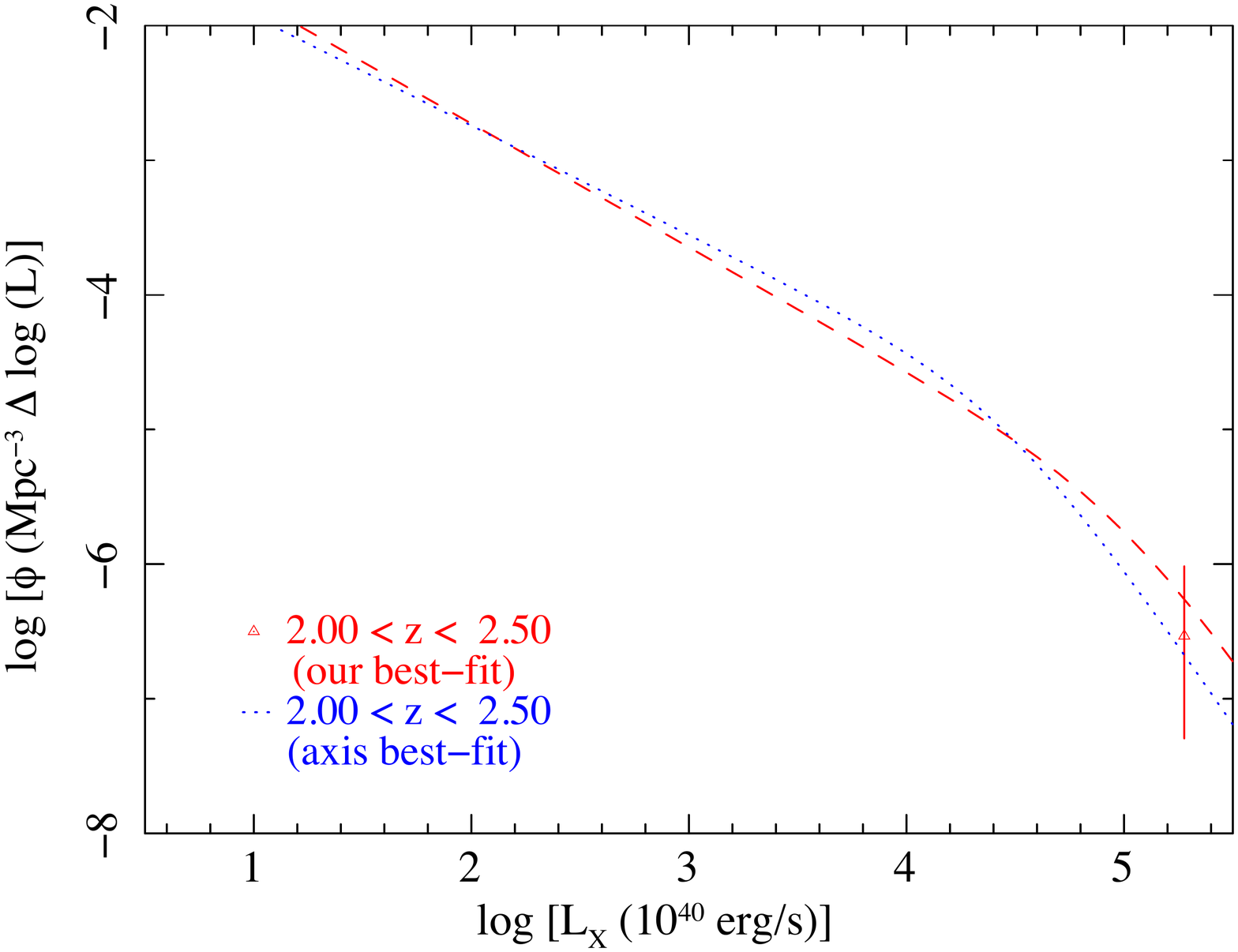}
      		\label{fig:loglumi_2000ev_8000ev_z5}
    	\end{subfigure}%
    	\begin{subfigure}{0.5\textwidth}
     		\centering
      		\includegraphics[width=\linewidth, trim={0 0 3cm 2cm}, clip]{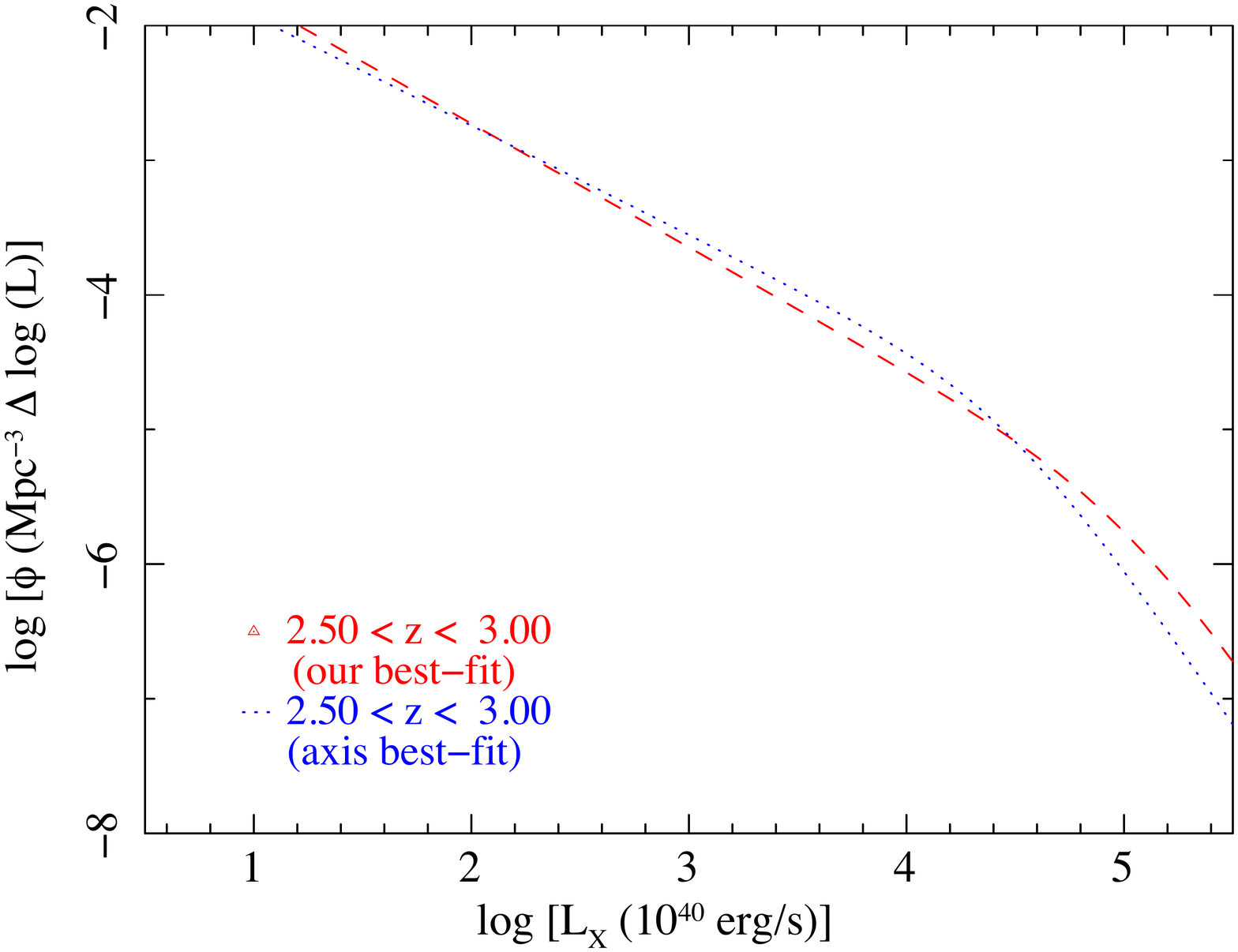}
      		\label{fig:loglumi_2000ev_8000ev_z6}
    	\end{subfigure}%
    \caption{X-ray Luminosity Functions in the fixed 2$-$8 keV observer-frame band for $0 < z < 3$. The data points represent the results from the binned XLF and the dashed lines are the curves of the analytical PLE model (eq. \ref{eq:xlf_powerlaw}). The red dashed lines are the model curves produced using our own ML best-fit parameters. The blue dotted lines are the model curves produced using the \protect\cite{ebrero2009xmm} PLE best-fit parameters, corrected from 2$-$10 keV to 2$-$8 keV.}
    \label{fig:obfsepxlfs}
    \end{figure*}

\begin{figure*}
    \centering
    	\begin{subfigure}{0.5\textwidth}
    		\centering
      		\includegraphics[width=\linewidth, trim={0 0 3cm 2cm}, clip]{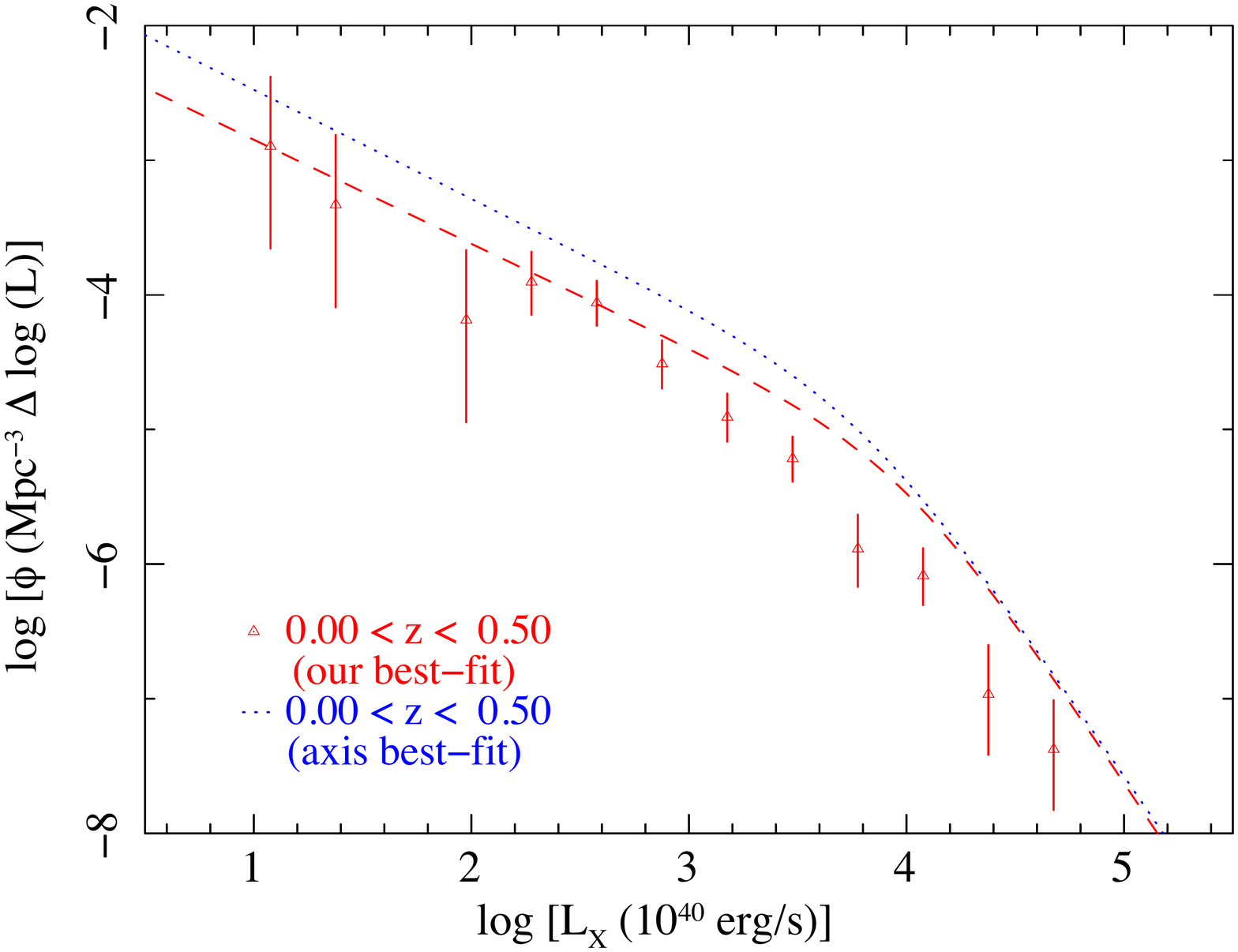}
      		\label{fig:loglumi1_ml}
    	\end{subfigure}%
    	\begin{subfigure}{0.5\textwidth}
    		\centering
      		\includegraphics[width=\linewidth, trim={0 0 3cm 2cm}, clip]{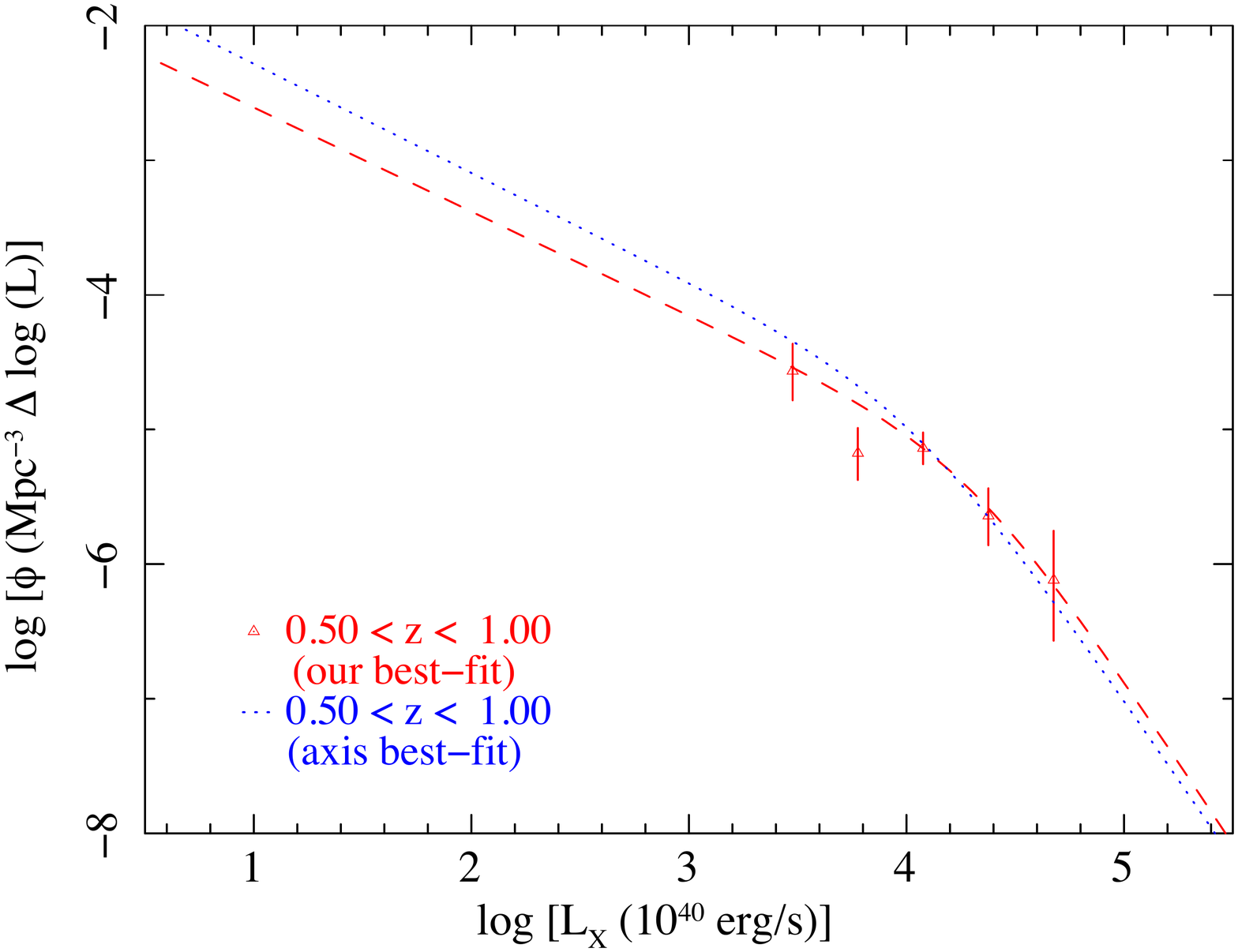}
      		\label{fig:loglumi2_ml}
    	\end{subfigure}%
    
    	\begin{subfigure}{0.5\textwidth}
     		\centering
      		\includegraphics[width=\linewidth, trim={0 0 3cm 2cm}, clip]{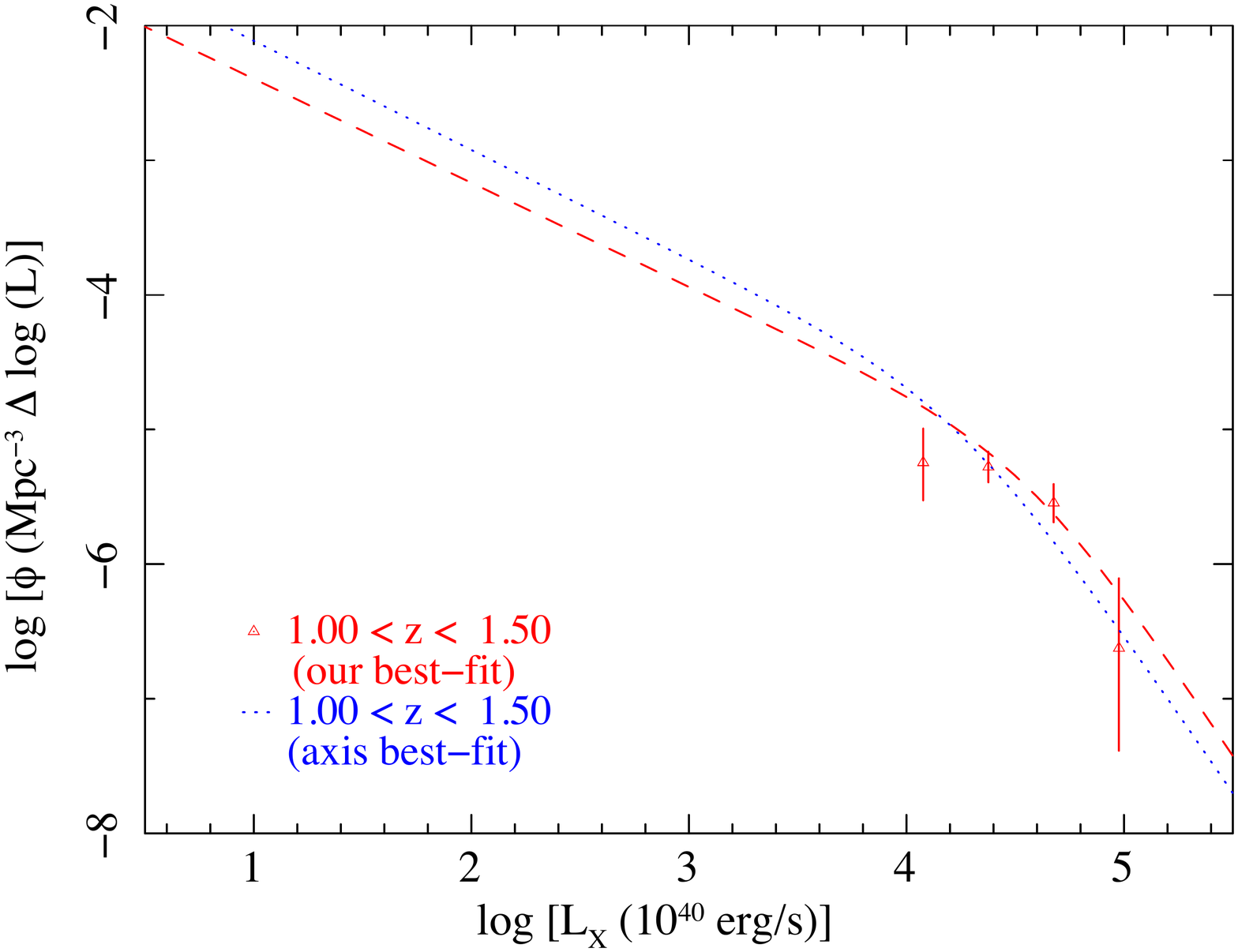}
      		\label{fig:loglumi3_ml}
    	\end{subfigure}%
    	\begin{subfigure}{0.5\textwidth}
     		\centering
      		\includegraphics[width=\linewidth, trim={0 0 3cm 2cm}, clip]{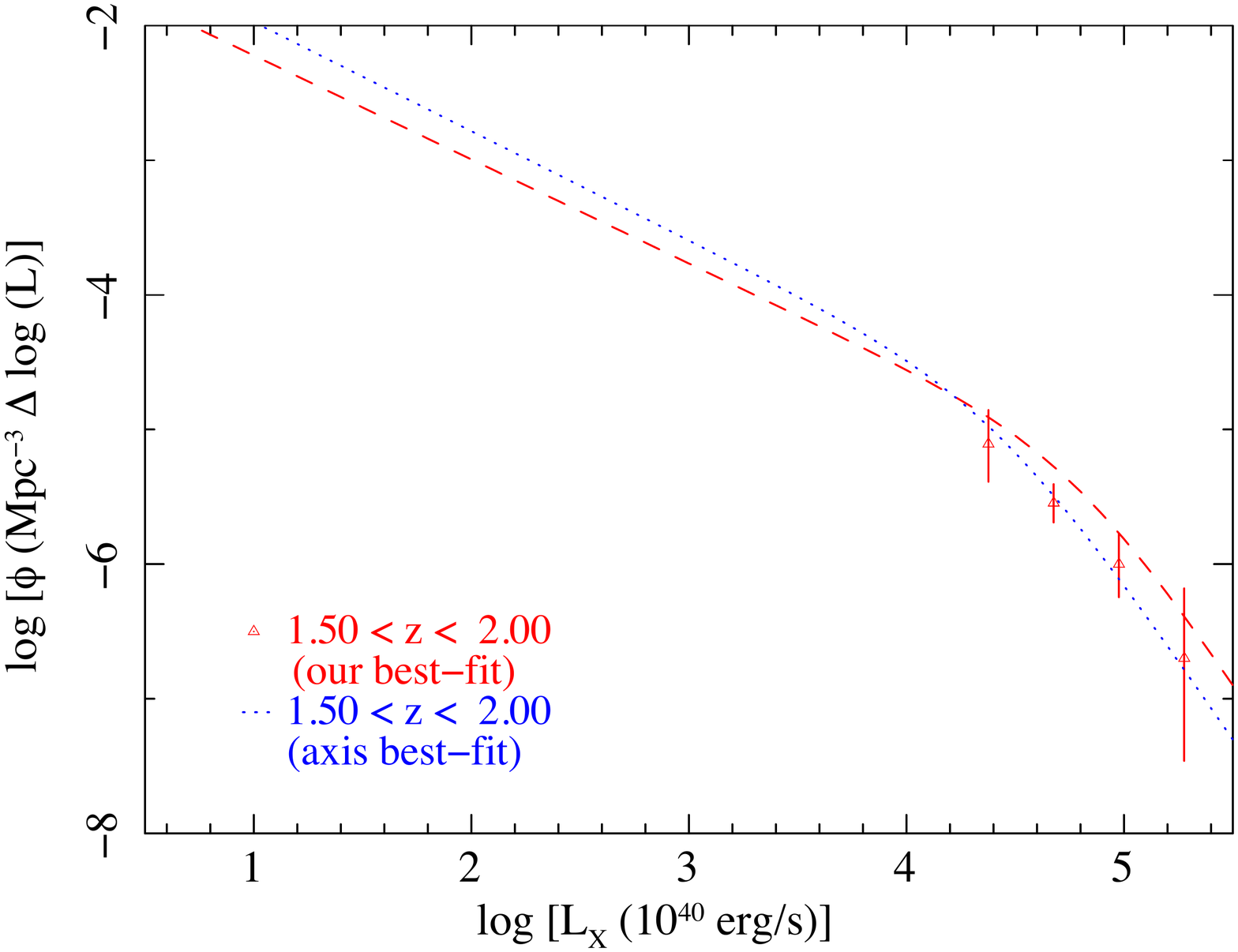}
      		\label{fig:loglumi4_ml}
    	\end{subfigure}%
    
    	\begin{subfigure}{0.5\textwidth}
     		\centering
      		\includegraphics[width=\linewidth, trim={0 0 3cm 2cm}, clip]{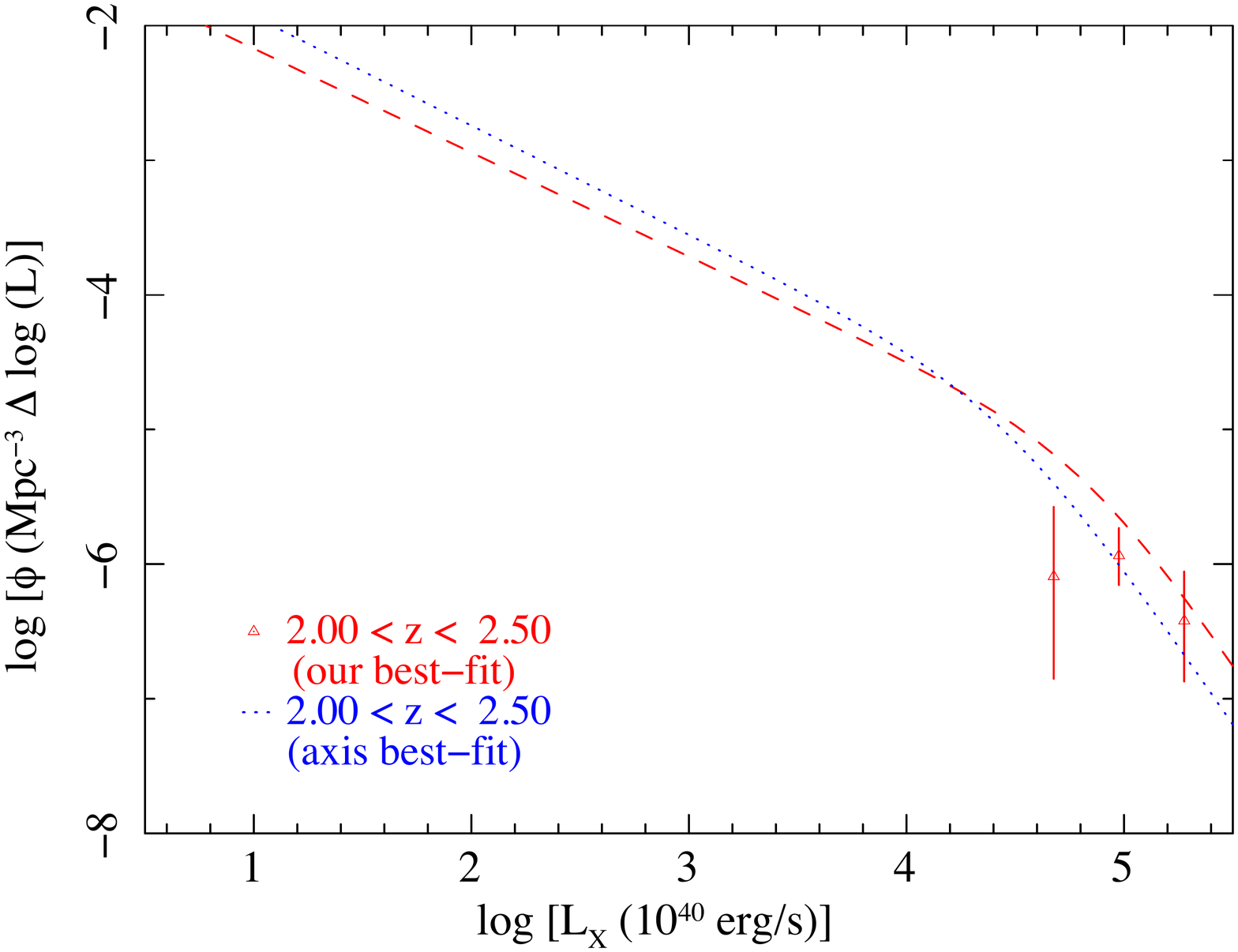}
      		\label{fig:loglumi5_ml}
    	\end{subfigure}%
    	\begin{subfigure}{0.5\textwidth}
     		\centering
      		\includegraphics[width=\linewidth, trim={0 0 3cm 2cm}, clip]{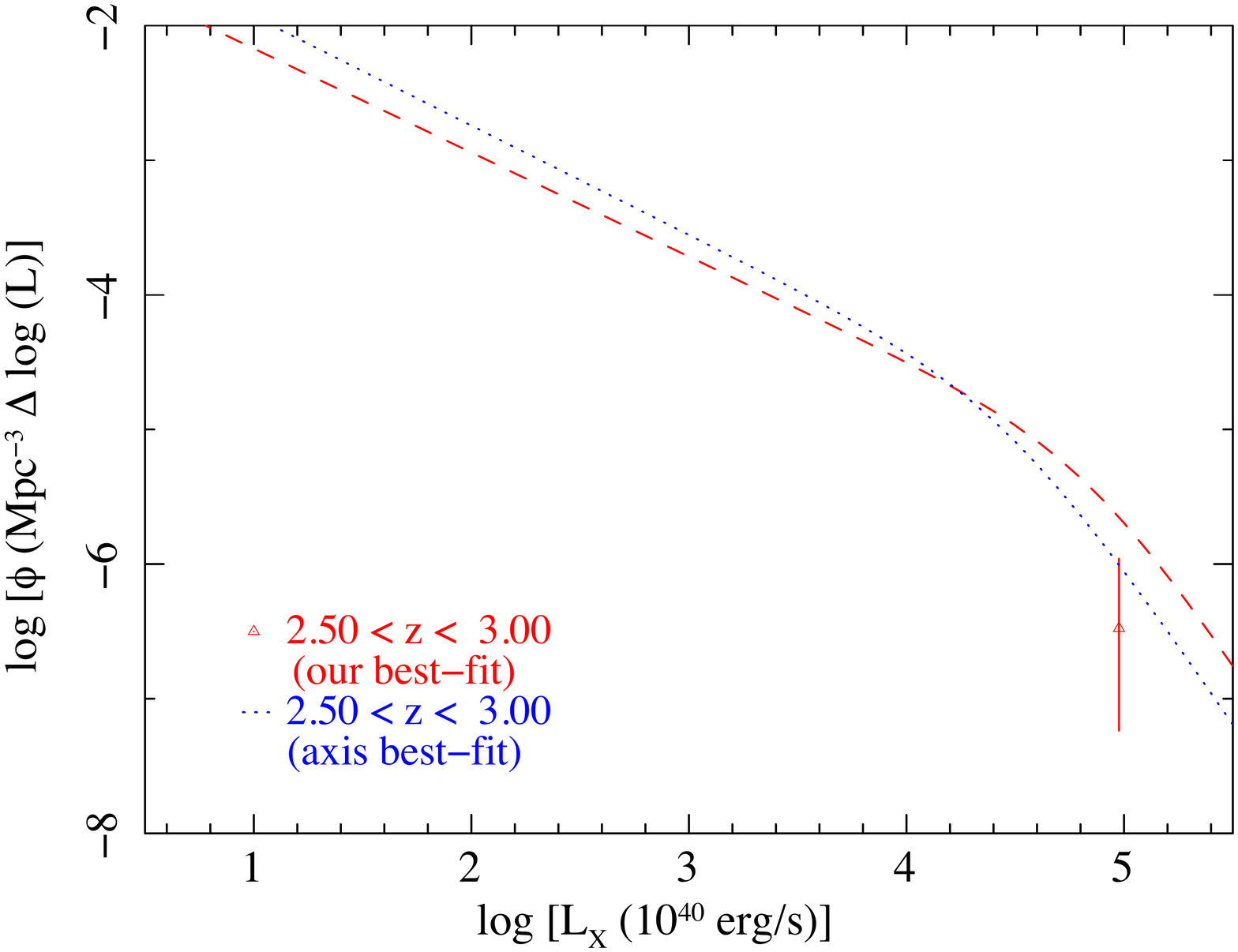}
      		\label{fig:loglumi6_ml}
    	\end{subfigure}%

    \caption{X-ray Luminosity Functions in the fixed 2$-$8 keV rest-frame band for $0 < z < 3$. The data points represent the results from the binned XLF and the dashed lines are the curves of the analytical PLE model (eq. \ref{eq:xlf_powerlaw}). The blue dotted lines are the model curves produced using the \protect\cite{ebrero2009xmm} PLE best-fit parameters, corrected from 2$-$10 keV to 2$-$8 keV. The red dashed lines are the model curves produced using our own ML best-fit parameters.}
    \label{fig:rfsepxlfs}
    \end{figure*}

Fig.~\ref{fig:obfsepxlfs} displays the fixed observer-frame XLFs for each $z$ bin in a separate plot for clarity, and we additionally plot the model curves from the \cite{ebrero2009xmm} PLE best-fit parameters for comparison (marked as magenta dashed lines), as described in Section~\ref{subsec:analyticalmodel}.
We display the same plots for clarity for the fixed rest-frame XLFs in Fig.~ \ref{fig:rfsepxlfs}.

Our ML model fitting results are obtained using the methods described in Section~\ref{subsec:mlfit}.
The results of the ML best-fit parameters for the fixed rest-frame and fixed observer-frame XLFs are summarized in Table \ref{table:mlbestfitpars}.
    
    \begin{table*}
	\caption{PLE best-fit parameters for the X-ray luminosity functions in the fixed observer-frame and fixed rest-frame 2$-$8 keV bands.}
	\centering
	    \setlength{\extrarowheight}{7pt}
    	\begin{tabular}{cccccc}
    	\hline
                        &                                 $A$ &              $\gamma_1$ &              $\gamma_2$ &            $\log_{10}(L_{0})$ &                   $p_1$ \\
                        & [10$^{-6}$ $h^{3}_{70}$ Mpc$^{-3}$] &                         &                         & [$h^{-2}_{70}$ erg s$^{-1}$] &                         \\
        \addlinespace[4pt]
        \hline
         Observer-frame &              $3.08^{+0.2}_{-0.2}$   &  $0.92^{+0.13}_{-0.11}$ &  $2.42^{+0.26}_{-0.29}$ &      $43.82^{+0.23}_{-0.29}$ &   $2.4^{+0.21}_{-0.21}$ \\
             Rest-frame &              $7.06^{+0.46}_{-0.46}$ &   $0.77^{+0.12}_{-0.1}$ &  $2.45^{+0.23}_{-0.28}$ &      $43.63^{+0.18}_{-0.22}$ &  $2.31^{+0.16}_{-0.15}$ \\
        \addlinespace[4pt]
        \hline
    	\label{table:mlbestfitpars}
    	\end{tabular}
	\end{table*}

\section{Discussion}\label{sec:discussion}

In this section, we compare the performance of our fixed rest-frame method with the standard method.
We then compare our results with the ones obtained by the AXIS survey and some other works.
We finally describe the future prospects for the fixed rest-frame method.

	\subsection{Comparison between the fixed rest-frame and the fixed observer-frame XLFs}\label{subsec:rfvsobf}  
    
    In both the fixed observer-frame and fixed rest-frame XLFs, adding the \textit{HEAO 1} data allowed us to improve the coverage of the higher luminosity sources at low redshifts. 
    This significantly improved the constraints on our best-fit parameters.
    Fig.~\ref{fig:xlfplots} shows that for both methods, the binned data points are reasonably consistent with the PLE model curves.
    
    Our ML best-fit results find that the parameters that define the evolution of the XLF ($p_1$), and the parameters that describe the shape of the XLF ($\gamma_1$ and $\gamma_2$, $\log_{10}(L_{0})$), are in agreement within the 1$\sigma$ confidence intervals for both methods.
    This means that using the new method with a fixed rest-frame band does not appear to have a significant effect on the results compared with the standard fixed observer-frame method.
    It is important to note that at high redshifts, our data set only spans high-luminosity sources.
    In comparing the two methods to each other in this work, the results suggest that the presence of absorbed AGN within the population does not have a very significant effect on the observed evolution of the XLF for the situation in which at high redshift, only high-luminosity sources are sampled.

    We have shown that it is practical to produce luminosity functions using the new method, and that it has produced results that are consistent with expectations in the luminosity-redshift regime in which we have tested it. 
    Our finding that both methods are consistent with each other, and that there appears to be no significant difference in using the new and standard methods, is consistent with the expected outcome from our data used in this paper. 
    Fig.~\ref{fig:rfsepxlfs} shows that our data are limited to bright AGN at higher redshifts. 
    The PLE model assumes that the XLF evolves only with the $\log_{10}(L_{0})$ luminosity break shifting with redshift. 
    Hence, the fitting of the XLF evolution is mainly constrained by the bright end of the XLF, while keeping the shape of the XLF fixed. 
    Even at high redshifts, the bright X-ray sources are not heavily absorbed \citep[e.g.][]{ebrero2009xmm}.
    As a consequence, performing XLF evolution studies for only high-luminosity AGN sources is less sensitive to the choice of the fixed observer-frame or rest-frame luminosity.
    If our data went to fainter fluxes at high redshift, we would be observing sources at or below the break in the luminosity function at high redshift.
    Many of those sources are expected to be absorbed \citep[e.g. see Figure 5 in][]{ebrero2009xmm}. 
    Then, the fixed rest-frame and fixed observer-frame methods are not expected to give the same results.
    We expect to see the benefits of the new method when lower luminosity AGN are included at high redshift, and the absorption distribution becomes a key factor in modelling the luminosity function.
    
    Comparing the fixed rest-frame XLF with the fixed observer-frame XLF from Fig.~ \ref{fig:xlfplots}, we also found that the error bars on the binned XLF data points are smaller in the fixed rest-frame, especially in the high-redshift bins.
    This effect is due to increased number of sources in the sample.
    At higher redshifts in the fixed rest-frame, we are sampling the softer $E_{obs}(z)$ energy bands. 
    The softer X-ray sources are easier to identify in the optical, and therefore a larger sample of them are included at higher redshifts in this study.

	\subsection{Comparison with other XLF studies}\label{subsec:otherXLFscomparison}  
    
\begin{figure*}
    \centering
    	\begin{subfigure}{0.5\textwidth}
    		\centering
      		\includegraphics[width=\linewidth, trim={0 0 3cm 2cm}, clip]{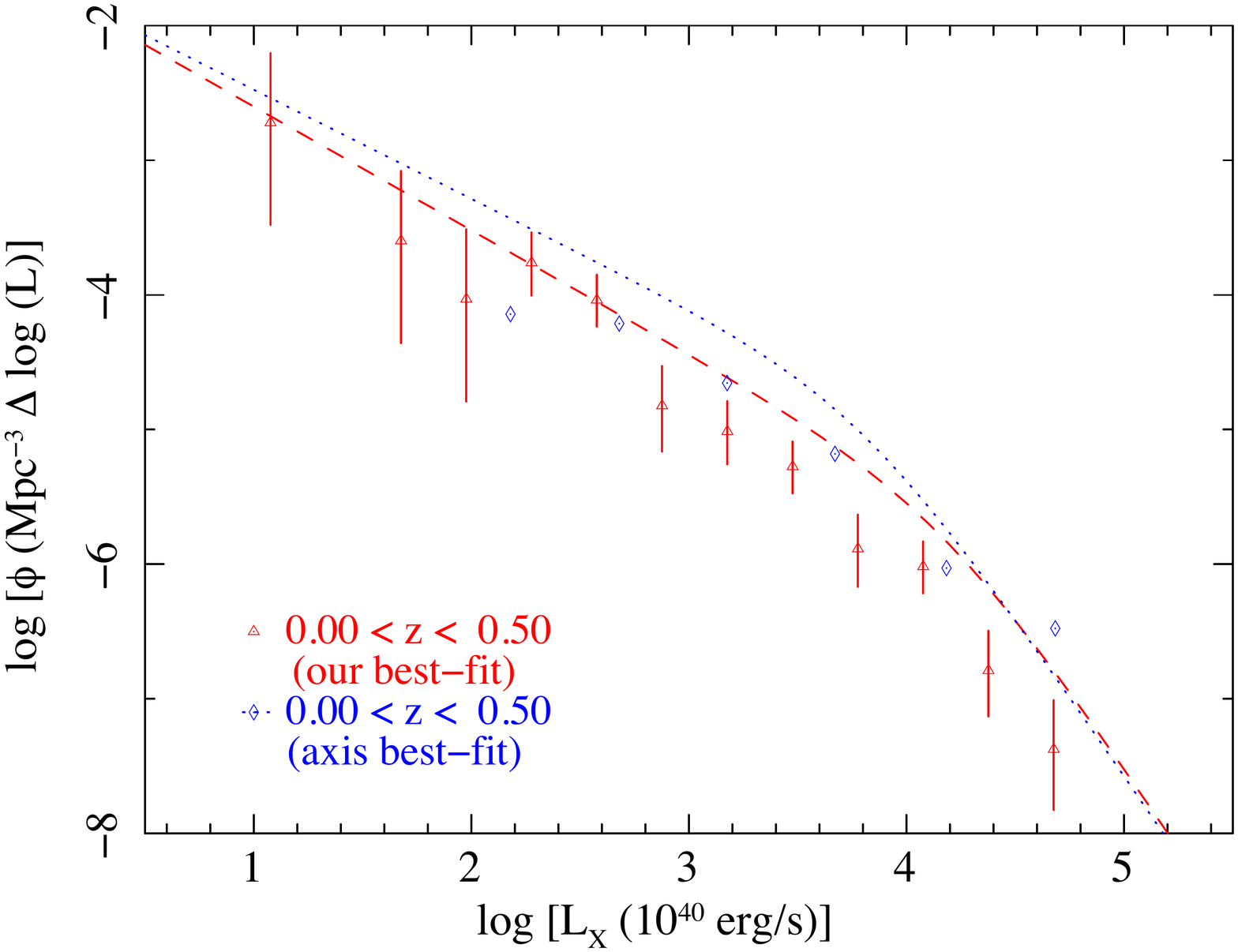}
      		\label{fig:loglumi_2000ev_8000ev_z1_axis}
    	\end{subfigure}%
    	\begin{subfigure}{0.5\textwidth}
    		\centering
      		\includegraphics[width=\linewidth, trim={0 0 3cm 2cm}, clip]{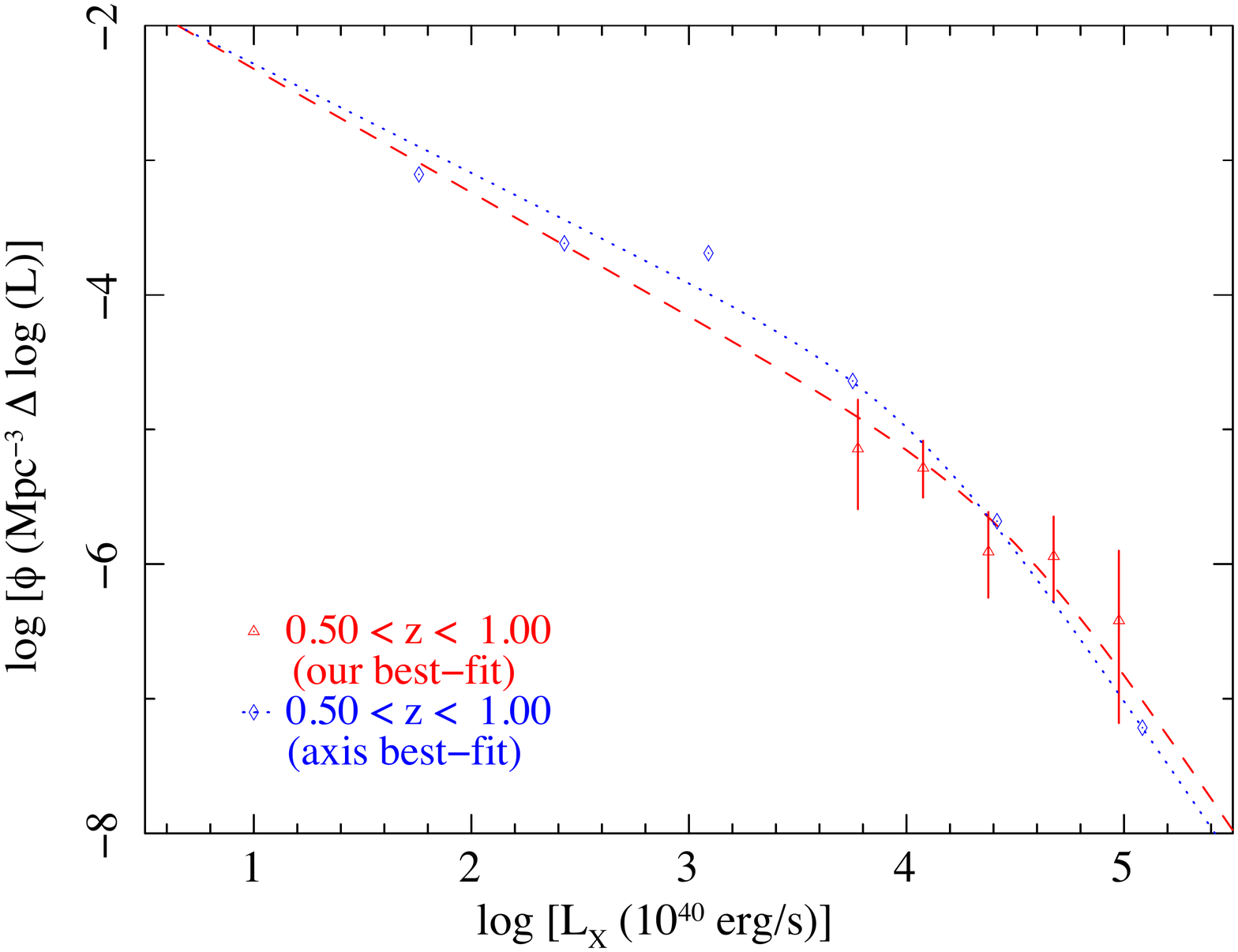}
      		\label{fig:loglumi_2000ev_8000ev_z2_axis}
    	\end{subfigure}%
    \caption{X-ray Luminosity Functions in the fixed 2$-$8 keV observer-frame band in the $0 < z < 0.5$ and $0.5 < z < 1$ bins. The red data points represent our results from the binned XLF and the dashed lines are the curves of the analytical PLE model (eq. \ref{eq:xlf_powerlaw}). The red dashed lines are the model curves produced using our own ML best-fit parameters. The blue dotted lines are the model curves produced using the \protect\cite{ebrero2009xmm} PLE best-fit parameters, converted from 2$-$10 keV to 2$-$8 keV. We also include the data points from their hard XLF in the $0 < z < 0.5$ and $0.5 < z < 1$ bins, converted to the 2$-$8 keV band.}
    \label{fig:axisobfxlfs}
    \end{figure*}

    We checked to make sure our fixed observer-frame XLF yields results that are in agreement with \cite{ebrero2009xmm}, given that we are using a subset of their data sample. 
    We find that our fixed observer-frame best-fit results (see Table \ref{table:mlbestfitpars}) are consistent with the values reported in Table 2 of \cite{ebrero2009xmm} for the PLE fit in the hard band.
    We also plot in Fig.~ \ref{fig:axisobfxlfs} the binned hard XLF data points from AXIS in the $0 < z < 0.5$ and $0.5 < z < 1$ bins in comparison with our own in the fixed observer-frame band for reference.
    As done before when reproducing their PLE curves, we converted the data points from 2$-$10 keV to 2$-$8 keV.
    We did not plot the binned XLF data points in the other redshift bins as the AXIS XLFs bin their data differently past $z = 1$.
    This plot, along with the plots in Fig.~ \ref{fig:obfsepxlfs}, also show that our fixed observer-frame PLE model results (which are in agreement with the fixed rest-frame results) are consistent with the \cite{ebrero2009xmm} curves.
    It is difficult to compare our results to more recent relevant works in XLF studies as many of them only report best-fit results for LADE or LDDE model fits.
    However, \cite{ueda2003hardxlf} construct a hard band (2-10 keV) AGN XLF and perform PLE fits, which we can compare with. 
    We find that our PLE results are consistent with the values reported in their Table 3 for the best-fit PLE parameters.
    
    Another interesting avenue we looked into was the comparison between our results and optical QSO surveys. 
    Many of these surveys, even to recent times, have used the PLE model to describe the LF evolution between $0 < z < 3$, which aligns well with the work done in this paper. 
    Optical QSO surveys rely on the detection of UV radiation from the QSOs, and so they are very sensitive to dust extinction: dusty QSOs will disappear from their samples, just as absorbed AGN will disappear from soft X-ray samples. 
    It is reasonable to question whether the evolution measured in optical surveys is affected by dust extinction, and just like in X-ray surveys, the rest-frame band is shifting with redshift (and so the effects of dust extinction affect QSOs at different redshifts differently). 
    We have introduced a new method in this paper that disentangles the effects of absorption from the measurement of evolution in the XLF. 
    Hence, we explore how the evolution we measure with our fixed rest-frame method compares to the evolution measured in optical surveys, and we do this using the quasar LF studies in \cite{ross2013quasarlf}, \cite{croom2004qso} and \cite{croom2009sdss}.
    
    The \cite{ross2013quasarlf} study uses the SDSS/BOSS DR9 data in their work, and they describe that their quasar LF is similar to the XLF.
    To compare with our results, we use the PLE best-fit values reported in Table 8 in their paper in the $0.3 < z < 2.2$ range. 
    From these best-fit results, they found a decrease in magnitude of 3.715 when looking at the evolution of the break magnitude in the LF for $0 < z < 2$.
    Converting our luminosity increase of $L_0$ to the decrease in magnitude between $0 < z < 2$, we find a decrease in magnitude by 2.67. 
    This indicates that the redshift evolution of the break luminosity for optical QSO LFs is stronger than in X-ray selected AGN LFs.
    
    The other studies of optical QSO LF also present evidence of strong evolution in the break luminosity. 
    The PLE model fitting results of the 2dF QSO redshift (2QZ) survey data in \cite{croom2004qso} show a change in magnitude of 4.35 when looking at the evolution of the peak luminosity in the LF for $0 < z < 2$.
    Similarly, the PLE results of the 2dF-SDSS LRG and QSO (2SLAQ) in \cite{croom2009sdss} indicate a 3.95 mag evolution of the peak luminosity for $0 < z < 2$ (when the sample is limited to brighter QSOs with a cut off at $-23$ mag rather than $-21.5$ mag).
    
    For the optical studies, extinction should in principle make the measured evolution smaller than the real evolution, because dust extinction gets worse further into the UV range, and so becomes worse at higher redshift.
    Hence, the optical QSO LF in these studies should suffer from the stronger absorption due to the shorter wavelength used for higher redshift samples, which would lead to a weaker evolution of the LF.
    However, we find a stronger evolution in the optical QSO LF compared to our results.
    Since our measurement is robust to the effects of absorption in the fixed rest-frame method, this indicates that the optical QSO LF has an intrinsically stronger evolution than the X-ray selected AGN LF in our work.

	\subsection{Future prospects for the fixed rest-frame method}\label{subsec:futureprospects}  

    With the current data used in this work, we are limited in our capability of solving the discrepancy on what the best model is to describe the XLF evolution.
    In that regard, this method proves to be a useful tool in checking whether this holds true when dealing with larger data sets \citep[e.g.][who include almost 3000 AGN sources in the hard band]{aird2015evolution}.
    Presently, we have not yet analyzed deep X-ray data at the faintest fluxes (e.g. from \textit{Chandra} surveys), which might be useful to fully leverage the capabilities of this method at present.
    For this to work for our new method, the X-ray data will need to be treated in an analogous fashion to the AXIS data used in our paper.
    We would need to reduce the data from scratch since we are sampling over many energy bands, and published studies do not provide X-ray fluxes for the required energy bands for each redshift bin, which would be needed to easily incorporate more surveys into this work.
    
    Within the next few years, we expect that \textit{eRosita} will have done the full survey of the X-ray sky. 
    Currently, the largest scale of observations of the whole X-ray sky comes from \textit{ROSAT}, which cannot be used for a 2$-$8 keV band XLF since it only covers the 0.1$-$2.0 keV band (hence, only covering the rest-frame 2$-$8 keV band at $z \geq 3$).
    Previous XLF studies have made use of wide-area \textit{Chandra} surveys, but these studies use low detection limits that carry only a couple of counts, which make observations in the 2$-$8 keV band difficult.
    On the other hand, the \textit{eRosita} mission will be able to provide the resolution equivalent to \textit{XMM-Newton} data, covering the full energy range from 0.5$-$8 keV over the entire sky.
    This is very promising in improving our detections of AGN and understanding their behavior over a large scale \citep{kolodzig2013erosita}.
    \textit{ATHENA} will also be able to exceed the capabilities of current X-ray observatories with its enhanced performance in X-ray spectroscopy and deep wide-field X-ray imaging.
    For this upcoming era of X-ray observatories, our new method could be applied on a huge scale.
    The potential value of our new fixed rest-frame method is greater for \textit{eRosita} and \textit{ATHENA}, than \textit{Chandra} and \textit{XMM-Newton}, because it can harness data from all-sky-shallow to very deep (and still quite wide, by today's deep survey scales).

\section{Summary and Conclusions}\label{sec:conclusions}

In this paper, we have introduced a new fixed rest-frame method for constructing X-ray luminosity functions of AGN, which aims to give a clearer description of how they evolve over cosmic time.
Our method fixes the X-ray energy band in the rest-frame by varying the observed energy band with redshift.
We tested our method against two X-ray samples in the hard band: 29 \textit{XMM-Newton} observations following the targets from the XMS survey \citep{barcons2007xmm}, and the 1st scan X-ray sources from the \textit{HEAO 1} experiment A2 survey \citep{piccinotti1982heao1xlf}.
The \textit{XMM-Newton} data were used to produce images and sourcelists in 6 X-ray energy bands, corresponding to 6 redshift ranges, to account for a target rest-frame band of 2$-$8 keV.
We used the spectroscopic optical identifications from the AXIS scheme for the redshift measurements of the \textit{XMM-Newton} data, as well as 4 extra published optical IDs of bright sources that were not included in the AXIS optical identifications.
We also used the optical identifications and redshift measurements from \cite{piccinotti1982heao1xlf} for the \textit{HEAO 1} X-ray data.

We constructed XLFs of AGN using two techniques; one using the method of \cite{page2000improved} to make a binned XLF, and one using an ML fit, which makes use of the full unbinned source sample \citep{page2021uvlf}. 
Both techniques were computed for the standard method (fixed observer-frame band) and our new method (fixed rest-frame band). 
We then used an analytical model described by a pure luminosity evolution (PLE) to fit the XLF data points. 
The model consists of a smoothly connected double power-law with a factor accounting for how it evolves with redshift.

The new method presented here eliminates the need to model the effects of intrinsic AGN absorption on the observed XLF behavior.
We were able to demonstrate the viability of this method in constructing XLFs.
We found that for both the fixed rest-frame and observer-frame methods, the binned data points are reasonably consistent with the PLE model curves.
We also find that our PLE best-fit results were consistent with \cite{ebrero2009xmm} and \cite{ueda2003hardxlf}.
Furthermore, we found an intrinsically stronger evolution in optical QSO LF studies \citep{ross2013quasarlf, croom2004qso, croom2009sdss} compared to our results of the X-ray selected AGN LF in our work.

As was shown by our comparison of the fixed observer-frame and fixed rest-frame XLFs in Section~\ref{subsec:rfvsobf}, the ML-fit results for both methods were consistent with each other (in the case where only high-luminosity sources are sampled at high redshift).
Even though the two methods produce similar results for the data that we have tested them on in our paper, we do not expect that to remain true if we were able to reach better coverage of the luminosity-redshift plane by going fainter in X-ray flux.
The power of the new method will be important if we include the fainter sources at high redshift, as the evolution of the shape of the XLF at the fainter sources is indicated in \citet{ebrero2009xmm, aird2015evolution}.
Hence, we expect to see the benefits of the new method when lower luminosity AGN are included at high redshift, and the absorption distribution becomes a key factor in modelling the XLF.
Encouraged by the success of the pilot study of this paper, this is an obvious direction of our future work, which could be applied on a huge scale with the upcoming era of X-ray observatories.

\section*{Acknowledgements}

AAQ acknowledges the funding bodies, the UAE Ministry of Presidential Affairs and the UAE Space Agency, for their support through the PhD scholarship.
MJP acknowledges support from the Science and Technology Facilities Council (STFC) grant numbers ST/N000811/1 and ST/S000216/1.
We thank the referee for their thoughtful and helpful feedback.
We thank Daisuke Kawata for his valuable comments and input on the paper.
We thank Francisco Carrera and Silvia Mateos for their help with redshifts for AXIS and BUXS sources.

\section*{Data Availability}
The data and reduction scripts used in this work can be made available upon request.



\bibliographystyle{mnras}
\bibliography{bibliography} 



\appendix
\section{Supplementary Tables}
    \begin{table*}
    \caption{List of the \textit{XMM-Newton} fields used in this work, with their general properties. This includes the \textit{XMM-Newton} observation number, the target name, the center (RA$_{Target}$,Dec$_{Target}$) and radius (R$_{Target}$) used to exclude the area around the target source with a circular region in \texttt{ds9}\protect\footnotemark, and the center (RA$_{OOT}$,Dec$_{OOT}$), angle (OOT$_{angle}$), width (OOT$_{width}$) and height (OOT$_{height}$) used to exclude the area around the OOT streaks with a rectangular region in \texttt{ds9}.}
    \centering
    \setlength{\extrarowheight}{5pt}
    \begin{tabular}{llllllllll}
        \hline
        Observation &     Target Name &          RA &         Dec &                   R &         RA &         Dec &          OOT &                  OOT&                 OOT \\
                    &                 & $_{Target}$ & $_{Target}$ &         $_{Target}$ &   $_{OOT}$ &    $_{OOT}$ &   $_{angle}$ &          $_{width}$ &         $_{height}$ \\
                    &                 &       [deg] &       [deg] & [$^{\prime\prime}$] &      [deg] &       [deg] & [$^{\circ}$] & [$^{\prime\prime}$] & [$^{\prime\prime}$] \\
        \addlinespace[4pt]
        \hline
         0012440301 &          PB5062 &  331.292917 &   -1.922328 &                 140 &   331.2614 &  -1.8289096 &        161.2 &                  40 &               765.0 \\
         0081340901 &    IRAS22491-18 &  342.955833 &  -17.873617 &                  32 &         -- &          -- &           -- &                  -- &                  -- \\
         0092850201$^a$ &    PKS 2135-147 &  324.437917 &  -14.548672 &                 120 &  324.40925 &  -14.459297 &        163.7 &                  44 &               765.0 \\
         0100240801 &          UZ LIB &    233.0975 &   -8.534811 &                 140 &  233.12562 &  -8.4441865 &         17.7 &                  40 &               765.0 \\
         0100440101 &        PHL 5200 &  337.126667 &   -5.314756 &                  16 &         -- &          -- &           -- &                  -- &                  -- \\
         0102040201 &      B2 1128+31 &      172.79 &   31.235006 &                 140 &  172.84967 &   31.315799 &         32.7 &                  44 &               765.0 \\
         0102040301 &      B2 1028+31 &  157.747083 &   31.048911 &                 140 &  157.78327 &   31.139418 &         19.0 &                  72 &               765.0 \\
         0103060101 &    PKS 2126-158 &  322.300417 &  -15.644567 &                 120 &   322.2726 &  -15.554186 &        163.5 &                  40 &               765.0 \\
         0106460101 &      Cl0939+472 &    145.7575 &   46.995658 &                 160 &         -- &          -- &           -- &                  -- &                  -- \\
         0109910101 &          A 1837 &  210.402083 &   -11.12865 &                 440 &         -- &          -- &           -- &                  -- &                  -- \\
         0111000101 &      CL 0016+16 &    4.638333 &   16.435547 &                 148 &         -- &          -- &           -- &                  -- &                  -- \\
         0111220201 &     Markarian 3 &     93.9025 &   71.037764 &                  76 &  93.816661 &   71.125597 &        162.6 &                  32 &               765.0 \\
         0112260201 & \parbox{1cm}{\vspace{1.5mm} A 399 \\ A 401 \vspace{1.5mm}} & \parbox{2cm}{\vspace{1.5mm} 44.472831 \\ 44.704176 \vspace{1.5mm}} & \parbox{2cm}{\vspace{1.5mm} 13.110249 \\ 13.489401 \vspace{1.5mm}} & \parbox{1cm}{\vspace{1.5mm} 307 \\ 405 \vspace{1.5mm}} & -- & -- &           -- &                  -- &                  -- \\ 
         0112370301 &           SDS-2 &          -- &          -- &                  -- &         -- &          -- &           -- &                  -- &                  -- \\
         0112371001 &           SDS-1 &          -- &          -- &                  -- &         -- &          -- &           -- &                  -- &                  -- \\
         0112620101 &     S5 0836+716 &   130.35125 &   70.894739 &                 160 &  130.23089 &   70.805634 &         24.0 &                  52 &               765.0 \\
         0112650401 &   G133-69 Pos$\_$1 &       -- &          -- &                  -- &         -- &          -- &           -- &                  -- &                  -- \\
         0112650501 &   G133-69 Pos$\_$2 &       -- &          -- &                  -- &         -- &          -- &           -- &                  -- &                  -- \\
         0112880301 &          EQ Peg &  352.969583 &   19.938461 &                 160 &  352.91519 &   20.028585 &        151.5 &                  48 &               765.0 \\
         0124110101$^b$ &          Mkn205 &    185.4325 &   75.310856 &                 140 &  185.11684 &   75.258375 &         56.8 &                  36 &               765.0 \\
         0124900101 &   MS1229.2+6430 &      187.88 &    64.23835 &                 140 &  187.71887 &   64.174053 &         47.0 &                  40 &               765.0 \\
         \parbox{2cm}{\vspace{1.5mm} 0112370401 + $^c$ \\ 0112371501 \vspace{1.5mm}} &           SDS-3 &          -- &          -- &                  -- &         -- &          -- &           -- &                  -- &                  -- \\
         \parbox{2cm}{\vspace{1.5mm} 0123100101 + $^c$ \\ 0123100201 \vspace{1.5mm}} &   MS0737.9+7441 &  116.017917 &   74.565156 &                 120 &  115.96813 &   74.469032 &        188.0 &                  40 &               765.0 \\
         \parbox{2cm}{\vspace{1.5mm} 0100240101 + $^c$ \\ 0100240201 \vspace{1.5mm}} &       HD 117555 &  202.699167 &   24.230853 &                 160 &  202.72695 &   24.322476 &         18.0 &                  40 &               765.0 \\
         \parbox{2cm}{\vspace{1.5mm} 0106660101 + $^c$ \\ 0106660601 \vspace{1.5mm}} &  LBQS 2212-1759 &          -- &          -- &                  -- &         -- &          -- &           -- &                  -- &                  -- \\
        \hline
        \multicolumn{10}{l}{$^a$ Different exposures were merged within the same \textit{XMM-Newton} observation (see Table \ref{table:mergedevtlist}).}\\
        \multicolumn{10}{l}{\multirow{2}{*}{\parbox{14cm}{$^b$ Different exposures (with different frame modes) within the same \textit{XMM-Newton} observation were reduced separately and the final images were summed (see Table \ref{table:summedobs}).}}}\\
        \\
        \multicolumn{10}{l}{$^c$ Different \textit{XMM-Newton} observations were merged for the same target (see Table \ref{table:mergedobs}).}\\
        \end{tabular}
    \label{table:exclusionareas}
    \end{table*}
    
    \footnotetext{\protect\url{https://ds9.si.edu/}}

    \begin{table*}
    \caption{Pairs of EPIC exposures (marked as Exposure$_{A}$ and Exposure$_{B}$ per instrument) used when merging extra eventlists within the same \textit{XMM-Newton} observation. Instruments that only had one exposure were left blank for Exposure$_{B}$.}
    \centering
    \setlength{\extrarowheight}{7pt}
    \begin{tabular}{lllllll}
        \hline
           Observation & M1 Exposure$_{A}$ & M1 Exposure$_{B}$ & M2 Exposure$_{A}$ & M2 Exposure$_{B}$ & PN Exposure$_{A}$ & PN Exposure$_{B}$ \\
        \addlinespace[4pt]
        \hline
            0092850201 &              S001 &              U003 &              S002 &              U003 &              S003 &                -- \\
        0112370401$^a$ &              S002 &              U002 &              S003 &              U003 &              U002 &                -- \\
        0123100101$^a$ &              S002 &                -- &              S003 &                -- &              S001 &              U014 \\
        \addlinespace[4pt]
        \hline
        \multicolumn{7}{l}{$^a$ Merged with another \textit{XMM-Newton} observation (see Table \ref{table:mergedobs}).}\\
    \end{tabular}
    \label{table:mergedevtlist}
    \end{table*}

    \begin{table*}
    \caption{Two sets of properties (Set A and Set B) corresponding to different M1, M2 and PN science exposures within the same the \textit{XMM-Newton} observation (0124110101). These sets were reduced separately due to having different frame modes, which prevented us from directly merging the event lists. The final images, exposure maps and background maps from Set A and Set B were summed together before making the final sourcelist in the reduction process. The properties listed in this table include the M1, M2 and PN rate thresholds used to filter out intervals of flaring particle background rate lightcurves (produced at $E$ > 5 keV using a time bin size of 20 s). The science exposures used for each set are also listed, along with the frame mode for each EPIC instrument.}
    \centering
    \setlength{\extrarowheight}{7pt}
    \begin{tabularx}{\textwidth}{XXX}
        \hline    
                            &                Set A &          Set B \\
        \addlinespace[4pt]
        \hline    
         M1 Threshold$^{a}$ &                  1.2 &            3.7 \\
         M2 Threshold$^{a}$ &                  1.2 &            3.7 \\
         PN Threshold$^{a}$ &                 29.0 &            5.7 \\
                M1 Exposure &                 S004 &           S008 \\
                M2 Exposure &                 S005 &           S009 \\
                PN Exposure &                 S003 &           S001 \\
              M1 Frame Mode &           Full Frame &   Large Window \\
              M2 Frame Mode &           Full Frame &   Large Window \\
              PN Frame Mode &  Extended Full Frame &     Full Frame \\
        \addlinespace[4pt]
        \hline
        \multicolumn{3}{l}{$^a$ In units of count s$^{-1}$.}\\
    \end{tabularx}
    \label{table:summedobs}
    \end{table*}

    \begin{table*}
    \caption{Details regarding the 4 sets of observations that were merged together in this work. The table lists the target name of the \textit{XMM-Newton} field and the two \textit{XMM-Newton} observations that were merged together for each given target field.}
    \centering
    \setlength{\extrarowheight}{8pt}
    \begin{tabularx}{\textwidth}{XXX}
        \hline
        Merge Set &     Target Name &     Observation \\
        \addlinespace[4pt]
        \hline
          SET 1 &           SDS-3 & \parbox{2cm}{\vspace{1.5mm} 0112370401$^a$ \\ 0112371501$^b$ \vspace{1.5mm}} \\
          SET 2 &   MS0737.9+7441 & \parbox{2cm}{\vspace{1.5mm} 0123100101$^a$ \\ 0123100201$^b$ \vspace{1.5mm}} \\ 
          SET 3 &       HD 117555 & \parbox{2cm}{\vspace{1.5mm} 0100240101 \\ 0100240201$^b$ \vspace{1.5mm}} \\     
          SET 4 &  LBQS 2212-1759 & \parbox{2cm}{\vspace{1.5mm} 0106660101$^b$ \\ 0106660601 \vspace{1.5mm}} \\ 
        \hline
        \multicolumn{3}{l}{$^a$ Different exposures were merged within the same \textit{XMM-Newton} observation (see Table \ref{table:mergedevtlist}).}\\
        \multicolumn{3}{l}{$^b$ Observation chosen by AXIS for the specified target.}\\
    \end{tabularx}
    \label{table:mergedobs}
    \end{table*}


\bsp	
\label{lastpage}
\end{document}